\documentclass[12pt,a4paper] {report}

\usepackage[dvips]{graphicx}
\usepackage{psfrag}
\usepackage{fancyhdr}
\usepackage{slashed}
\usepackage{verbatim}
\usepackage{simplewick}
\usepackage{amssymb}
\usepackage{doublespace} 
\usepackage[bf,small]{caption2}
\fancypagestyle{plain}{%
\fancyhead{} 
\fancyfoot[C]{\bfseries \thepage}
}
\begin{document}

%
\bibliographystyle{unsrt}
%

\raggedbottom

\newcommand{\nc}{\newcommand}
\nc{\trace}{\mathrm{Tr}}
\nc{\realtrace}{\mathrm{Re\; Tr}}
\nc{\maxrealtrace}{\mathrm{max\, Re\; Tr}}
\nc{\ud}{\mathrm{d}}
\nc{\nn}{\nonumber}
\nc{\darrow}{\stackrel{\leftrightarrow}{\partial}}
\nc{\darrows}{\stackrel{\leftrightarrow}{\slashed{\partial}}}
\nc{\Darrows}{\stackrel{\leftrightarrow}{\slashed{D}}}

%
%
\pagenumbering{roman}

\begin{titlepage} 
\begin{center}

\vspace{\stretch{1}}
\includegraphics[scale=0.2]{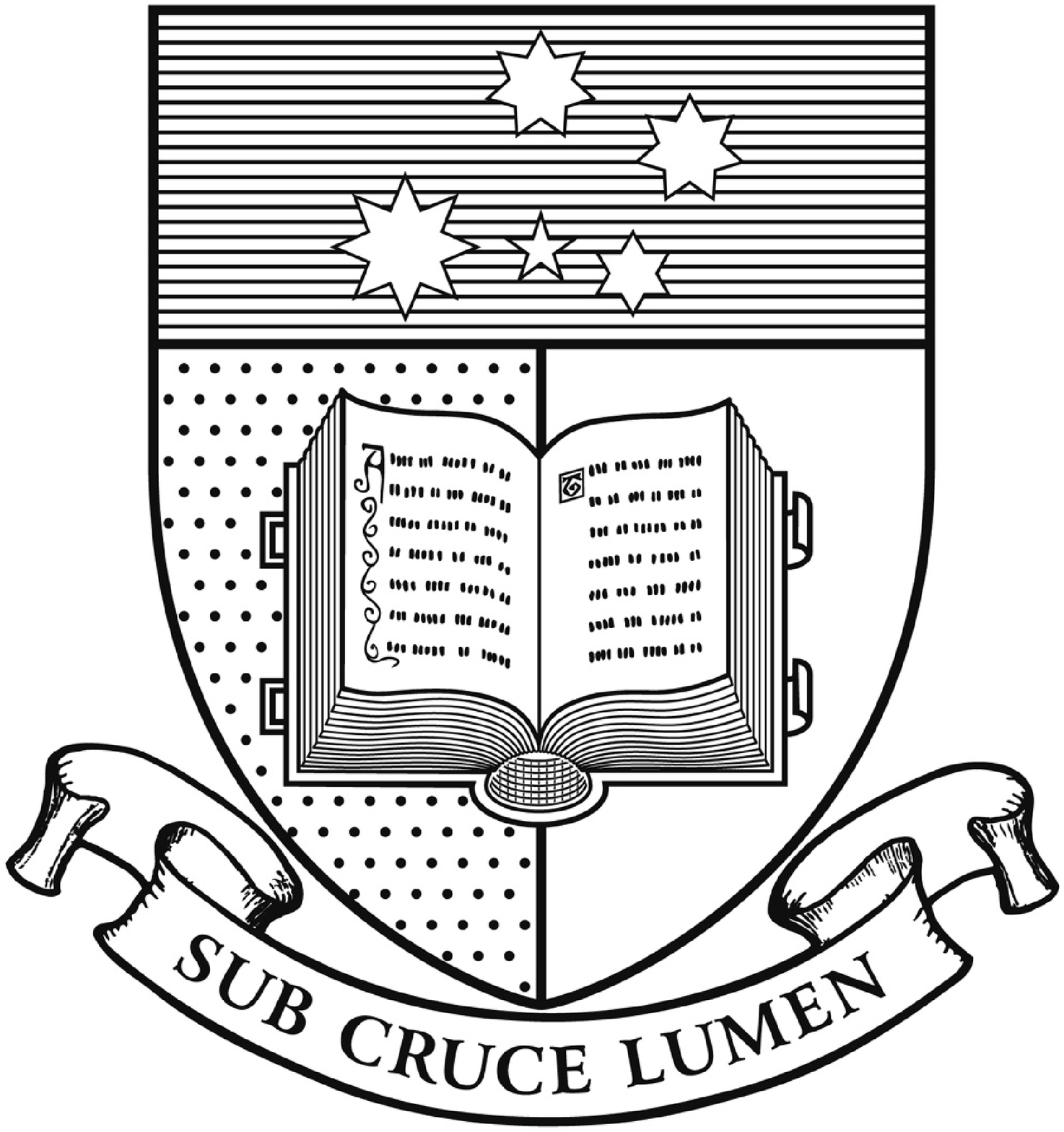} \\
{\Large \scshape The University of Adelaide} \\
{\Large \scshape School of Chemistry and Physics} \\
{\Large \scshape Discipline of Physics} \\

\vspace{\stretch{6}}
{\LARGE \bf The Vector Meson Mass in Chiral \linebreak Effective Field Theory} \\

\vspace{\stretch{1}}
{\Large Jonathan Michael MacGillivray Hall} \\

\vspace{\stretch{1}}
{\large Supervisor: Derek B. Leinweber} \\

\vspace{\stretch{12}}
{\Large  November 2007} \\

\pagebreak

\end{center} 
\end{titlepage}

\pagestyle{empty}

\begin{abstract}

A brief overview of Quantum Chromodynamics (QCD) as a  non-Abelian
gauge field theory, including symmetries and formalism of interest,
will precede a focused discussion on the use of an Effective Field 
Theory (EFT) as a low energy perturbative expansion technique. Regularization
schemes involved in Chiral Perturbation Theory ($\chi$PT)
will be reviewed and compared with EFT. Lattices will be discussed as a useful procedure for studying large mass particles.

 An Effective Field Theory will be formulated, and the self energy of the $\rho$ meson for a Finite-Range Regulated (FRR) theory will be calculated. This will be performed in both full QCD and the simpler quenched approximation (QQCD). Finite-volume artefacts, due to the finite box size on the lattice, will be quantified.

 Currently known lattice results will be used to calculate the $\rho$ meson mass, and the possibility of \emph{unquenching} will be explored. The aim of the research was to determine whether a stable unquenching procedure for the $\rho$ meson could be discovered. The results from the original research indicate that there is no such procedure because the $\rho$ mesons are unstable. Unless additional data involving lighter quark masses is available, an element of modelling is needed for successful unquenching.

\end{abstract}

\newpage

\begin{center}
\subsubsection*{Acknowledgments}
\end{center}

Thank you to Associate Professor Derek Leinweber, for his patient and intelligent supervision and his sense of humour expressed during our enjoyable discussions. I congratulate Derek Leinweber on his recent award: the AIP Boas Medal 2007. Also thank you to Doctor Ross Young for his explanations of theory and our pleasant conversations. 

Doctor Rod Crewther has been very informative in the fields of physics and education, and I appreciate our discussions. Doctor Ay\c{s}e K$\imath$z$\imath$lers\"{u} has made a valuable contribution through her lectures.

Thank you to Professor Robert Vincent and the staff of the School of Chemistry and Physics for their general assistance, and more specifically to Professor Anthony Williams and the staff of the Special Research Centre for the Subatomic Structure of Matter. I also really appreciated the daily companionship of my fellow students.

I give my thanks to my loving family for their support.

It is with hope and faith that we endeavour to extend our learning to reach new insights just beyond our present reach.

\begin{singlespace} 

\newpage
\setcounter{page}{1}
\pagestyle{plain} 
\tableofcontents

\newpage
\listoffigures  

\newpage
\listoftables

\end{singlespace} 

%
%
\newpage
\pagestyle{fancy}
\pagenumbering{arabic}
\setcounter{page}{1}

\renewcommand{\topfraction}{0.85}
\renewcommand{\textfraction}{0.1}
\renewcommand{\floatpagefraction}{0.75}

\chapter{Introduction}
\label{chpt:Introduction}

\textit{``One measure of the depth of a physical theory is the extent to which it poses serious challenges to aspects of our worldview that had previously seemed immutable.''}
(Greene, B. 1999. The Elegant Universe p.386 \cite{Greene})

\section{Prologue}
The theoretical physicist challenges previous theory, using original research that enables alternative coherence to emerge, as outlined by Bohm \cite{Bohm} (p.223). On the basis of a literature review, original research has been completed and presented in this thesis, in the theoretical framework of Quantum Chromodynamics principally using the tool: Chiral Effective Field Theory. 

 Before the body of the theory is discussed, it is necessary to consider briefly the notion of convergence. Outside the radius of convergence, an infinite series expansion of a function of some variable is not valid. Nevertheless, it is sometimes tempting to assume that an invalid result of this type is still useful, and that much important information can be gleaned from applying formulae well outside their radii of convergence. But unfortunately, such results are not quantifiably inaccurate by some absolute measure. For example, the ``approximation'' need not even yield sensible results at all. Consider this example of Guido Grandi's  binomial expansion \cite{Kline}, which could also be thought of as the sum of a geometric series outside convergence:

\begin{eqnarray}
 \frac{1}{1+x} \Bigg|_{x = 1} &=& \left (1 - x + x^2 - x^3 + \cdots \right )\Bigg|_{x = 1} \nn \\
\nn \\
\frac{1}{2} &=& 1 - 1 + 1 - 1 + \cdots \nn
\end{eqnarray}

Although odd at first sight, one could potentially find this result acceptable. Clarifying this result has often been the subject of tremendous philosophical effort by Grandi, Euler, Borel \textit{et al}. \cite{Kline}.  However, it was noted by Callet and Lagrange \cite{Lagrange} that:

\begin{eqnarray}
\frac{1 + x}{1+ x + x^2} \Bigg|_{x = 1} &=& \frac{1-x^2}{1-x^3} \quad = \quad \left \{ (1-x^2)(1 + x^3 - x^6 + \cdots)\right \} \Bigg|_{x = 1} \nn \\
\nn \\
\frac{2}{3} &=& 1 - 1 + 1 - 1 + \cdots \nn
\end{eqnarray}

Consider also that the total summation can be grouped differently to give different results.  Surely some fundamental aspect of arithmetic has been lost in this approximation; even if one decides, after much philosophizing, that one of them is correct. There is no reason that applying formulae outside their applicable zone should yield any sort of sensible result at all.

\section{Overview of Theory and Aims}

Quantum Chromodynamics (QCD) is a special kind of gauge field theory. It is similar to Quantum Electrodynamics (QED), but introduces quarks as the elementary particles, spin $1/2$ fermions, which also have the properties of flavour and colour. These enter by virtue of the ``non-Abelian'' nature of QCD. The gauge connection, which is a generalized tensor potential for a Yang-Mills field, is non-commutative.  The strictly conserved quantum number \textit{colour} is mediated by the gauge particles: gluons. The colours are conventionally called red, green and blue, but they represent the ``charge'' acted on by the strong force (there is no connection to the \emph{actual} colour). Therefore, the quarks form a representation SU$(3)_{colour}$, with eight group generators. It was necessary to suppose such an additional quantum number so that the non-integer spin quarks would obey Fermi-Dirac statistics correctly.

There are also six flavours of quark currently known, and it is seldom conceived that there should be more to be discovered, seeing that the heaviest of the flavours, $top$, is a massive $175$ GeV/$c^2$ (Table ~\ref{table:quarks}).

\begin{table}
  \newcommand\T{\rule{0pt}{2.8ex}}
  \newcommand\B{\rule[-1.4ex]{0pt}{0pt}}
  \begin{center}
    \begin{tabular}{lll}
      \hline
      \hline
      \T \B Quark Flavour & Mass(GeV/$c^2$) & Electric Charge \\
      \hline
      up \T &  0.003 & \,\,2/3 \\  
      down & 0.006 & -1/3 \\
      charm \T  & 1.3   & \,\,2/3 \\
      strange & 0.1       & -1/3 \\
      top \T & 175  & \,\,2/3 \\
      bottom \B & 4.3  & -1/3\\
      \hline
    \end{tabular}
  \end{center}
  \caption{\footnotesize{Mass and Electric Charge of Quarks in their Generations}}
  \label{table:quarks}
\end{table}

 In the discussion of QQCD in this thesis, $N_{f} = 3$ will be considered, but only $N_{f} = 2$ in the full QCD section, for simplicity.

The Nambu-Goldstone mode is a very important mechanism whereby massless particles can be created out of the vacuum by spontaneous symmetry breaking. The QCD Lagrangian has an approximate chiral symmetry associated with the fact that the masses of the $up$, $down$ (and to some extent $strange$) quarks are much less than the mass of a nucleon, so that the mass term in the QCD Lagrangian is negligible. The Goldstone bosons created for  $N_{f} = 3$ QCD are the three types of pions ($\pi^{0}$,$\pi^{+}$,$\pi^{-}$), the four kinds of kaons ($K^{0}$,$\bar{K^{0}}$,$K^{+}$,$K^{-}$), and the $\eta$ meson, which form an octet. The ninth meson is the $\eta$' and is special because it is a flavour singlet (it is also very heavy: $m_{\eta'} = 958$ MeV/$c^2$). These Goldstone bosons have mass because the chiral symmetry is only approximate, and so it is explicitly broken.

The $\rho^+$ meson is the principal subject of this discourse and it, like the pion, consists of an up quark and an anti-down quark. It has a mass of $770$ MeV/$c^2$, which is significantly larger than the pion at $140$ MeV/$c^2$. The $\rho$ meson has its constituent quarks aligned in their spin, and thus has vector properties in simple models.

When dealing with either full QCD or QQCD,  Effective Field Theory will be employed. The Chiral Lagrangian can be expanded in terms of effective degrees of freedom including all possible hadrons. The self energy of a hadron can be expressed as a polynomial expansion in the quark mass. 

When the self energy of a hadron is considered, Feynman diagrams can be drawn for any process where the hadron may transform into one or many mesons.  At higher orders in the Perturbation Theory, more complex processes can occur, though the heavy meson dressings do not contribute significantly to the overall self energy. This is because the denominators of their propagators are very large.  Therefore, they can often be neglected.

For computational reasons, it is convenient to ignore all closed loop contributions to the self energy of a hadron.  This approximation is called \emph{quenching} and can result in some bizarre behaviour. In particular to the $\rho$ meson, the self energy contribution of the $\eta'$ meson is large in quenched QCD, but small in full QCD. The $\eta'$ is the only meson which contributes to the $\rho$ self energy in QQCD.

The lattice technique will be used to obtain results through simulations. This provides a non-perturbative technique for QCD, as the lattice spacing acts as a regulator.  Lattice QCD works easily for heavy masses, and calculations are carried out from first principles in a finite volume with discrete values of momenta. Obviously for small boxes, finite volume effects can produce inaccurate results, but these effects can be quantified.

\chapter{Quantum Chromodynamics and Symmetries}
\label{chpt:QCD and Symmetries}

\section{Lagrangian Formalism}
From this point onwards, it will be convenient to adopt the simplification $c = 1 = \hbar$, and the Einstein summation notation, whereby repeated indices are automatically summed.

In modern field theories, the action is defined for a Lagrangian density, with kinetic and potential field terms. For scalar quantum fields \footnote{Scalar fields take the form: $\varphi(x) = \int \! \frac{\ud^3p}{(2\pi)^3}\frac{1}{\sqrt{2\omega_{\vec{p}}}}\bigg( a_{\vec{p}}e^{-ip\cdot x} + a_{\vec{p}}^{\dagger}e^{ip\cdot x} \bigg)$\,.}, the action appears as follows:

\begin{equation}
S = \int \! \ud^4x \, \mathcal{L}(\varphi_1, \partial\varphi_1, \cdots ,\varphi_i, \partial\varphi_i, \cdots)\,,
\label{eqn:scalaraction}
\end{equation}
an integral over the Special Lorentz-invariant four-volume $d^{4}x$. Thus the Euler-Lagrange equations of motion can be defined:
\begin{equation}
\frac{\partial\mathcal{L}}{\partial\varphi_{i}} = \partial_\mu \frac{\partial\mathcal{L}}{\partial\partial_\mu \varphi_{i}}\,.
\label{eqn:eulerlagrange}
\end{equation} 

These spinless fields are interpreted as bosons, because interchangeability of the fields came out naturally from our choice of Hamiltonian in terms of creation and annihilation operators representing independent oscillators \cite{Crewther}:

\begin{eqnarray}
\mathcal{H} &=&  \frac{\partial\mathcal{L}}{\partial\partial_{0} \varphi_i} \partial_{0} \varphi_{i} - \mathcal{L} \nn \\
\nn \\
&=& \int \! \frac{\ud^3p}{(2\pi)^3}\omega_{\vec{p}}\, a_{\vec{p}}^\dagger\,  a_{\vec{p}} \quad = \quad \int \! \ud^3x :T_{00}: \,. \footnotemark
\label{eqn:scalarhamiltonian}
\end{eqnarray}

For the Stress-Energy Tensor:
\begin{equation}
T_{\mu\nu} = \frac{\partial\mathcal{L}}{\partial\partial_\mu \varphi_{i}}\partial_\nu \varphi_{i} - g_{\mu\nu}\mathcal{L}\,.
\label{eqn:stressenergy}
\end{equation}

\footnotetext{``Normal Ordered'' notation (e.g. $:T_{00}:$) is somewhat deprecated. Strictly speaking, it is redundant with the proper formalism of the renormalization of ``infinities'' in the theory. Nevertheless, it denotes that the creation operators $a^{\dagger}$ be put to the left of all annihilation operators $a$.}

 The free fields satisfy the Klein-Gordon Equation for non-interacting relativistic scalar particles:
\begin{equation}
(\Box + m^2)\, \varphi(x) = 0 \,, \footnotemark
\label{eqn:kleingordon}
\end{equation}
\footnotetext{the d'Alembertian wave operator $\Box$ is taken to be consistent with the ``particle physicist's'' choice of Minkowski Metric where the energy component is positive. $\Box \equiv \frac{\partial^2}{\partial t^2} - \nabla^2$.}
and the canonical commutators of the fields define the space propagator, often called the Pauli-Jordan Function (which contains the Bessel Function of order 1,  $J_1(x)$):

\begin{eqnarray}
\left[\varphi(x),\varphi(y) \right] &=&  \Delta(x-y\,;m) \quad = \quad D(x-y) - D(y-x) \nn \\
\nn \\
&=& -\frac{1}{2\pi} \epsilon(x_0) \left\{ \delta(x^2) - \frac{m}{2\sqrt{x^2}} \theta(x^2) J_1 (m\sqrt{x^2}) \right \}\,.
\label{eqn:bosoncanonical}
\end{eqnarray}

It is also useful to define the Feynman Propagator: 
\begin{equation}
\Delta_{F} = \int \! \frac{\ud^4p}{(2\pi)^4} \frac{i}{p^2 - m^2 + i\epsilon} e^{-ip.(x-y)}\,,
\end{equation}
which is a Green's function, \footnote{A Green's function or ``Integrating Kernel'' $G(x,y)$ is defined with a differential operator $D_{x}G(x,y) = \delta(x-y)$. To calculate some function $u(x) = \int \! f(x) \ud x$, the integral can often be made tractable by observing: $\int \! D_{x}G(x,y)f(y)\ud y = f(x) = D_{x}u(x)$. Thus $u(x) = \int \! G(x,y)f(y)\ud y$.} defined using the Feynman Prescription for pole integration.  The poles at $p^0 =\pm(E_{\vec{p}} - i\epsilon) $ are displaced slightly from the real axis as per the Feynman Prescription \cite{P&S}.

Fermions satisfy the Dirac Equation, which is consistent with the Klein-Gordon Equation, but was derived from considering a Special Relativistic evolution equation for spinor fields $\psi$:

\begin{equation}
(i \gamma^\mu \partial_\mu - m)\,\psi(x) = 0\,.
\label{eqn:dirac}
\end{equation}

The $\gamma^\mu$ matrices form a Clifford Algebra, each of which is a $4 \times 4$ matrix containing the Pauli spin matrices (Appendix \ref{app:spin}). The Feynman slash notation also will be adopted; for example, $\gamma^\mu \partial_\mu = \slashed{\partial}$.


\section{Symmetries}

\subsection{Noether's Theorem}

In modern physics, it is important to be able to construct conserved quantities under symmetry of the action. By Hamilton's Principle of Least Action, the variation in this Lagrangian of $N$ fermion fields can be calculated:

\begin{eqnarray}
\delta S &=& 0 \nn \\
\nn \\
\Rightarrow \delta \mathcal{L} &=& \sum_{i}^{N} \left \{ \frac{\partial \mathcal{L}}{\partial \psi_{i}}\delta \psi_{i}  + \frac{\partial \mathcal{L}}{\partial\partial_{\mu} \psi_{i}} \delta \left ( \partial_{\mu}\psi_{i} \right )  + \mbox{h. c. }\right \} \quad = \quad 0\,.
\end{eqnarray}

Using the Euler-Lagrange equations of motion, together with expressions for the variations in the fields under a particular transformation, an expression of the form $\partial_{\mu} j^{\mu}$ can be found. The conserved Noether current $j^{\mu}$ has a corresponding conserved charge:

\begin{equation}
Q = \int_{\sigma} \! \ud \sigma_{\mu} j^{\mu}\,,
\end{equation}
for a space-like four-surface $\sigma$.

As an example, recall the Stress Energy Tensor in Eq. (\ref{eqn:stressenergy}), which is constructed in exactly this way. The corresponding generators are:

\begin{equation}
P_{\nu} = \int_{\sigma} \! \ud \sigma^{\mu} T_{\mu\nu}\,.
\label{eqn:conservedmtm}
\end{equation}

\section{Gauge Field Theories}

\subsection{Abelian Theories}

In Abelian theories such as QED, the free Lagrangian for fermions (Dirac particles) is invariant under a global phase transformation \cite{Kizilersu}:

\begin{eqnarray}
\mathcal{L}_{0} &=& \sum_{i}^{N} \left \{ \frac{i}{2}\bar{\psi_{i}}\darrows \psi_{i} - \bar{\psi_{i}}\,\vec{m}\,\psi_{i} \right \} \nn \\
\nn \\
 &=& \sum_{i}^{N} \left \{ \frac{i}{2}\bar{\psi'_{i}}\darrows \psi'_{i} - \bar{\psi'_{i}}\,\vec{m}\,\psi'_{i} \right \} \nn \\
\nn \\
&=& \mathcal{L}'_{0}\,, \quad \mbox{for} \quad \psi'_{i}(x) = e^{-iq\theta}\psi_{i}(x)\,,
\end{eqnarray}
where q is the eigenvalue of the generator Q of the gauge group U$(1)$, and $\vec{m}$ is a diagonal matrix of fermion masses constructing the `potential' or `mass' term of the Lagrangian.

Notice that the Lagrangian would not generally be invariant under a local gauge transformation
$\psi'_{i}(x) = e^{-iq\theta(x)}\psi_{i}(x)$, for a space-time dependent phase $\theta$. It would be convenient if this were not so, such as in Classical Electrodynamics, where there is a local gauge symmetry $\vec{A}' = \vec{A} + \nabla\chi$. Applying the idea of using a gauge field under transformation, a new vector field $A_{\mu}$ is introduced to preserve local gauge symmetry. Defining the covariant derivative $D_{\mu} = \partial_{\mu} + iqA_{\mu}$ (in the fundamental representation), such that $[D_{\mu}\psi(x)]' = e^{-iq\theta(x)}[D_{\mu}\psi(x)]$, the transformation law for the new gauge field $A$ can be defined:

\begin{equation}
\quad A'_{\mu} = A_{\mu} + \partial_{\mu}\theta(x) \,,
\end{equation}
and as a consequence, an interaction term for the Lagrangian is generated:

\begin{equation}
\mathcal{L}_{int} = -q \,\bar{\psi_{i}}\,\slashed{A}\,\psi_{i} \quad = \quad j^{\mu}A_{\mu}\,.
\end{equation}

Unfortunately, not all terms in the Lagrangian can be derived this way, so constructing a full Lagrangian for a physical theory is achieved in a more intuitive way \cite{Kizilersu}.

\subsection{Non-Abelian Theories}

In non-Abelian gauge theories such as QCD, a whole range of symmetry groups can be considered,
based on number of flavours, colours and how the quarks bond together to form hadrons. Consider that the fermion field $\psi$ in SU$_{f}(3)$ can be expressed as a spinor with entries $u$, $d$ and $s$. Yet it is also possible to express all the colours in SU$_{c}(3)$ in a very compact form \cite{Kizilersu}:

\begin{equation}
\Psi = \pmatrix{\psi_{r} \cr \psi_{g} \cr \psi_{b}}\,, \quad \psi_{c} \, = \, \pmatrix{u_{c} \cr d_{c} \cr s_{c}}\,.
\end{equation}

Because of this extra colour index, the gauge fields do not \textit{commute} (as per the name ``non-Abelian'') and so the Field Strength Tensor is defined thus:

\begin{eqnarray}
F^{\mu\nu}_{a} &=& \partial^{\mu}A^{\nu} - \partial^{\nu}A^{\mu} - g \,c_{abc}A^{\mu}_{b}A^{\nu}_{c} \nn \\
\nn \\ 
&=& -\frac{i}{g} \left[ D^{\mu}_{a}, D^{\nu}_{a} \right]\,,
\end{eqnarray}
where $g$ is the charge eigenvalue relating to the gauge transformation $U = e^{-ig\tilde{\alpha}(x)}$, and the structure constants $c_{abc}$ are equal to the Levi-Civita pseudo-tensor $\epsilon_{abc}$ in the fundamental representation of SU$(2)$, or the totally anti-symmetric constants $f_{abc}$ in SU$(3)$. The generator matrices $J_{i}^{(T)}$ for a representation $T$ are very useful for defining matrix quantities where the colour index is summed, hence the notation $\tilde{\alpha} = \alpha^{a} J_{a}^{(T)}$. Examples of generators include the Pauli matrices $J_{i} = \tau_{i} / 2$ in SU$(2)$ and the Gell-Mann matrices $J_{a} = \lambda_{a} / 2$ in SU$(3)$.  

Therefore, the many terms of the QCD Lagrangian can be expressed in neat form:

\begin{equation}
\mathcal{L}_{QCD} = \bar{\Psi}(x)\left ( i \Darrows - \vec{m} \right)\Psi(x) - \frac{1}{2}\trace{\tilde{F}_{\mu\nu}\tilde{F}^{\mu\nu}} \,.
\label{eqn:qcdlag}
\end{equation}

The non-Abelian gauge field matrix transforms as follows:

\begin{equation}
\tilde{A}_{\mu}'(x) = \tilde{A}_{\mu}(x) + \partial_{\mu}\tilde{\alpha}(x) + ig \left[\tilde{A}_{\mu}(x), \tilde{\alpha}(x) \right] \,.
\end{equation}

\subsection{(Nambu-) Goldstone's Theorem}

As can be seen in Eq. (\ref{eqn:qcdlag}), there is a so-called \textit{chiral} symmetry in $\mathcal{L}_{QCD}$ associated with the transformation $\Psi' = e^{i\theta\gamma_{5}}\Psi$, only when the masses of the matrix $\vec{m}$ are vanishingly small ($\theta$ here is a non-local phase parameter).

If this is true,  Eq. (\ref{eqn:qcdlag}) can be re-expressed as two pure helicity states (left/right handed), and separate transformations occur for each, thus forming a group in SU$(3)_L \otimes$ SU$(3)_R$:

\begin{equation}
\mathcal{L}_{H} = i \bar{\Psi}_{L} \slashed{\partial} \Psi_{L} + i \bar{\Psi}_{R} \slashed{\partial} \Psi_{R} \,.
\end{equation}

Noether's Theorem applied to this Lagrangian finds the conserved vector and axial currents \cite{Stewart}:

\begin{eqnarray}
V^{\mu}_{a} &=& \bar{\Psi} \gamma^{\mu}\frac{\lambda_{a}}{2} \Psi \,, \\
\nn \\
A^{\mu}_{a} &=& \bar{\Psi} \gamma^{\mu}\gamma_{5}\frac{\lambda_{a}}{2} \Psi \,.
\end{eqnarray}

Assuming the symmetry group SU$_{f}(3)$ means that the vector charge operators will annihilate the QCD ground state \cite{Stewart}, it follows that:

\begin{equation}
Q^{V}_{a} \mid \! 0\rangle = 0\,.
\end{equation}

But evidence shows that the dynamically broken chiral symmetry leaves the Wigner-Weyl mode (where the QCD vacuum is also chiral symmetric) unrealized \cite{Stewart}, and thus:

\begin{equation}
Q^{A}_{a} \mid \! 0\rangle \neq 0\,.
\end{equation}

Goldstone's Theorem states that spontaneously broken symmetries yield massless Goldstone pseudo-scalar bosons. In $N_{f} = 3$ QCD, the eight pseudo-scalar mesons created are the three types of pions ($\pi^{0}$,$\pi^{+}$,$\pi^{-}$), four kinds of kaons ($K^{0}$,$\bar{K^{0}}$,$K^{+}$,$K^{-}$), and the $\eta$ meson. These particles are not massless in nature, since the symmetry SU$(3)_L \otimes$ SU$(3)_R$ is explicitly broken by $m_{i} \neq 0$, which explains why flavour charge is not conserved (in for example \emph{weak} interactions) by the Adler$-$Bell$-$Jackiw Anomaly \cite{'tHooft:1972fi}. However, physically  $m_{\pi} = 140$ MeV (pion) and  $m_{p} = 940$ MeV (proton), the pion mass being almost an order of magnitude lighter.  Making the chiral approximation is therefore sometimes called the \textit{heavy baryon limit}.  The ninth meson, the $\eta'$, is special and will be discussed further in Chapter  \ref{chpt:Results}.

The state of the Goldstone bosons is defined as $\mid\!\pi_{b}(q)\rangle$, and normalized as \\
 $\langle \pi_{a}(p)\!\mid\!\pi_{b}(q)\rangle = 2 E_{p} \delta_{ab}(2\pi)^{3}\delta^{3}(\vec{p} - \vec{q})$. The divergence of the Partially Conserved Axial Current (PCAC) is \cite{Stewart}:

\begin{equation}
\partial_{\mu}A^{\mu}_{a} = i\bar{\Psi} \left \{\vec{m},\frac{\lambda_{a}}{2} \right \} \gamma_{5} \Psi\,.
\end{equation}

The eight Goldstone bosons decay to the vacuum via the axial current:

\begin{eqnarray}
\langle0 \!\mid A_{\mu}^{a}(x) \mid \!\pi^{b}(q)\rangle &=& i f_{\pi}q_{\mu}\delta^{ab}e^{-iq.x} \,;\\
\nn \\
\mbox{thus} \quad \langle0 \!\mid Q^{A}_{a}(t = 0) \mid \!\pi^{b}(q)\rangle &=& i \delta_{ab} f_{\pi}  E_{q} (2\pi)^{3} \delta^{3} (\vec{q})\,.
\end{eqnarray}

Using these equations, the Goldstone bosons can be related directly to the properties of quarks in the Gell-Mann$-$Oakes$-$Renner Relation \cite{Stewart}:

\begin{equation}
m_{\pi}^2 = - \frac{1}{2f_{\pi}^2} (m_{u} + m_{d})\langle\bar{u}u + \bar{d}d\rangle + O(m_{u,d}^2)\,.
\end{equation}

These results are vital considerations in discussions involving 
chiral symmetry. In the following chapter, Perturbation Theory in the 
chiral regime will be discussed, and Effective Field Theory as a useful scheme will be investigated.

\chapter{Chiral Effective Field Theory}
\label{chpt:EFT}

\section{Chiral Perturbation Theory}

In any Effective Field Theory, the key idea is to write down a general Lagrangian as an expansion of fields representing effective hadronic degrees of freedom, which transform under arbitrary symmetry groups \cite{Rossphd}.  As long as the symmetries of QCD (as discussed in Chapter \ref{chpt:QCD and Symmetries}) are preserved, the correct physics of QCD must be incorporated into it. All possible couplings representing interactions which preserve such symmetry must be included, and thus a low energy theory (near the chiral limit) can be constructed. This means that an appropriate \emph{counting regime} must be selected, whereby terms in the expansion are summed to a given order of the expansion scale \cite{Rossphd}.

Let the QCD Lagrangian be rewritten as the sum of a chirally symmetric part $\mathcal{L}_{0}$ (invariant in the Lie group SU$(3)_L \otimes$ SU$(3)_R$), and a symmetry breaking part $\mathcal{L}_{G}$:

\begin{equation}
\label{eqn:split}
\mathcal{L}_{QCD} = \mathcal{L}_{0} + \mathcal{L}_{G}\,.
\end{equation}
$\mathcal{L}_{G}$ is generally small and can be treated perturbatively. For Grassmann variables $\psi$ (Appendix \ref{app:spin}), and effective field sources of vectors ($v$) axial-vectors ($a$) scalars ($s$) and pseudo-scalars ($p$), the generating functional can be written as:

\begin{equation}
\label{eqn:PTfunctional}
\mathcal{Z}[v,a,s,p] = \int\! \mathcal{D}A_{\mu}\mathcal{D}\bar{\psi}\mathcal{D}\psi \exp\left\{i \! \! \int \ud^4 x \mathcal{L}_{QCD}(A_{\mu},\bar{\psi},\psi;v,a,s,p) \right\}\,, 
\end{equation}
where: 
\begin{equation}
\mathcal{D}[\psi] \equiv \lim_{N \rightarrow \infty} [d\psi_{1}]\cdots [d\psi_{N}] \,.
\end{equation}
The Lagrangian with new source terms included is \cite{Rossphd}:

\begin{equation}
\mathcal{L}_{QCD}(A_{\mu},\bar{\psi},\psi;v,a,s,p) = \mathcal{L}_{QCD}^{0} + \bar{\psi}(\gamma^{\mu}v_{\mu} + \gamma{5}\gamma^{\mu}a_{\mu} - s + i\gamma_{5} p)\psi \,.
\end{equation}
An important example of this technique is the \emph{$\sigma$-model} construction.

\subsection{The Linear $\sigma$-Model}

It is instructive to express the  $\mathcal{L}_{0}$ in terms of couplings 
between a fermion field of nucleons, $\psi = (p , n)^{T}$, a 
three-dimensional pion field $\vec{\pi} \equiv \bar{\psi}\vec{\tau}\gamma_{5}\psi$, and a scalar field $\sigma \equiv \bar{\psi}\psi$ \cite{Ross}. 

Quarks are not included as fundamental particles. Instead, these effective degrees of freedom are employed. The Lagrangian is thus:

\begin{equation}
\label{eqn:sigmalag}
\mathcal{L}_{\sigma} = i\bar{\psi}\slashed{\partial}\psi + \frac{1}{2}\partial_{\mu}\vec{\pi}\cdot\partial^{\mu}\vec{\pi} - g \bar{\psi}(\sigma - i\vec{\tau}\cdot\vec{\pi}\gamma_{5})\psi + \frac{\mu^2}{2}(\sigma^2 + \vec{\pi}^2) - \frac{\lambda}{2}{(\sigma^2 + \vec{\pi}^2)}^{2}\,,
\end{equation}
where the $\vec{\pi}$ field transforms as a rotation in isospin space \cite{Ross}:

 \begin{eqnarray}
\vec{\pi} &\rightarrow& \vec{\pi} + \vec{\alpha} \times \vec{\pi} \quad = \quad \vec{\pi} + \vec{\alpha}\,\sigma\,, \\
\sigma &\rightarrow& \sigma - \vec{\alpha} \cdot \vec{\pi}\,.
\end{eqnarray}

 It is often useful to identify the last two terms of Eq. (\ref{eqn:sigmalag}) as the potential energy term $V(\sigma,\vec{\pi})$. It can be shown that $(\sigma^2 + \vec{\pi}^2)$ is invariant under chiral transformations.

\subsection{The Non-Linear $\sigma$-Model}

The $\sigma$ mentioned above is of limited usefulness unless the expectation value is restricted to the pion decay rate $\langle\sigma\rangle = f_{\pi}$, thus forcing the $\sigma$ field (and potential $V(\sigma,\vec{\pi})$) to become infinitely massive with no excitations \cite{Ross}. Now a choice of parameterization $\vec{\phi}(x)$ can be chosen for both the $\vec{\pi}$ and $\sigma$ fields:

\begin{eqnarray}
\sigma(x) &=& f_{\pi}\cos\left( \frac{\mid\!\vec{\phi}\!\mid}{f_{\pi}}\right)\,,  \\
\nn \\
\vec{\pi}(x) &=& f_{\pi}\hat{\phi} \sin \left( \frac{\mid\!\vec{\phi}\!\mid}{f_{\pi}} \right)\,.
\end{eqnarray}

Therefore, a very important complex unitary matrix (ie. $U^{\dagger}U = \mathbb{I}$) can be constructed:

\begin{equation}
U(x) = \exp\left (i\frac{\vec{\tau}\cdot\vec{\phi(x)}}{f_{\pi}} \right)\,,
\end{equation}
where the relation $(\frac{1}{2}\trace{U^{\dagger}U} = 1)$ is chirally invariant \cite{Ross}. Ideas in $\sigma$-models have been used to construct chiral quark models such as the Cloudy Bag Model, where the internal structure of baryons is explicitly modelled \cite{Emily}. However, this model will not be discussed in this discourse.

\section{Effective Fields in the Meson Sector}

In an Effective Field Theory for mesons, a Lagrangian with effective degrees of freedom is constructed in a similar way as the $\sigma$-model, and the fields are \emph{asymptotic} (specific to some characteristic energy scale) \cite{Ross}. This formalism follows from Bernard, Kaiser and Meissner \cite{Bernard:1995dp}, based on the work of Gasser and Leutwyler \cite{Gasser:1984gg}.

An Effective Lagrangian $\mathcal{L}_{eff}$ can be split into two components just as in Eq. (\ref{eqn:split}):

\begin{equation}
\label{eqn:split2}
\mathcal{L}_{eff} = \mathcal{L}_{eff,\,0} + \mathcal{L}_{eff,\,G}\,.
\end{equation}

As before, the meson fields can be gathered up into a complex matrix:

\begin{equation}
U(x) = \exp\left (i\frac{\vec{\lambda}\cdot\vec{\pi(x)}}{f_{\pi}} \right) \in SU(3)\,,
\end{equation}
which transforms in a non-linear fashion \cite{Ross}:

\begin{equation}
U(x) \rightarrow T_{R}\,U(x)\,T_{L}^{\dagger}\,, \quad \mbox{where} \quad T_{L,R}\in SU(3)_{L,R}\,.
\end{equation}

To construct the most general chirally symmetric Lagrangian for the Meson Sector, the meson field $U(x)$ must be expanded out as a Taylor series in powers of derivatives of $U(x)$ \cite{Emily}. Hence an expansion of momenta and mass for all possible couplings can be obtained, which is suitable for the low energy region:

\begin{equation}
\label{eqn:split3}
\mathcal{L}_{eff} = \mathcal{L}_{eff}^{(2)} + \mathcal{L}_{eff}^{(4)} + \mathcal{L}_{eff}^{(6)} + \cdots\,.
\end{equation}

The leading order term, once the QCD symmetries have been included, can be written:

\begin{equation}
\mathcal{L}_{eff}^{(2)} = \frac{1}{4} f_{\pi}^2 \left(\trace{[\nabla_{\mu}U^{\dagger}\nabla^{\mu}U + 2B\vec{m}(U + U^{\dagger})]}  \right)\,, \quad \mbox{(for  constant B)}\,,\,
\end{equation}
and the symmetry breaking part of it $\mathcal{L}_{G}^{(2)}$ can be expanded into a familiar form \cite{Ross}:

\begin{eqnarray}
\mathcal{L}_{G}^{(2)} &=& (m_{u,d} + m_{s}) Bf_{\pi}^2 - \frac{1}{2}m_{u,d}\,\vec{\pi}\cdot\vec{\pi} - \frac{1}{3}(m_{u,d} + 2m_{s})\eta^2  \nn \\
&-& \frac{1}{2}(m_{u,d} + m_{s})\vec{K}\cdot\vec{K} + \cdots
\end{eqnarray}

\subsection{Regularization and Renormalization}

When a diagram of an interaction process  is calculated (using the Feynman Rules \cite{P&S}), the resulting loop integral is often divergent. In order to quantify the asymptotically divergent components of a loop integral, a regularization scheme must be chosen. 

Consider the result for the nucleon mass in $\chi$PT (Heavy Baryon Limit) to leading order \cite{Leinweber:2003dg}:

\begin{equation}
\label{eqn:mN}
m_{N} = a_0 + a_{2} m_{\pi}^2 + \chi_{\pi}I_{\pi} + O(m_{\pi}^4) \,.
\end{equation}

This is a polynomial expression in $m_{\pi}^2$ obtained by summing the Feynman diagrams in a geometric series (see Chapter \ref{chpt:Results}), but with an extra \emph{non-analytic} one-loop (pion) additive correction. This $I_{\pi}$ is the loop integral representing such a dressing, and $\chi_{\pi}$ is known experimentally. The simplest way of renormalizing this expression is to shift the coefficients $a_0$ and $a_2$ by an (infinite) amount equal to the divergent part of the integral; incorporating the ultraviolet behaviour into them, and thus rescaling  Eq. (\ref{eqn:mN}) \cite{Rossphd}. This process is used for the nucleon mass specifically by Young \cite{Young:2005tr}. The loop integral itself can be regulated in a number of ways, as long as local gauge invariance is not broken. 

\subsection{Dimensional Regularization}
Dimensional Regularization (DR), first developed by 't Hooft and Veltman \cite{'tHooft:1972fi}, is an important procedure whereby loop integrals are extrapolated to generalized fractional dimensions $\epsilon$ and shown to converge. Since there is no intrinsic scale dependence in the interaction, DR is the most suitable scheme for ``point-like'' particles \cite{Rossphd}.

Consider this undetermined divergent four-dimensional loop integral as a test subject (what it represents is not important for this discussion):

\begin{equation}
\int \! \frac{\ud^4 k}{(2\pi)^4}  \frac{1}{k^2 + m_{\pi}^2} \rightarrow \int \! \frac{\ud k}{(2\pi)^{4 - \epsilon}} \frac{k^{3-\epsilon}}{k^2 + m_{\pi}^2} \frac{(2\pi)^{2 - \epsilon/2}}{\Gamma(2 - \epsilon/2)}\,. \footnotemark
\end{equation}

\footnotetext{The function $\Gamma$ is the generalized factorial function, defined on the complex numbers $\mathbb{C}$ as: $\Gamma(z) = \int_{0}^{\infty} \! \ud s\,  e^{-s} s^{z-1}$.}

The limit as $\epsilon \rightarrow 0$ is then taken. Thus the minimal subtraction scheme of the renormalized Eq. (\ref{eqn:mN}) is recovered correctly \cite{Rossphd}.

\subsection{Finite-Range Regularization}
An alternative to DR is to introduce a functional regulator $u(k^2;\Lambda)$, which controls the divergent integral at high momentum values. 

Consider again the test subject from the previous example:

\begin{eqnarray}
\int \! \frac{\ud^4 k}{(2\pi)^4}  \frac{1}{k^2 + m_{\pi}^2} &\rightarrow& \int \! \frac{\ud^4 k}{(2\pi)^4}  \frac{u^2(k^2)}{k^2 + m_{\pi}^2}\,, \\
\nn \\
\mbox{for a dipole regulator: } u(k^2) &=& {\left(\frac{\Lambda^2}{k^2 + \Lambda^2} \right)}^2\,.
\end{eqnarray}

The aim is to find a regularization scheme which allows sensible extrapolation \emph{outside} the chiral convergence radius. The choice of parameter $\Lambda$ determines how fast the integral will now converge (it is expected that  $u(k^2;\Lambda) \rightarrow 0$ as $k \rightarrow 0$). There are two major concerns with doing this: 

(\emph{a}) Forcing the integral to converge may yield unphysical results by ignoring large momenta; and

(\emph{b}) Applying a \emph{regulator-dependent} regime outside the chiral radius is modelling, and model dependent results may not preserve the physics.

In addressing (\emph{a}), it is important to realize that allowing hard momenta to flow through the integral yields unphysical results. The high de Broglie frequency resolves the internal structure of the hadrons, which are the quarks. However, quarks are not present in the low energy $\chi$EFT Lagrangian. For (\emph{b}) it can be argued that the application of Dimensional Regularized $\chi$PT in this region is not even modelling, but wrong (as presented in the Introduction). The polynomial expansion of hadron mass is not expected to converge, and indeed it does not in $\chi$PT as mentioned by Young \cite{Young:2002ib}. The Power Counting Regime of ignoring higher order terms in the expansion totally breaks down outside the chiral radius, because series expansion is truncated without an attempt to estimate the higher order contributions, as mentioned by Leinweber \cite{Leinweber:2005cm} \cite{Leinweber:2005xz}.

It should be noted that the results of calculations using FRR are consistent with using DR (within the chiral radius). The infinite series is resummed so that leading order terms are large and the series converges, as is explicitly demonstrated with real data in Chapter \ref{chpt:Results}. The result obtained from $\chi$PT can be recovered in $\chi$EFT by taking $\Lambda$ to infinity.

Another form of FRR is the sharp cut-off form factor. Loop integrals are calculated for limits $k \rightarrow \Lambda$, and the momentum not allowed to go to infinity. Both these FRR techniques are employed in Chapter \ref{chpt:Results}.

All hadrons have meson clouds contributing to their \emph{self energy}. Since the $\rho$ meson is the principal subject of this discourse, observe that $m_{\rho} = 770$ MeV, which is almost as heavy as the proton ($m_{p} = 940$ MeV). So expressions in Chapter \ref{chpt:Results} will be evaluated for stationary $\rho$ mesons. A thorough discussion of the non-relativistic approximation is provided in Appendix \ref{app:loopint}.


\chapter{Lattice QCD}
\label{chpt:Lattice QCD}








As discussed in Chapter \ref{chpt:EFT}, a perturbative calculation is often divergent in the strong coupling region (low energy) because of quark confinement. Consider an approach to QCD at high energy/heavy masses. In QED, the interaction coupling approaches infinity at short distances, in what is called the \emph{Landau Pole}. But in QCD the opposite is true because of \emph{asymptotic freedom} in the high energy region. This means that quarks behave as non-interacting (free) particles for high energy interactions, and can thus be written as a perturbative expansion of its $n$-point Green's functions \cite{Stewart}. 

Lattice QCD provides a non-perturbative technique for QCD (applicable not just in the high energy region). It involves the construction of a finite-volume box of discrete momenta, and calculations are performed from first principles. The lattice spacing acts as regulator for the theory. 

\section{Functional Method}
Consider the generating functional technique, choosing a set of fields $\Phi (x)$ to stand in for $A_{\mu}$ ,$\bar{\psi}$ and $\psi$ fields, and integrating over all possible paths. For a Lagrangian $\mathcal{L}(\Phi, \partial^{\mu}\Phi)$, the action can be written as follows:

\begin{equation}
S[\Phi] = \int \! \ud^4 x \mathcal{L}(\Phi(x), \partial^{\mu}\Phi(x))\,,
\end{equation}
and the generating functional with source terms $J(x_{i})$ expressed in the same notation as Eq. (\ref{eqn:PTfunctional}):

\begin{equation}
\label{eqn:genfunc}
\mathcal{Z}[J(x_{i})] = \frac{1}{Z} \! \int \! \mathcal{D}[\Phi] e^{iS[\Phi] - \int\!\ud^4 x J(x_{i})\Phi(x_{i})}\,, 
\end{equation}
with normalization:
\begin{equation}
 Z = \int \! \mathcal{D}[\Phi]e^{iS[\Phi]}\,.
\end{equation}

This normalization factor is the \emph{partition function} used when calculating expectation values of observables.
The $n$-point Green's functions can then be calculated as the time-ordered vaccuum expectation values of the fields \cite{Stewart} \cite{Zuber}:

\begin{equation}
G^{(n)}(x_{1},\cdots,x_{n}) = \langle 0\!\mid\mathcal{T}[\Phi(x_{1})\cdots\Phi(x_{n})] \mid\! 0\rangle \,,
\end{equation}
up to a normalization constant. This calculation is performed by differentiating the generating functional (Eq. (\ref{eqn:genfunc})) with respect to sources $J(x_{i})$, and then setting them to zero \cite{Stewart}:

\begin{equation}
G^{(n)}(x_{1},\cdots,x_{n}) = \frac{1}{Z} \! \int \! \mathcal{D}[\Phi]\Phi_{1}\cdots\Phi_{n}e^{iS[\Phi]}\,.
\end{equation}

From this formalism, all physical observables of a system can be obtained. To evaluate expectation values $\langle\mathcal{O}\rangle$ numerically, it is common practice to remove difficulties in Minkowski space-time by an analytic continuation to imaginary Euclidean time, or a \emph{Wick Rotation}: $t \rightarrow -it$, and $S = iS_{Eucl}$ \cite{Stewart} \cite{Emily}. Now the expectation values will be numerically soluble, since the highly oscillatory behaviour of the Green's functions have been exponentially damped \cite{Stewart}. Thus:

\begin{equation}
\langle 0 \rangle = \frac{\int \! \mathcal{D}[\Phi]\mathcal{O}e^{-S_{Eucl}[\Phi]}} {\int\! \mathcal{D}[\Phi]e^{-S_{Eucl}[\Phi]}} \,,
\end{equation}
which is of the same form as the correlation function in statistical mechanics \cite{Emily}. Using the Euclidean Action, the fermionic part of the partition function can be calculated explicitly, leaving an expression in terms of a fermion correlation matrix $M$ \cite{Emily}:

\begin{equation}
Z = \int \! \mathcal{D} A_{\mu} \mathrm{det}M e^{F_{\mu\nu}F^{\mu\nu}/4}\,.
\end{equation}

The Quenched Approximation can be then summarized by setting $\mathrm{det}M$ to be constant, since the vacuum polarization effects of the QCD vacuum are suppressed in QQCD \cite{Emily}.

\section{Lattice Construction} 

A Euclidean (as opposed to Minkowski) hypercube can be constructed with length $L$ and lattice spacing $a$. The Quantum Field Theory can then be represented by the functional integrals defined on such a box \cite{Stewart}. The ideal physics is recovered in the limit as $a \rightarrow 0$ \cite{Stewart}. The momenta are discretized, and so can only take certain values in our four-box:

\begin{equation}
k_{\mu} = \frac{2\pi}{a N_{\mu}}n_{\mu}\quad \mbox{(component-wise)}\,,
\end{equation}
where $n_{\mu}$ is an integer array and $N_{\mu}$ is the number of lattice sites, such that $-N_{\mu}/2 < n_{\mu} \leq N_{\mu}/2$ \cite{Stewart}. Thus the maximum value $k$ can take is $\pi/a$. This means that the ultraviolet physics included in our lattice is entirely determined by the lattice spacing $a$, which thus acts as a regulator.

This lattice technique will be employed in Chapter \ref{chpt:Results} to solve self energy contributions to the mass of the $\rho$ meson. Since the self energies are now also discretized on the finite-volume lattice, three-dimensional loop integrals encountered (after the time component has been integrated out) can be replaced by summations over all possible momentum values using this procedure \cite{Armour:2005mk}:

\begin{equation}
\label{eqn:discr}
\int \! \ud^3k \, \,   \approx \frac{1}{L^3} \left( \frac{2 \pi}{a} \right )^3 \sum_{k_{x},k_{y},k_{z}}\,.
\end{equation}

This will be an important step when the lattice approach is used to calculate and graph the self energy $m_{\rho}$ for different values of quark mass $m_{q}$. Lattice simulations results can then be compared to infinite-volume direct integral calculations, and therefore the finite-volume effects can be quantified, which come into play in the original research in Chapter \ref{chpt:Results}.

\subsection{Applicable Region}
It is important to identify clearly the constraints of Lattice Gauge Theory, even though it is well defined over all possible lattice sizes, spacings and quark masses, and also infinitely scalable \cite{Stewart}. To avoid major finite-volume effects the lattice size should be about $2.5$ to $3.0$ fermi in length \cite{Stewart} \cite{Dong:1995ec} \cite{Fukugita:1994ba} \cite{Labrenz:1996jy} \cite{Sheikholeslami:1985ij} \cite{Leinweber:1999ig}. This is a large lattice size and computationally intense to perform calculations. The cost of calculation is proportional to the square of the lattice volume and inversely proportional to the sixth power of the lattice spacing \cite{Stewart}.  Therefore, in real simulations, finite-volume effects play a non-negligible role.

 Ideally, quarks of their correct observed physical mass could be used in lattice simulations. Quarks this light exhibit non-locality and are thus sensitive to finite-volume effects. They also critically slow down the fermion matrix inversion algorithms \cite{Stewart}. Even though lattice QCD is non-perturbative and successful for large quark masses, calculations with physical quark masses are not feasible with current computing power as of November 2007. Hence extrapolations are necessary in order to make predictions about the physical region. 

\section{Improved Actions}
When constructing an action on the lattice as originally formulated by Wilson \cite{Wilson:1974sk}, there is difficulty in implementing the fermion field due to the \emph{fermion doubling problem} \cite{Rossphd}. This problem occurs when solving the kinetic part of the fermion equations of motion on the lattice: $(i\slashed{D} -m)\psi = 0$. The covariant derivative is taken as an average (or a forward-backward average so that the result is Hermitian), and the propagator derived is of the form: $\sin(\slashed{p} + m)^{-1}$.  The correct behaviour of this Green's function is exhibited as $p \rightarrow 0$; however, as $p \rightarrow \pi$ the propagator also vanishes (at the edge of the Brillouin Zone).  Therefore for $\sin(\slashed{p}) = 0$ there are $2^d$ degenerate quarks for each flavour, which corresponds to $16$ degenerate quarks in Minkowski space. In order to amend this, Wilson introduced a five-dimensional operator which increases the mass of the doubler species proportional to lattice spacing $a$ \cite{Rossphd} \cite{Wilson:1974sk}. Note that as $a \rightarrow 0$ in the continuum limit, the Wilson term disappears and correct QCD is recovered.

The Wilson Action defines the Wilson Loop (where fermions are \emph{parallel transported} around a closed loop), which is the \emph{plaquette} from which the gauge connection can be derived \cite{Kizilersu}.  However, chiral symmetry is violated by the Wilson Action, and large scaling violations occur \cite{Rossphd}. These errors (of $\mathcal{O}(a)$) can be removed, and higher order errors ($\mathcal{O}(a^2)$) suppressed by the use of non-perturbatively \emph{improved} actions \cite{Rossphd} \cite{Luscher:1996sc} \cite{Zanotti:2001yb} \cite{Narayanan:1994gw}. In order to fix chiral symmetry, a \emph{Clover} term is often added to the action, which takes the form $\bar{\psi}F_{\mu\nu}\psi$. This is also a five-dimensional object, but as long as extra terms added into the action are polynomial in $a$, the continuum limit can be recovered.

Examples of $N_{f} = 2$ clover-improved Wilson actions appear in work carried by Eriji \cite{Ejiri:2006ft}, Maezawa \cite{Maezawa:2007fd} and Aoki \cite{Aoki:2005et}. The gauge action and the quark action are defined following Maezawa  \cite{Maezawa:2007fd}:

\begin{eqnarray}
  S   &=& S_g + S_q, \\
\nn \\
  S_g &=& 
  -{\beta}\sum_x\left(
   c_0\sum_{\mu<\nu;\mu,\nu=1}^{4}W_{\mu\nu}^{1\times1}(x) 
   +c_1\sum_{\mu\ne\nu;\mu,\nu=1}^{4}W_{\mu\nu}^{1\times2}(x)\right), \\
\nn \\
  S_q &=& \sum_{f=1,2}\sum_{x,y}\bar{\psi}_x^f D_{x,y}\psi_y^f\,,
\end{eqnarray}
where $\beta$, $c_{0}$, $c_{1}$ are constants and $K$ is the hopping parameter. $W^{1\times 1}_{\mu\nu}(x)$ and $W^{1\times 2}_{\mu\nu}(x)$ are $1\times 1$ and $1\times 2$ dimension Wilson Loops, respectively.
The lattice field strength is defined as $F_{\mu\nu} = -i/8 (f_{\mu\nu} \ f_{\mu\nu}^{\dagger})$, where $f_{\mu\nu}$ is the clover-shaped gauge-link combination. It is now chosen that:

\begin{eqnarray}
 D_{x,y} = \delta_{xy}
   -{K}\sum_{\mu}\{(1-\gamma_{\mu})U_{x,\mu}\delta_{x+\hat{\mu},y}
    &+&(1+\gamma_{\mu})U_{x,\mu}^{\dagger}\delta_{x,y+\hat{\mu}}\} \nn \\
   &-&\delta_{xy}{c_{SW}}{K}\sum_{\mu<\nu}\sigma_{\mu\nu}F_{\mu\nu}\,.
\end{eqnarray}

Thus the partition function is defined as:
\begin{equation}
Z(\beta,K,\mu) = \int \! \mathcal{D} U (\mathrm{det}M)^{N_{f}} e^{-S_{g}}\,.
\end{equation}

\subsection{Gauge Smoothing}
Short-range topological defects occur in the QCD vacuum, and it is sometimes necessary to remove them; either for topological investigation, or to reduce the exceptional configuration problem \cite{Rossphd}. This problem relates to lattice artefacts associated with instantons (topological solutions which minimize the energy), represented by eigenmodes of the Dirac operator ($\slashed{D}$), which invalidate the calculation of the fermion propagator \cite{DeGrand:1998mn}.

Gauge smoothing can be achieved in a number of ways, such as \emph{smearing} or \emph{cooling}. However, the smeared lattice link is the only type relevant in this discourse. Smearing involves an averaging procedure of neighbouring links on the lattice \cite{Leinweber:2006ug}. The original procedure called APE smearing produced \emph{fat-links}, which were awkward to use because they break SU$(3)$ symmetry \cite{DeGrand:1998mn} \cite{Peter}. A more analytic approach is called \emph{stout-link} smearing, which utilizes the exponential function in order to remain in the group \cite{Peter}.

\subsection{Fat-Link Irrelevant Clover Fermionic Actions}
Another action-improvement scheme is the Fat-Link Irrelevant Clover (FLIC) Action \cite{Zanotti:2001yb}. FLIC fermions couple the smoothed gauge configurations to the quark fields \cite{Rossphd} \cite{Leinweber:2006ug}.

As described by Boinepalli \cite{Boinepalli:2004fz}, a mean-field improved FLIC Action may take the form:

\begin{eqnarray}
S^{FL}_{SW} &=& S^{FL}_W - \frac{igC_{SW}\kappa}{2(u^{FL}_{0})^2} \bar{\psi}(x)\sigma_{\mu\nu} F_{\mu\nu}\psi(x)\,, \\
\nn \\
S^{FL}_{W} &=& \sum_{x} \bar{\psi}(x)\psi(x) + \kappa \sum_{x,\hat{\mu}}\bar{\psi}(x) \Big[ \gamma_{\mu} \Big( \frac{U_{\mu}(x)}{u_{0}}\psi(x + \hat{\mu}) \nn \\
 &-& \frac{U_{\mu}^{\dagger}(x - \hat{\mu})}{u_{0}}\psi(x - \hat{\mu}) \Big) - \Big(\frac{U^{FL}_{\mu}(x)}{u^{FL}_{0}}\psi(x + \hat{\mu}) \nn \\
&+& \frac{U^{FL\dagger}_{\mu}(x - \hat{\mu})}{u^{FL}_{0}} \psi(x - \hat{\mu}) \Big) \Big] \,,
\end{eqnarray}
where $F_{\mu\nu}$ is improved to order $\mathcal{O}(a^4)$, and $u^{FL}_{0}$ is the plaquette measure of the mean link in fat-links. $\sigma_{\mu\nu}$ is the standard Hermitian Pauli representation of the Dirac matrices (in Appendix \ref{app:spin}) and $\kappa = 1/(2m + 8)$ \cite{Boinepalli:2004fz} \cite{Kamleh:2004aw} \cite{Hedditch:2003zx}. FLIC fermions feature in lattice QQCD data by Boinepalli \cite{Boinepalli:2004fz} and Zanotti \cite{Zanotti:2001yb}, both of which are used to produce the research presented in the following chapter. 

\chapter{Results for the $\rho$ Meson}
\label{chpt:Results}

\section{Non-Analytic Loop Integral Contributions}

The formalism has now been established, and investigations can begin, in order to discover the mass of the $\rho$ meson. Because all hadrons exhibit a cloud of mesons that surround them, the state is said to be \emph{fully dressed}. Each particular process, or \emph{dressing}, can be written down as a Feynman graph. For example, in full QCD, the two main processes which contribute to the $\rho$ mass are: the $\rho$ meson transforming into a pion ($\pi^{+}$) and an $\omega$ meson and then back into a $\rho$ meson, and the $\rho$ meson transforming into two pions ($\pi^{+}$, $\pi^{0}$) and then back into a $\rho$ meson. These processes are denoted $\rho \rightarrow \pi\omega$ and $\rho \rightarrow \pi\pi$, respectively.

In the quenched theory,  all disconnected loops in Feynman graphs are omitted. The $\rho \rightarrow \pi\omega$ and $\rho \rightarrow \pi\pi$ dressings cannot
be constructed without a disconnected loop in the quark-flow diagram, and are thus omitted.
  This is simpler than the quenching of the baryon sector, whereby not
all the quark-flow diagrams contributing to a process are omitted, but rather, a selection.

The only non-trivial contribution to the $\rho$ meson in QQCD is a flavour-singlet
coupling:  the $\eta'$ meson.

\begin{figure}
\begin{minipage}[b]{0.5\linewidth} 
\centering
\includegraphics[height=55pt]{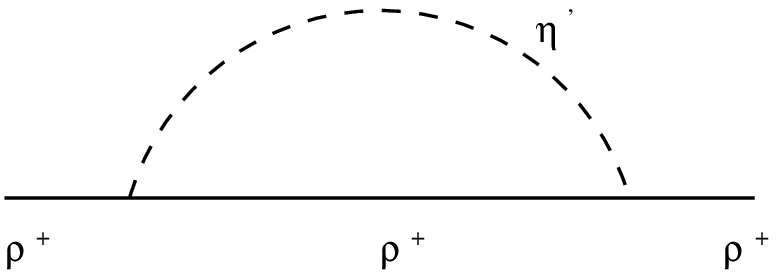}
\caption{\footnotesize{Flavour-singlet $\eta'$ contribution}}
\label{fig:simple}
\end{minipage}
\hspace{12mm}
\begin{minipage}[b]{0.5\linewidth}
\centering
\includegraphics[height=55pt]{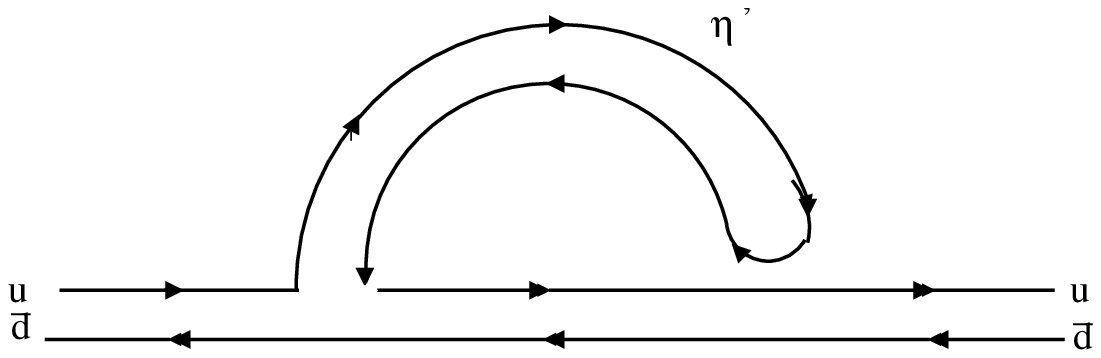}
\caption{\footnotesize{Single hairpin diagram}}
\label{fig:single2}
\end{minipage}
\end{figure}
\begin{figure}
\begin{center}
 \includegraphics[height=55pt,angle=0]{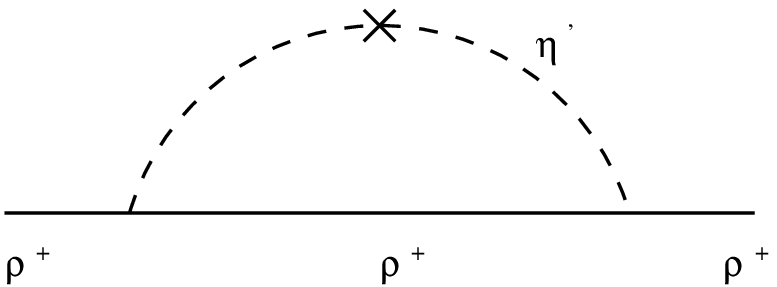}
 \caption{\footnotesize{Double hairpin $\eta'$ contribution}}
\label{fig:simplealt}
 \end{center}

\begin{minipage}[b]{0.5\linewidth} 
\centering
\includegraphics[height=55pt]{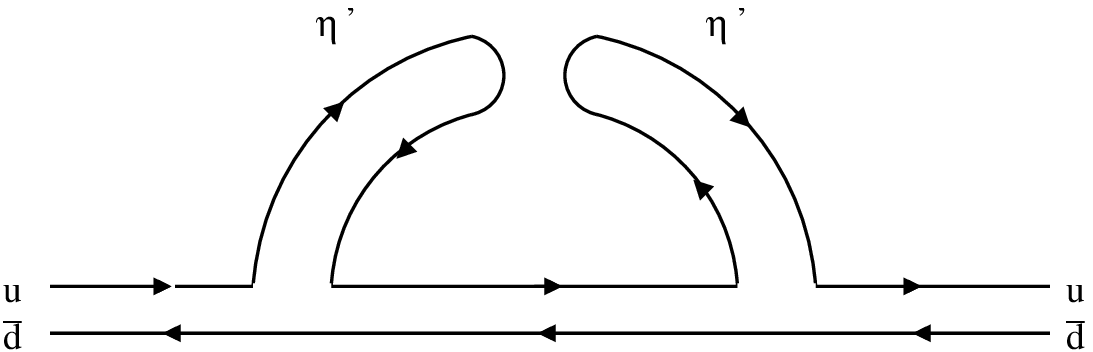}
\caption{\footnotesize{Double hairpin diagram}}
\label{fig:double}
\end{minipage}
\hspace{12mm}
\begin{minipage}[b]{0.5\linewidth}
\centering
\includegraphics[height=55pt]{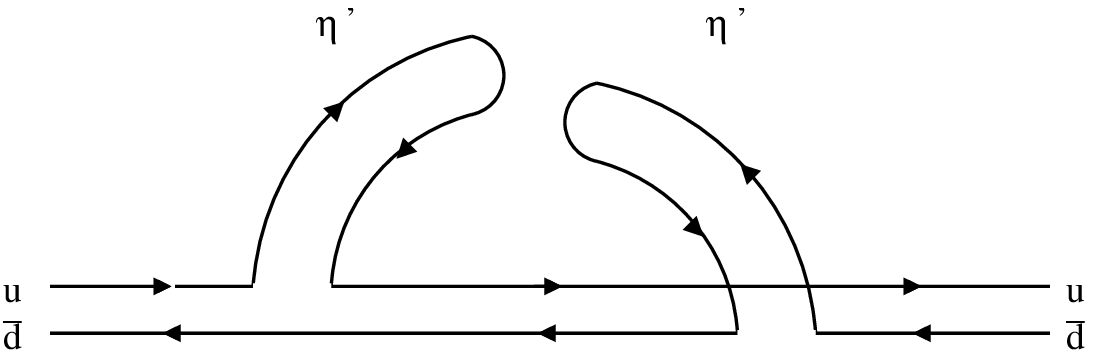}
\caption{\footnotesize{Alt. double hairpin diagram}}
\label{fig:doublealt}
\end{minipage}
\end{figure}

The contributions to the $\rho$ meson self energy in Q$\chi$PT are represented in Fig.~\ref{fig:simple} and Fig.~\ref{fig:simplealt}. The quark-flow diagrams for the single hairpin contribution are Fig.~\ref{fig:single2} and its mirror image. The quark-flow diagrams for the double hairpin contribution are Fig.~\ref{fig:double} and Fig.~\ref{fig:doublealt}.

%
Recall that the chiral Lagrangian can be expanded into terms that depend on the quark mass $m_{q}$. Therefore $m_{\rho}$ can be written as a polynomial expansion in $m_{\pi}^2$ (denoted as $P(m_{\pi}^2)$), with two non-analytic terms
representing the contributions from the two hairpin diagrams.  These terms represent the real physics from $\chi$PT \cite{Leinweber:2001ac}:

\begin{eqnarray}
  m_{\rho} &=& \chi_{1} m_{\pi} + \chi{_3} m_{\pi}^3 + P(m_{\pi}^2) \nn \\\nn \\
 &=& \chi_{1} m_{\pi} + \chi{_3} m_{\pi}^3 + (a_{0}^{\Lambda} + a_{2}^{\Lambda}m_{\pi}^2 + a_{4}^{\Lambda}m_{\pi}^4 + \cdots)  \, .
\label{eqn:massfit}
\end{eqnarray}

By calculating the loop integral expression using the Dipole FRR scheme,
the $m_{\pi}$ and $m_{\pi}^3$ coefficients suggested in \cite{Chow:1997dw} and \cite{Booth:1996hk} can be shown to be consistent choices for this regularization scheme.

\begin{figure}
\begin{center}
 \includegraphics[width=200pt,angle=0]{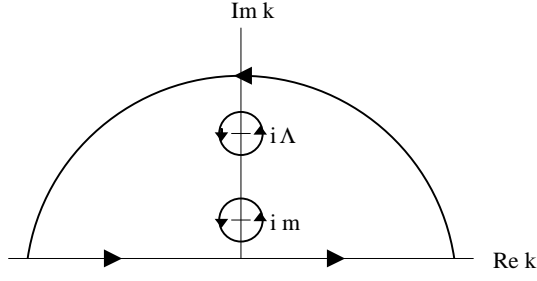}
\caption{\footnotesize{Pole contribution for closed contour}}
\label{fig:poleintegration}
\end{center}
\end{figure}

\vspace{\baselineskip}
\textbf{Original Research}
\\
For the $m_{\pi}^3$ term corresponding to the single hairpin diagram, Cauchy's Integration Formula for Derivatives is used. Two poles on the positive imaginary axis are enclosed by the contour of choice, using the Feynman Prescription, a formalism described in  \cite{P&S} and illustrated in  Fig.~\ref{fig:poleintegration}.
Choosing a dipole regulator with arbitrary parameter $\Lambda$ :

\begin{eqnarray}
 u(k^2) &=& \frac{\Lambda^4}{(k^2 + \Lambda^2)^2}\,. \\
\nn \\
\mbox{Then for} \, \,  \chi_{3} m_{\pi}^3  &:& \, \,  \frac{\chi_{3}}{2 \pi^2} \int \! \ud^3k \, \, \frac{k^2 u^2(k^2)}{k^2 + m_{\pi}^2} \nn \\
\nn \\
 &=&\frac{\chi_{3}}{\pi} \int_{-\infty}^{\infty} \! \ud k \, \, \frac{k^4 \Lambda^8}{(k - i m_{\pi})(k + i m_{\pi})(k - i \Lambda)^4(k + i \Lambda)^4} \nn \\
\nn \\
 &=&\chi_{3} \left ( \frac{m_{\pi}^3 \Lambda^8}{(\Lambda^2 - m_{\pi}^2)^4}  + \frac{2 i}{3!} f^{(3)}(i \Lambda) \right ) \\
\nn \\
&\to& \quad \chi_{3} m_{\pi}^3 \nn \\
 \mbox{as} \quad \Lambda \quad &\to& \quad \infty \, ,\nn \\
\nn \\
 \mbox{where}  \quad f(k) &=& \frac{k^4 \Lambda^8}{(k^2 + m_{\pi}^2)(k + i \Lambda)^4} \,\, ,\\
\nn \\
\mbox{and} \quad f^{(3)}(i \Lambda) &=& \frac{-3 i \Lambda^5 (m_{\pi}^6 -9 m_{\pi}^2 \Lambda^2 - 9 m_{\pi}^2 \Lambda^4 +  \Lambda^6)}{16 (m_{\pi}^2 - \Lambda^2)^4} \\
\nn \\
&\to& \quad P_{1}(m_{\pi}^2) \nn \\
 \mbox{as} \quad \Lambda \quad &\to& \quad \infty \, .\nn
\label{singpincauchy}
\end{eqnarray}
 
This polynomial expression $P_{1}(m_{\pi}^2)$ for $f^{(3)}(i \Lambda)$ can be absorbed into the coefficients of  $P(m_{\pi}^2)$, and  $P(m_{\pi}^2)$ then becomes the total analytic  contribution to the $\rho$ meson mass.

Alternatively, a sharp cutoff regulator can be realized:

\begin{eqnarray}
\frac{\chi_{3}}{2 \pi^2} \int \! \ud^3k \, \, \frac{k^2 u^2(k^2)}{k^2 + m_{\pi}^2} &=& \frac{2 \chi_{3}}{\pi} \int_{0}^{\Lambda} \! \ud k \frac{k^4 (-m_{\pi}^4 + m_{\pi}^4)}{k^2 + m_{\pi}^2} \nn \\
\nn \\
&=& \frac{2 \chi_{3}}{\pi} \left (\frac{\Lambda^3}{3} - m_{\pi}^2 \Lambda + m_{\pi}^4 \int_{0}^{\Lambda} \! \ud k \frac{1}{k^2 + m_{\pi}^2} \right ) \nn \\
\nn \\
&=& \frac{2 \chi_{3}}{\pi} \left (\frac{\Lambda^3}{3} - m_{\pi}^2 \Lambda + m_{\pi}^3 \tan^{-1}(\frac{\Lambda}{m_{\pi}}) \right )  \\
\nn \\
&\to& \quad \chi_{3} m_{\pi}^3 \nn \\ 
\mbox{as} \quad \Lambda \quad &\to& \quad \infty \, , \mbox{ (ignoring analytic terms),} \nn \\ 
\nn \\
 \mbox{where}  \quad \tan^{-1}(\frac{\Lambda}{m_{\pi}}) \quad &\sim& \quad \frac{\pi}{2} \, .\nn
\label{singpinalt}
\end{eqnarray}


Similarly, for the $m_{\pi}$ term corresponding to the double hairpin diagram:

\begin{eqnarray}
\mbox{For} \, \, \chi_{1} m_{\pi}  &:& \, \,  \frac{- \chi_{1}}{3 \pi^2} \int \! \ud^3k \, \, \frac{k^2 u^2(k^2)}{(k^2 + m_{\pi}^2)^2} \nn \\
\nn \\
 &=& \frac{-2 \chi_{1}}{3 \pi} \int_{-\infty}^{\infty} \! \ud k \, \, \frac{k^4 \Lambda^8}{(k - i m_{\pi})^2(k + i m_{\pi})^2(k - i \Lambda)^4(k + i \Lambda)^4} \nn \\
\nn \\
\label{doubpincauchy}
 &=& \frac{-2 \chi_{1}}{3 \pi} \left (2 \pi i \, g'(i m)  + \frac{2 \pi i}{3!} h^{(3)}(i \Lambda)\right )\, ,  \\
\nn \\
 \mbox{where}  \quad g(k) &=& \frac{k^4 \Lambda^8}{(k + i m_{\pi})^2(k^2 +  \Lambda^2)^4} \, \, , \\
\nn \\
  \quad h(k) &=& \frac{k^4 \Lambda^8}{(k^2 + m_{\pi}^2)^2(k + i \Lambda)^4} \, . \\
\nn \\
\nn \\
\mbox{It is found that}  \quad g'(i m) &=& \frac{- i m \Lambda^8 (5 m_{\pi}^2 + 3 \Lambda^2)}{4(m_{\pi}^2 - \Lambda^2)^5}  \\
\nn \\
&\to& \quad \frac{3 i m}{4}  \\
 \mbox{as} \quad \Lambda \quad &\to& \quad \infty \nn \, ,\\
\nn \\
\mbox{and} \quad h^{(3)}(i \Lambda) &=& \frac{3 i \Lambda^5(-m_{\pi}^6 + 15m_{\pi}^4 \Lambda^2 + 45m_{\pi}^2 \Lambda^4 + 5 \Lambda^6)}{16 (m_{\pi}^2 - \Lambda^2)^5} \\
\nn \\
&\to& \quad P_{2}(m_{\pi}^2) \nn \\
 \mbox{as} \quad \Lambda \quad &\to& \quad \infty \, .\nn \\
\nn \\
\mbox{Thus} \quad  \frac{- \chi_{1}}{3 \pi^2} \int \! &\ud^3k& \, \, \frac{k^2 u^2(k^2)}{(k^2 + m_{\pi}^2)^2} \quad \to \quad 
\chi_{1}m_{\pi} \nn \\
 \mbox{as} \quad \Lambda \quad &\to& \quad \infty \, .\nn
\end{eqnarray}

 This polynomial expression $P_{2}(m_{\pi}^2)$ for $h^{(3)}(i \Lambda)$ can also be absorbed into the coefficients of $P(m_{\pi}^2)$.

Again, a sharp cutoff regulator can be realized for this integral:

\begin{eqnarray}
\frac{-\chi_{1}}{3 \pi^2} \int \! \ud^3k \, \, \frac{k^2 u^2(k^2)}{(k^2 + m_{\pi}^2)^2} &=& \frac{4 \chi_{1}}{3 \pi} \int_{0}^{\Lambda} \! \ud k \frac{(k^2 + m_{\pi}^2)(k^2 - m_{\pi}^2) + m_{\pi}^4}{(k^2 + m_{\pi}^2)^2} \nn \\
\nn \\
&=& \frac{4 \chi_{1}}{3 \pi} \int_{0}^{\Lambda} \! \ud k \frac{k^2 - m_{\pi}^2}{k^2 + m_{\pi}^2} +  \int_{0}^{\Lambda} \! \ud k \frac{ m_{\pi}^4}{(k^2 + m_{\pi}^2)^2}\nn \\
\nn \\
&=& \frac{4 \chi_{1}}{3 \pi} \left (\Lambda + \frac{\Lambda m_{\pi}^2}{2(\Lambda^2 + m_{\pi}^2)} - \frac{3}{2} m_{\pi}\tan^{-1}(\frac{\Lambda}{m_{\pi}}) \right ) \nn \\
\nn \\
&\to& \quad \chi_{1} m_{\pi} \nn \\ 
\mbox{as} \quad \Lambda \quad &\to& \quad \infty \, ,\mbox{ (ignoring analytic terms).}
\label{doubpinalt}
\end{eqnarray}

Using these results, the key non-analytic contributions to the mass of the $\rho$ meson near the chiral limit can be included as an integral expression:

\begin{eqnarray}
  \delta m_{\rho} &=& - \frac{\chi_{1}}{3 \pi^2} \int \! \ud^3k \, \, 
\frac{k^2 u^2(k^2)}{(k^2 + m_{\pi}^2)^2} + \frac{\chi_{3}}{2 \pi^2} 
\int \! \ud^3k \, \, \frac{k^2 u^2(k^2)}{k^2 + m_{\pi}^2}\nn \\
\nn \\
&=& - \frac{4 \chi_{1}}{3 \pi} \int_{0}^{\infty} \! \ud k \, \, 
\frac{k^4 u^2(k^2)}{(k^2 + m_{\pi}^2)^2} + \frac{2 \chi_{3}}{\pi} 
\int_{0}^{\infty} \! \ud k \, \, \frac{k^4 u^2(k^2)}{k^2 + m_{\pi}^2}\,.
  \label{eqn:massintegral}
\end{eqnarray}

The three dimensional integral expression will be useful for lattice calculations, using Eq. (\ref{eqn:discr}) and the one dimensional integral expression for the infinite-volume limit.

For QQCD modelling of the $\rho$ meson, the result from Booth \textit{et al}. \cite{Booth:1996hk} will be used:

\begin{equation}
  \delta m_{\rho} =   C_{1/2}m_{\pi} + C_{3/2}m_{\pi}^3\,.
\label{eqn:deltaMQ}
\end{equation}

Thus values for the coefficients $ C_{1/2} = \chi_{1}$ and $ C_{3/2} = \chi_{3}$ are as follows:

\begin{eqnarray}
\chi_{1} &=& -\frac{g_{2}^2}{4 \pi f_{\pi}^2} \mu_{0}^2 \, , \\
\nn \\
\chi_{3} &=& -\frac{1}{12 \pi f_{\pi}^2}\left (2g_{2}(g_{1} + g_{4}) - 5 g_{2}^2 A_{0} \right ) \, .
\label{eqn:chi1chi2}
\end{eqnarray}

The coefficients $\chi_{1}$ and $\chi_{3}$ will be estimated using the convention for vector mesons \cite{Booth:1996hk}. The choice of constants will follow the convention set by \cite{Chow:1997dw}, where $A_{0} = 0.2$,  $\mu_{0} = (0.4 $ GeV$)^2$ and $g_{1}$ is initially set to zero. The constant $g_{2}$ will be taken to be $0.74$ as in \cite{Allton:2005fb}, with the first approximations of $g_{4} = 
\frac{g_{2}}{3}$. The pion coupling factor $f_{\pi}$ was taken to be 92.4 MeV as per the convention in \cite{Armour:2005mk}.

Now that the constants and coefficients have been determined, it will be highly convenient to express the perturbative expansion of the mass in terms of the $\rho$ self energies, $\Sigma^{\rho}_{Quenched}$.

The $\rho$ mass can be expressed as \cite{Leinweber:2001ac}:

\begin{equation}
m_{\rho}^2 = P(m_{\pi}^2) + \Sigma^{\rho}_{Quenched}\,.
\label{eqn:mrho}
\end{equation}
This can be shown from the Schwinger-Dyson Formula, as applied by \cite{Leinweber:1993yw}. Consider the effective Lagrangian of the form:

\begin{equation}
\mathcal{L}_{int} = -i g_{\rho\pi\pi}\rho^{\mu} \bigg(\pi \darrow_{\mu} \pi \bigg) + g_{\rho\pi\pi}^2 \pi^2 \rho^2\,.
\end{equation}
Assuming interactions occur exclusively through the $\rho$ channel, the Schwinger-Dyson equations for the $\rho$ propagator are as follows:

\begin{eqnarray}
G_{\mu\nu} &=& G_{\mu\nu}^0 + G_{\mu\sigma}^0 \,
 \Sigma^{\sigma\tau} \, G_{\tau\nu} \, ,\\
\nn \\
\mbox{where} \quad \Sigma^{\sigma\tau} &\equiv& \Sigma^{\rho} 
\left (g^{\sigma\tau} - {q^\sigma q^\tau \over q^2}  \right )\,  , \\
\nn \\
\mbox{and} \quad G_{\mu\nu}^0 &=& {-i \over q^2 - \mu_{\rho}^2 + i \epsilon}\,
 \left (g_{\mu\nu} - {q_\mu q_\nu \over q^2}  \right )  \quad \cite{Leinweber:1993yw} \,,
\label{eqn:Sch-Dy}
\end{eqnarray}
in the Landau Gauge. Thus the self energy $\Sigma^{\rho}$ is defined through the relation:

\begin{equation}
G_{\mu\nu} = {-i \over q^2 - \mu_{\rho}^2 - \Sigma^{\rho} + i
\epsilon}\, \left ( g_{\mu\nu} - {q_\mu q_\nu \over q^2} \right )\quad \cite{Leinweber:1993yw}  \, .
\label{eqnGmunu}
\end{equation}
Therefore $\mu_{\rho}^2 = \mu_{\rho}^2 + \Sigma^{\rho}$, consistent with (\ref{eqn:mrho}). The relation from \cite{Chow:1997dw} can also be obtained:
\begin{equation}
\Sigma^{\rho} \sim 2 \mu_{\rho} \delta m_{\rho} \quad \cite{Booth:1996hk}  \, .
\label{eqn:Sigmaconv}
\end{equation}

A dimensional analysis shows that $\Sigma^{\rho}$ has the dimensionality of mass squared, which is consistent with the overall expansion for $m_{\rho}^2$.

For full QCD, the results from Allton \textit{et al}. \cite{Armour:2005mk} will be used:

\begin{eqnarray}
\label{eqn:finalq}
m_{\rho}^2 &=& (a_{0}^{\Lambda} + a_{2}^{\Lambda}m_{\pi}^2 + a_{4}^{\Lambda}m_{\pi}^4 + \cdots)^2 + \Sigma^{TOT}   \\
\nn \\
&=& (a_{0}^{\Lambda} + a_{2}^{\Lambda}m_{\pi}^2 + a_{4}^{\Lambda}m_{\pi}^4 + \cdots)^2 + 2 \mu_{\rho}\delta m_{\rho}  \,, \\
\nn \\
\mbox{where } &\delta m_{\rho}& \,\, \mbox{is as (\ref{eqn:massintegral}).} \nn 
\end{eqnarray}

This means that
$\Sigma^{TOT}$ is taken to be $\Sigma^{\rho}_{\pi\pi} + \Sigma^{\rho}_{\pi\omega}$. These are the leading and next-to-leading non-analytic terms contributing to the $\rho$ meson self energy in full QCD. Processes involving other meson dressings of the $\rho$ meson do occur, but the masses of these mesons are large compared to the pion and $\omega$ meson, and therefore the propagators are suppressed due to their large denominators \cite{Stewart}.

\begin{figure}
\begin{center}
\includegraphics[height = 90pt]{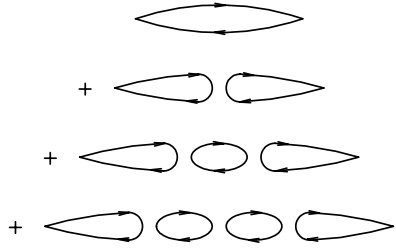}
\caption{\footnotesize{Diagrammatic  representation of $\eta'$ propagator terms}}
\label{fig:PQetaPrime}
\end{center}
\end{figure}

To demonstrate this, consider the $\eta'$ propagator in full QCD. The first two terms of the perturbative expansion represent the $\eta'$ propagator in QQCD (the diagrammatic representation is in Fig.~\ref{fig:PQetaPrime}). The terms can be summed as a geometric series and expressed in closed form, as argued in Allton \cite{Allton:2005fb}:

\begin{eqnarray}
\frac{g_2^2}{q^2 + {m_{\pi}}^2}  &-&  \frac{g_2^2 \, \mu_0^2}{\left ( q^2 + {m_{\pi}}^2 \right )^2} \left [ 1  - 
\frac{\mu_0^2}{q^2 + {m_{\pi}}^2} +\left ( \frac{\mu_0^2}{q^2 + {m_{\pi}}^2} \right )^2 - \cdots \right ] \nn \\
\nn \\
 &=& \frac{g_2^2}{q^2 + {m_{\pi}}^2} -  \frac{g_2^2 \, \mu_0^2}{ \left ( q^2 + {m_{\pi}}^2 \right )^2}  \left[1 +  \left( \frac{+\mu_{0}^2}{q^2 + {m_{\pi}}^2}\right) \right]^{-1}  \nn \\
\nn \\
&=& \frac{g_{2}^{2}}{q^2 +  m_{\pi}^2 + \mu_{0}^2}
\equiv \frac{g_{2}^{2}}{q^2 +  m_{\eta'}^2 }\, .
\end{eqnarray}

Observe that $m_{\eta'} \gg m_{\pi}$.
Therefore, the $\eta'$ meson, which contributed significantly in the quenched theory, is quite unimportant in the full theory.

\begin{figure}
\begin{minipage}[b]{0.5\linewidth} 
\centering
\includegraphics[height=75pt]{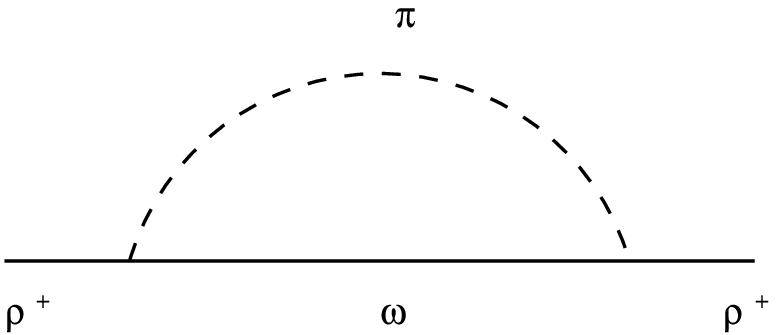}
\caption{\footnotesize{$\rho \rightarrow \pi\omega$ Feynman diagram}}
\label{fig:rpo}
\end{minipage}
\hspace{12mm}
\begin{minipage}[b]{0.5\linewidth}
\centering
\includegraphics[height=75pt]{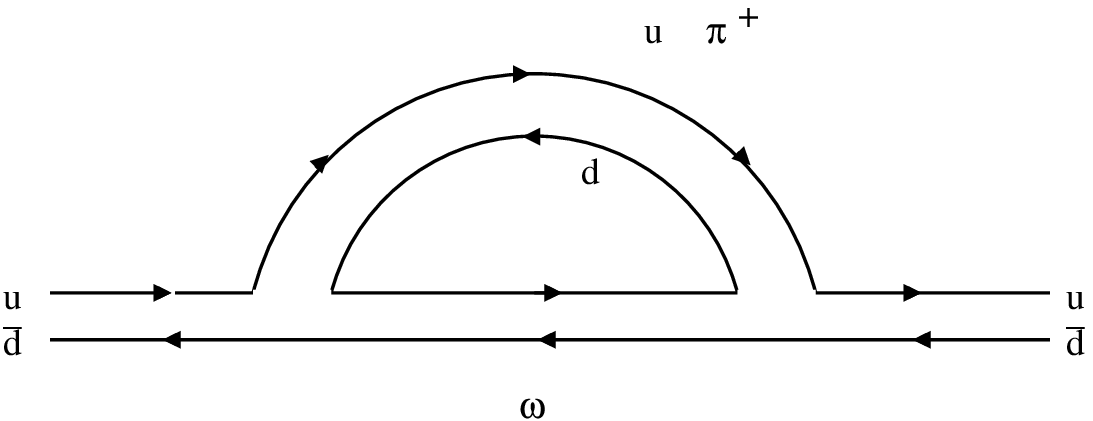}
\caption{\footnotesize{$\rho \rightarrow \pi\omega$ Quark-flow diagram}}
\label{fig:qfrpo}
\end{minipage}
\end{figure}

The diagrams corresponding to $\rho \rightarrow \pi\omega$ are Fig.~\ref{fig:rpo} and Fig.~\ref{fig:qfrpo}, with corresponding self energy given by Eq. (\ref{eqn:sigmapiomega}).

\begin{figure}
\begin{minipage}[b]{0.5\linewidth} 
\centering
\includegraphics[height=90pt]{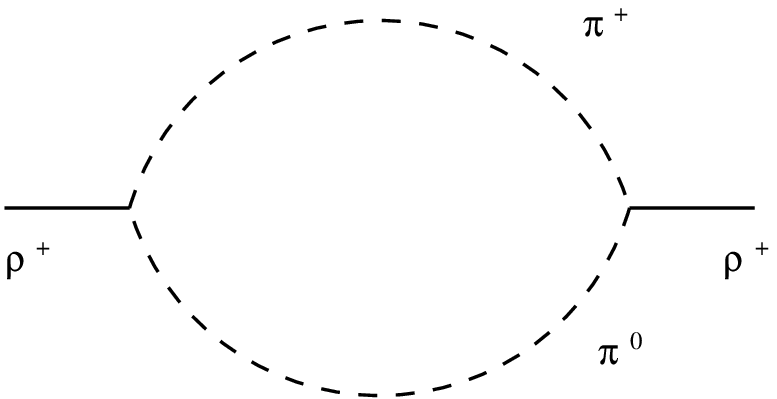}
\caption{\footnotesize{$\rho \rightarrow \pi\pi$ Feynman diagram}}
\label{fig:rpp}
\end{minipage}
\hspace{12mm}
\begin{minipage}[b]{0.5\linewidth}
\centering
\includegraphics[height=90pt]{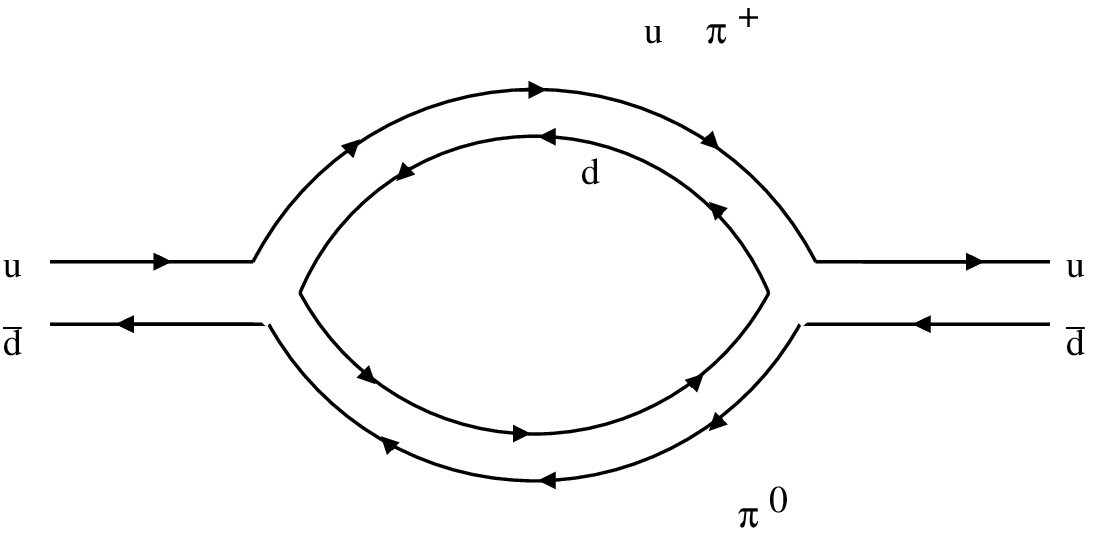}
\caption{\footnotesize{$\rho \rightarrow \pi\pi$ Quark-flow diagram}}
\label{fig:qfrpp}
\end{minipage}
\end{figure}

The diagrams corresponding to $\rho \rightarrow \pi\pi$ are Fig.~\ref{fig:rpp} and Fig.~\ref{fig:qfrpp}, with corresponding self energy given by Eq. (\ref{eqn:sigmapipi}).

\begin{equation}
\label{eqn:sigmapiomega}
\Sigma^\rho_{\pi\omega} = -\frac{f_{\rho\pi\omega}^2}{3\pi^{2}f_\pi^2}
    \int_{0}^{\infty} \frac{k^{4} \, u_{\pi\omega}^{2}(k)~dk}
            {\omega_{\pi}(k)\, ( \omega_\pi(k) +  (\mu_{\omega} - \mu_{\rho}) )}\,,
\end{equation}

\begin{eqnarray}
\label{eqn:sigmapipi}
\Sigma^\rho_{\pi\pi   }&=&-\frac{f_{\rho\pi\pi}^{2}}{6\pi^{2}}
    \int_{0}^{\infty} \frac{k^{4} \, u_{\pi\pi}^{2}(k)~dk} 
           {\omega_{\pi}(k)\, (\omega_{\pi}^{2}(k) - \mu_{\rho}^{2} / 4)}\,, \\
\nn \\
\label{eqn:omegapi}
\mbox{for} \quad \omega_{\pi}^2(k) &=& k^2 + m_{\pi}^2\,.
\end{eqnarray}

 Note that the denominator of the $\rho \rightarrow \pi\pi$ dressing has a singularity at  $k = \sqrt{\frac{\mu_{\rho}^2}{4} - m_{\pi}^2}$. Thus an infinite-volume integral would need to use Cauchy's Residue Theorem, as in Arfken \cite{Arfken} (p.400).

Each of these integrals is derived from first principles using Quantum Field Theory and presented in Appendix \ref{app:loopint}.

The regulators are set for correct on-shell normalization:

\begin{eqnarray}
u_{\pi\omega}(k)&=&u(k) \, , \nn \\
\nn \\
\label{eqn:regs}
u_{\pi\pi}(k)&=&u(k)\,\, u^{-1}\!\! \left ( \sqrt{\mu_\rho^2 / 4 - \mu_\pi^2}\right )\,.
\end{eqnarray}


\section{Finite-Volume Effects in QCD Simulations}

\textbf{Original Research}

Firstly, a simulation of finite-volume effects affecting the $\rho$ self energy in full QCD will be presented. 
It is useful to compare lattice simulation results on finite-sized boxes with the results of an infinite-sized box. Therefore, insight can be gained into how important finite-volume effects are in regards to the integrals, and what an appropriate box size might be.

\begin{figure}
\begin{minipage}[b]{0.5\linewidth} 
\centering
\includegraphics[height=210pt,angle=90]{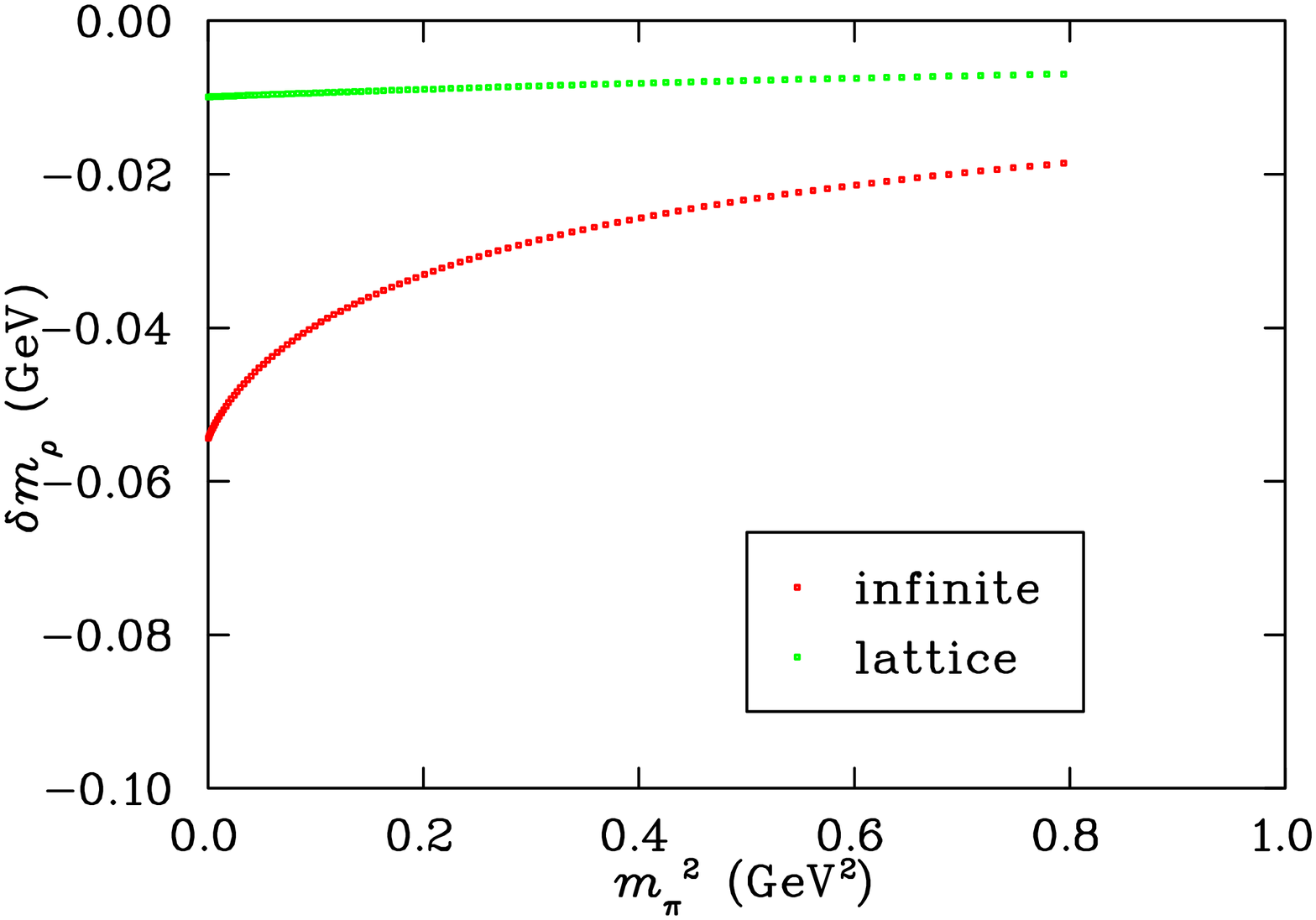}
\caption{\footnotesize{$\rho \rightarrow \pi\omega$, 1 fm box simulation}}
\label{fig:box1rpo}
\end{minipage}
\hspace{12mm}
\begin{minipage}[b]{0.5\linewidth}
\centering
\includegraphics[height=210pt,angle=90]{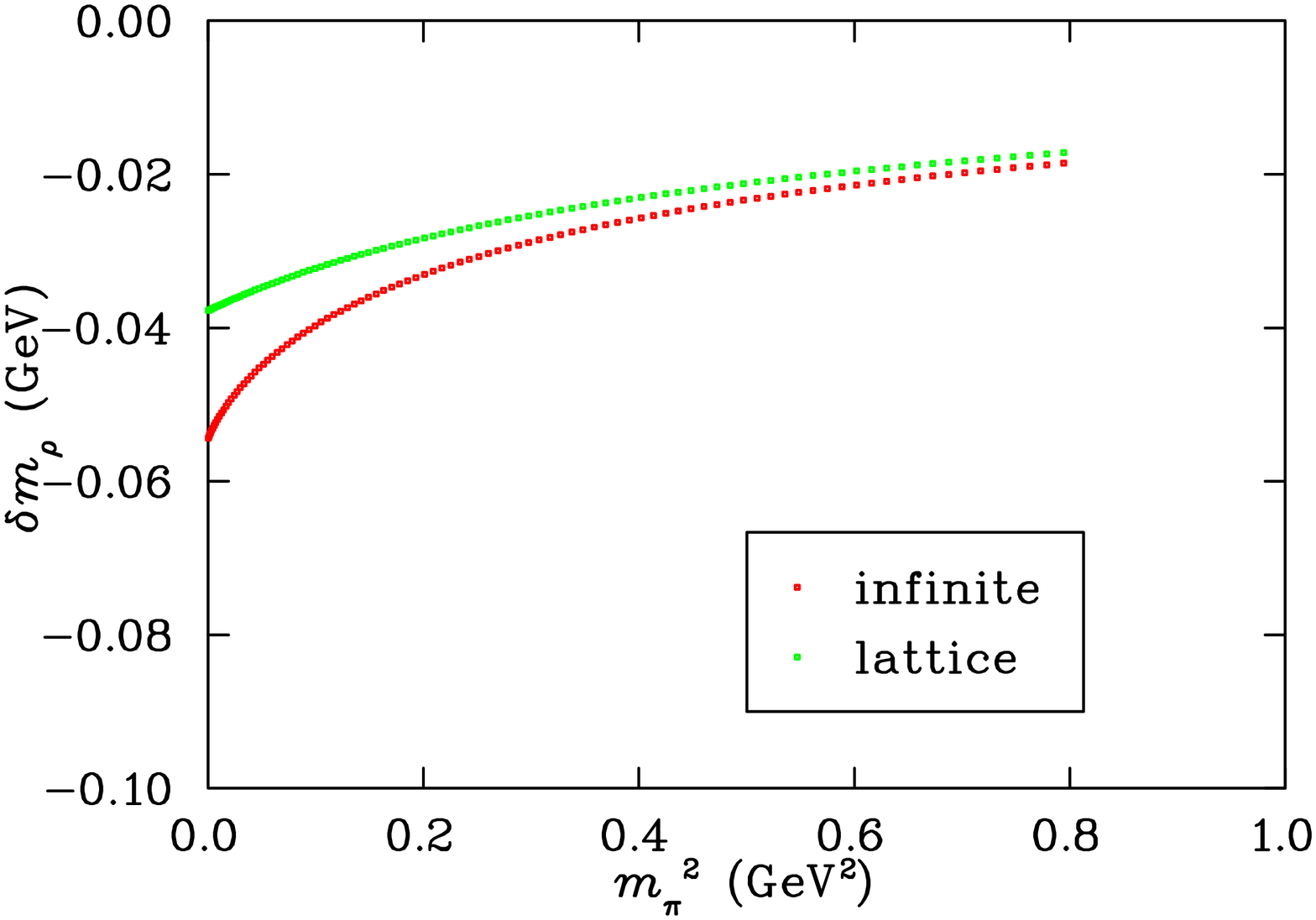}
\caption{\footnotesize{$\rho \rightarrow \pi\omega$, 2 fm box simulation}}
\label{fig:box2rpo}
\end{minipage}

\begin{minipage}[b]{0.5\linewidth} 
\centering
\includegraphics[height=210pt,angle=90]{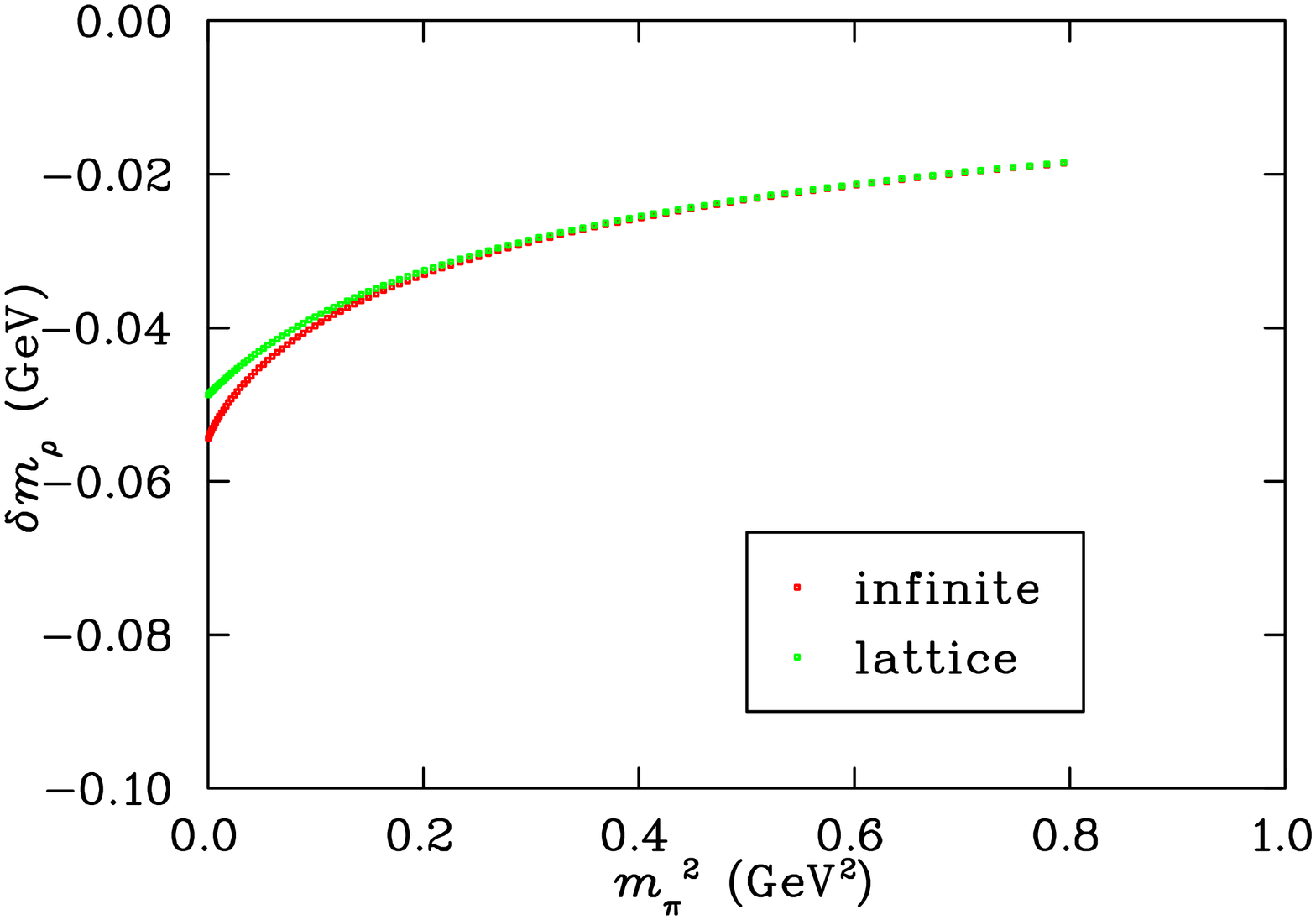}
\caption{\footnotesize{$\rho \rightarrow \pi\omega$, 3 fm box simulation}}
\label{fig:box3rpo}
\end{minipage}
\hspace{12mm}
\begin{minipage}[b]{0.5\linewidth}
\centering
\includegraphics[height=210pt,angle=90]{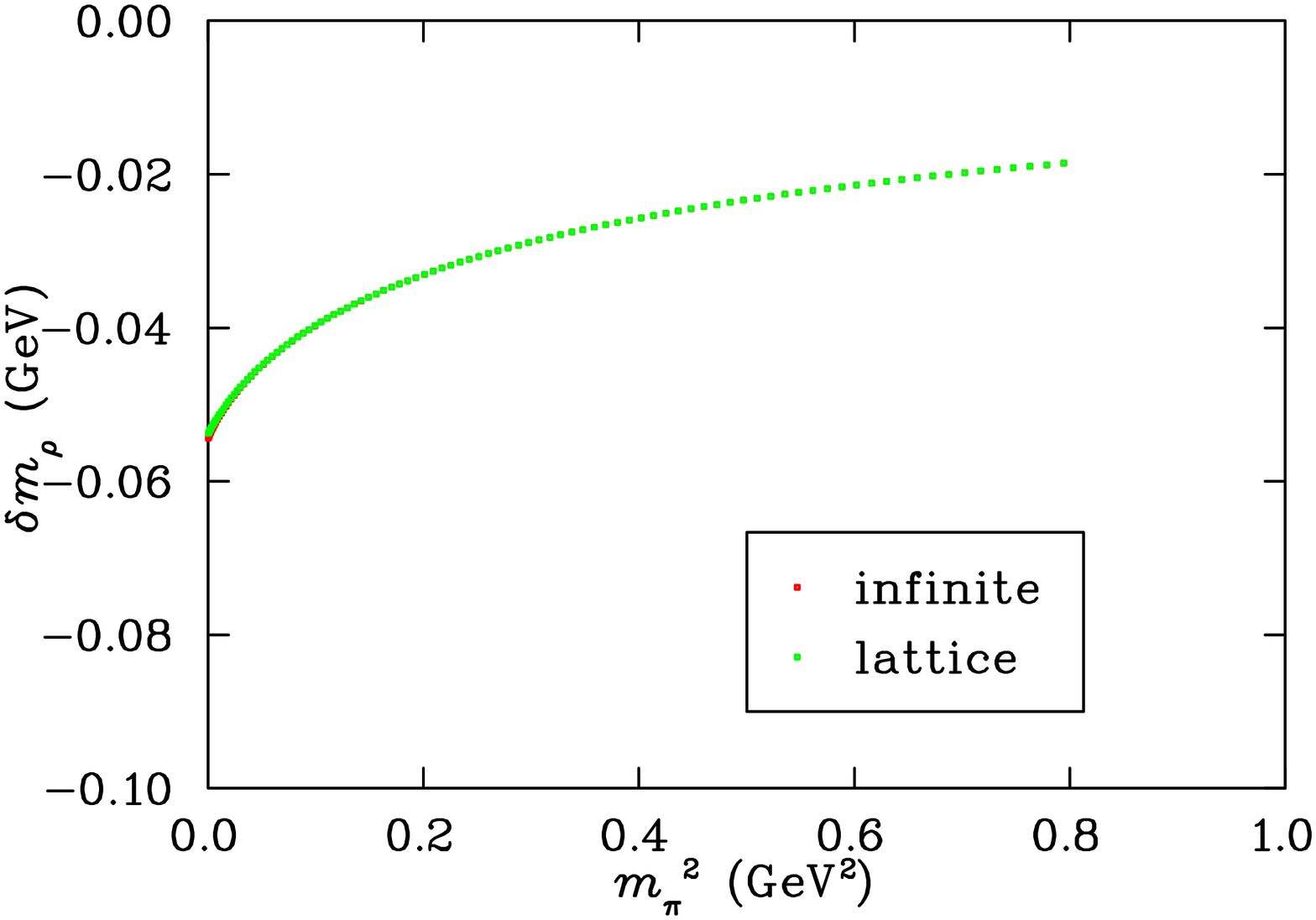}
\caption{\footnotesize{$\rho \rightarrow \pi\omega$, 6 fm box simulation}}
\label{fig:box6rpo}
\end{minipage}





\begin{minipage}[b]{0.5\linewidth} 
\centering
\includegraphics[height=210pt,angle=90]{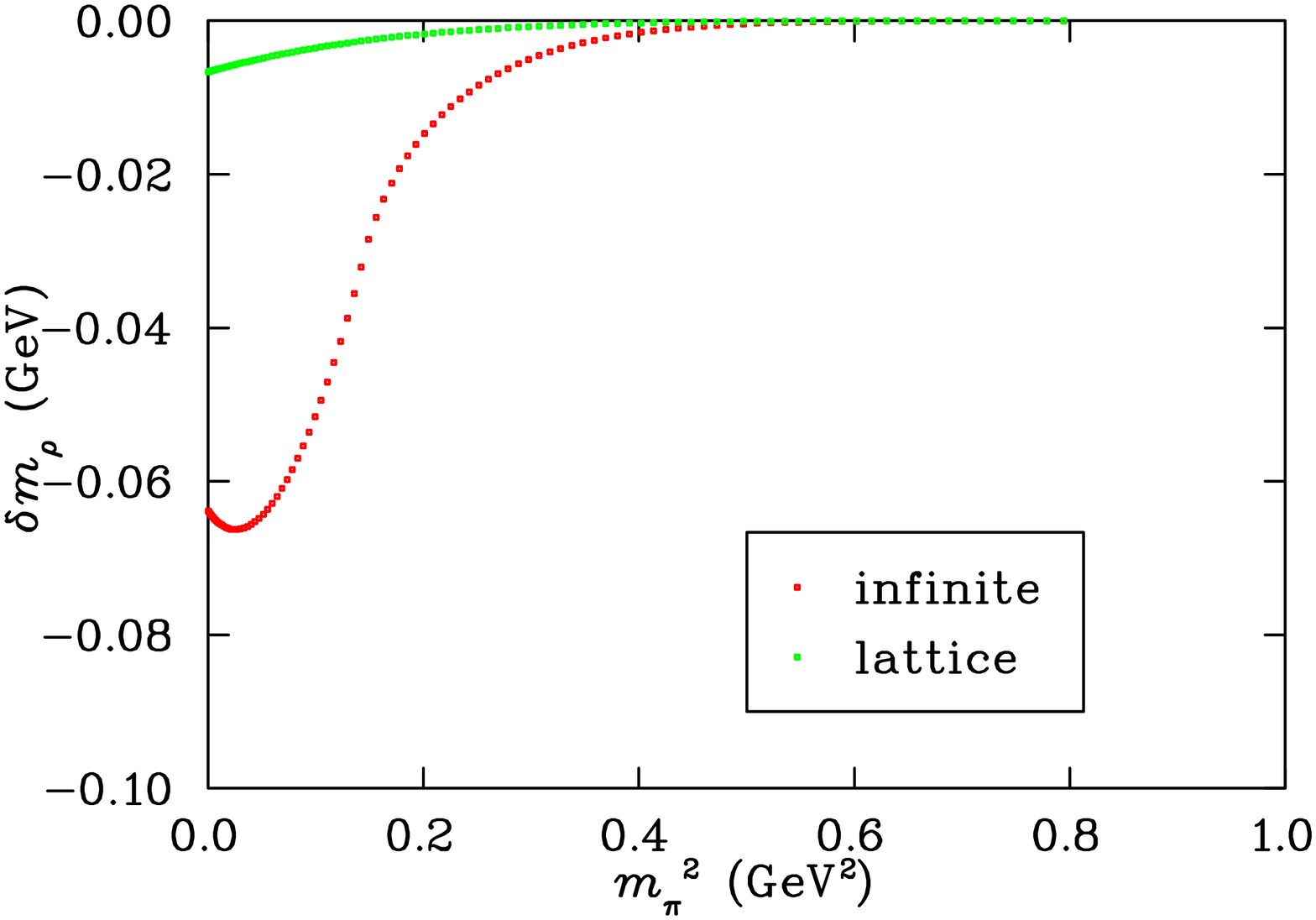}
\caption{\footnotesize{$\rho \rightarrow \pi\pi$, 1 fm box simulation}}
\label{fig:box1rpp}
\end{minipage}
\hspace{12mm}
\begin{minipage}[b]{0.5\linewidth}
\centering
\includegraphics[height=210pt,angle=90]{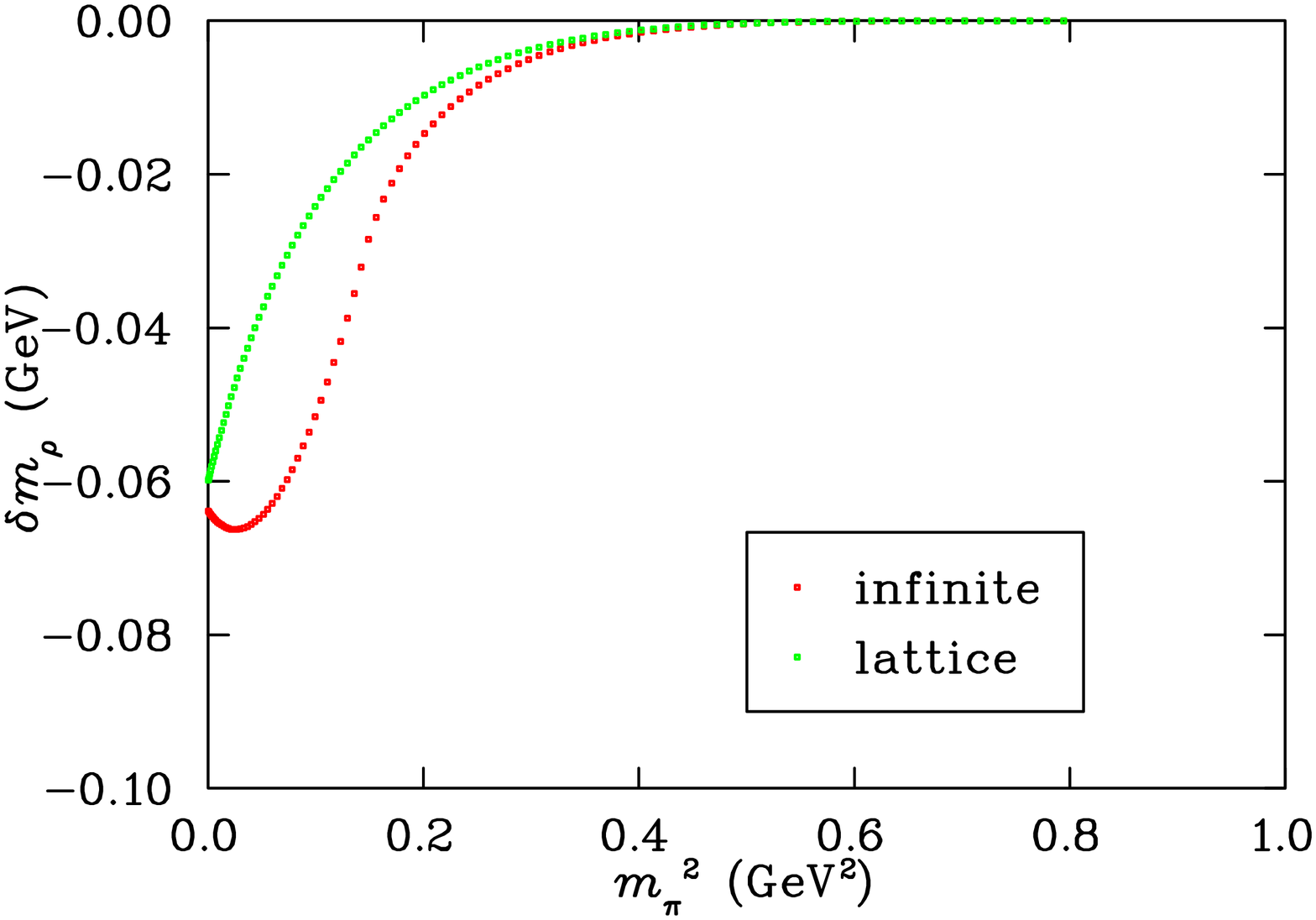}
\caption{\footnotesize{$\rho \rightarrow \pi\pi$, 2 fm box simulation}}
\label{fig:box2rpp}
\end{minipage}

\begin{minipage}[b]{0.5\linewidth} 
\centering
\includegraphics[height=210pt,angle=90]{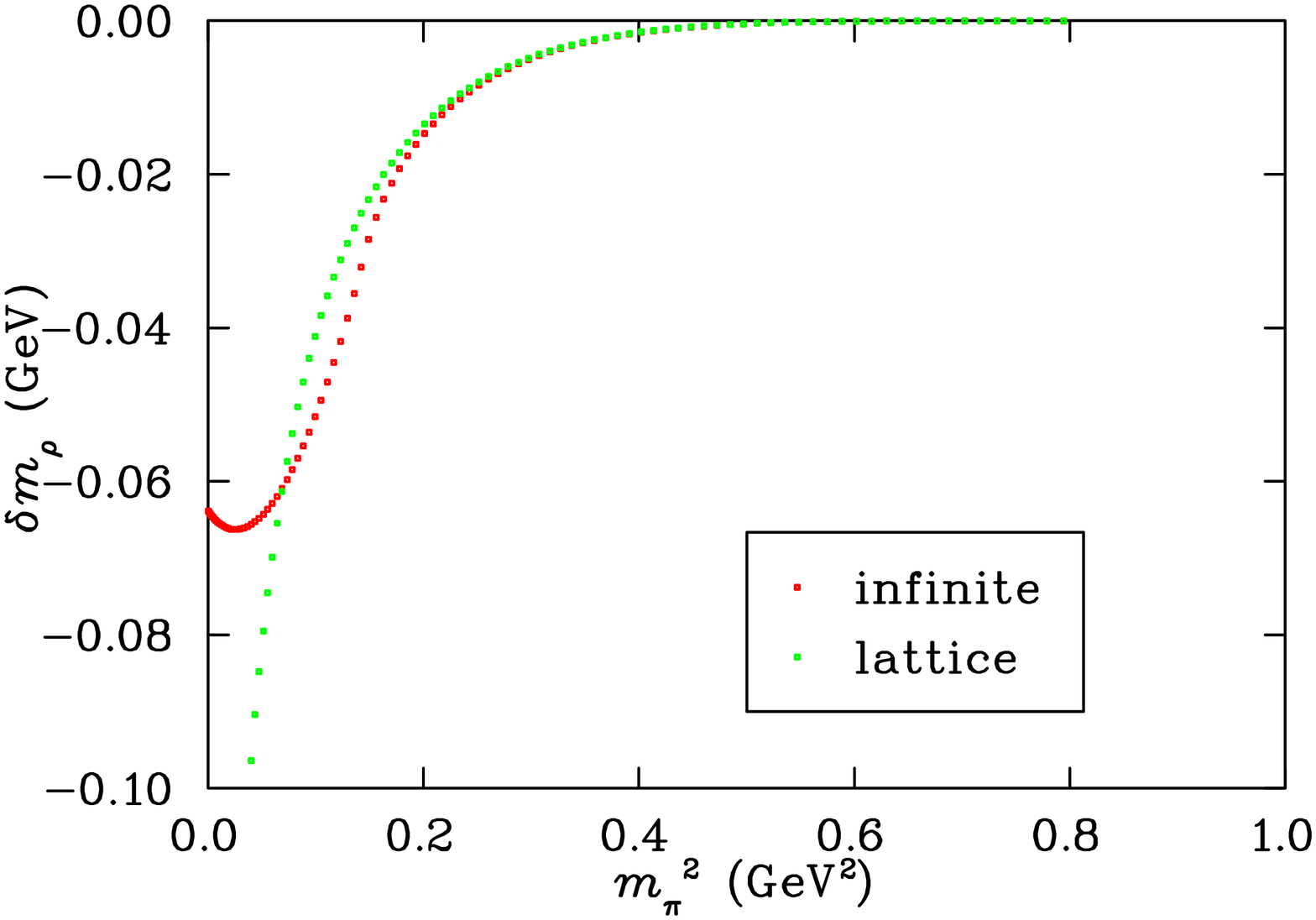}
\caption{\footnotesize{$\rho \rightarrow \pi\pi$, 3 fm box simulation}}
\label{fig:box3rpp}
\end{minipage}es
\hspace{12mm}
\begin{minipage}[b]{0.5\linewidth}
\centering
\includegraphics[height=210pt,angle=90]{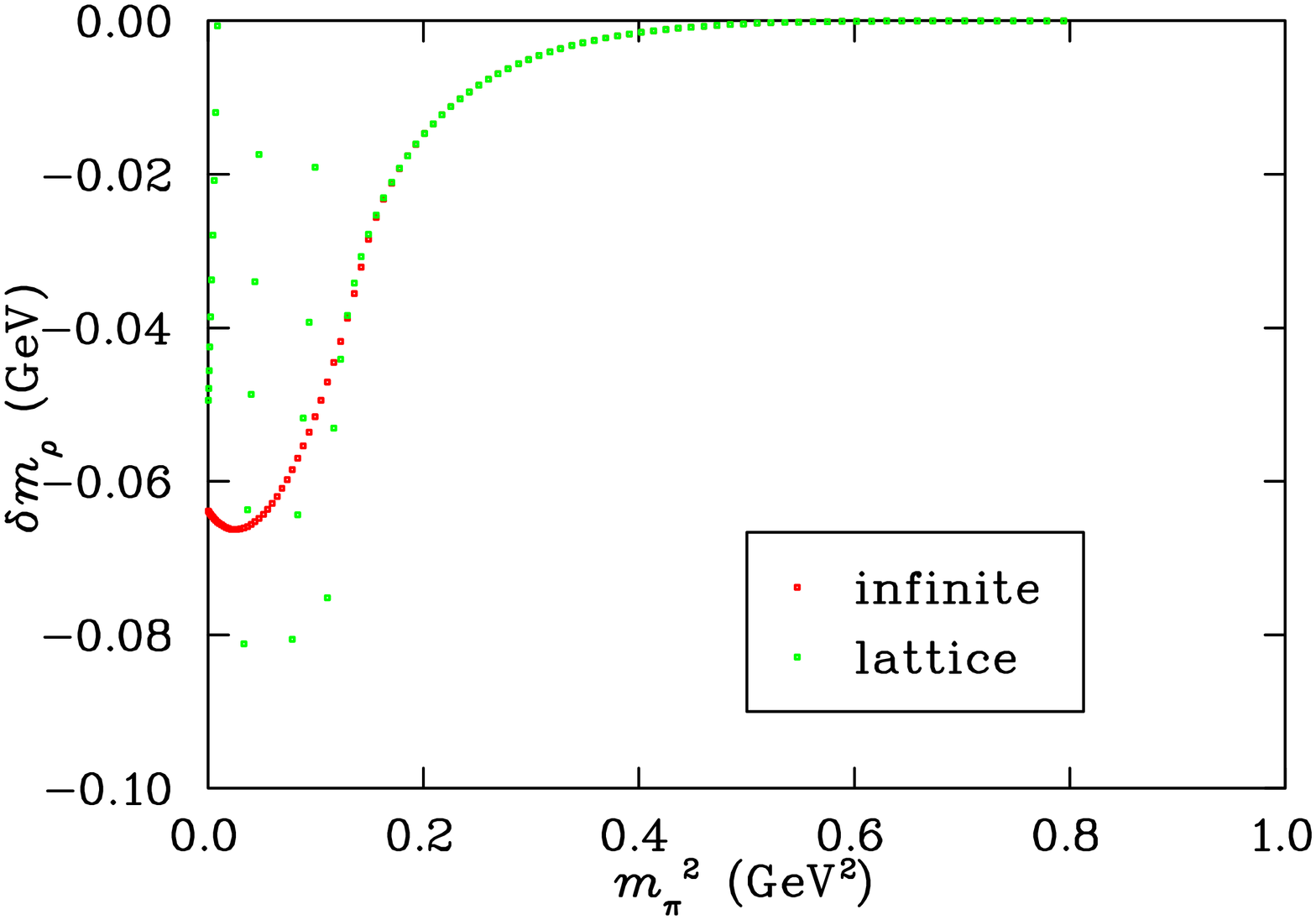}
\caption{\footnotesize{$\rho \rightarrow \pi\pi$, 6 fm box simulation}}
\label{fig:box6rpp}
\end{minipage}
\end{figure}

Comparing Fig.~\ref{fig:box1rpo} to Fig.~\ref{fig:box6rpo}, there is a marked increase in the accuracy of the finite-volume result. As a guide, lattice experiments are usually performed on boxes of approximate length $3$ fermi \cite{Boinepalli:2004fz}. Others suggest a lattice box length of $2.5$ to $3.0$ fermi \cite{Stewart} \cite{Dong:1995ec} \cite{Fukugita:1994ba} \cite{Labrenz:1996jy} \cite{Sheikholeslami:1985ij} \cite{Leinweber:1999ig}, as mentioned in Chapter \ref{chpt:Lattice QCD}.

When comparing Fig.~\ref{fig:box1rpp} to Fig.~\ref{fig:box6rpp}, it is clear that when the momentum values in the lattice simulation are close to the singularity in  Eq. (\ref{eqn:sigmapipi}), the expression exhibits incorrect behaviour. The discrete lattice spacing regulates the integral and information is lost, so that the full contribution around the Cauchy pole does not exactly cancel as it should in the continuum limit. 

\section{Analysis of Quenched Lattice Data}




Now that the finite-volume effects have been quantified, QQCD data for FLIC fermions by Boinepalli \cite{Boinepalli:2004fz} and Zanotti \cite{Zanotti:2001yb} will be considered. In the Boinepalli data, eight values of $m_{\rho}$ and $m_{\pi}^2$ were calculated. In the Zanotti data, five values were calculated. 

The loop integral contributions for the $\rho$ meson mass in QQCD, namely the single and double hairpin $\eta'$ contributions, were subtracted from each data point. These contributions were calculated in the infinite-volume box case, and also the finite-volume case with a box the same size as the original data. In Boinepalli,  $L_{x} = 0.128 \times 20$ fermi and in Zanotti,  $L_{x} = 0.116 \times 16$ fermi. The difference between the infinite-volume and finite-volume prediction of $m_{\rho}$ was also calculated, since it is expected that this quantity is invariant with respect to the regulator parameter $\Lambda$. This procedure was carried out for different values of $\Lambda$, from $\Lambda = 0.5$ GeV to $\Lambda = 2.0$ GeV. The results of these data are presented in Appendix \ref{app:qsigma} and \ref{app:qsigma2}. 

The coefficients of the polynomial expansion of the bare $\rho$ mass were calculated using the Singular Value Decomposition Fit algorithm from Press \cite{NumRec}. The coefficients from the Boinepalli and Zanotti data respectively for different values of $\Lambda$ are listed in Tables ~\ref{table:qcoeffsshort} and \ref{table:qcoeffs2short}.

\begin{table}
{\scriptsize
\verbatiminput{Misc/qcoeffsshort.data}
}
 \caption{\footnotesize{Sample coefficients based on Boinepalli}}
  \label{table:qcoeffsshort}

{\scriptsize
\verbatiminput{Misc/qcoeffs2short.data}
}
 \caption{\footnotesize{Sample coefficients based on Zanotti}}
  \label{table:qcoeffs2short}
\end{table}

\begin{figure}
\centering
\includegraphics[height=280pt,angle=90]{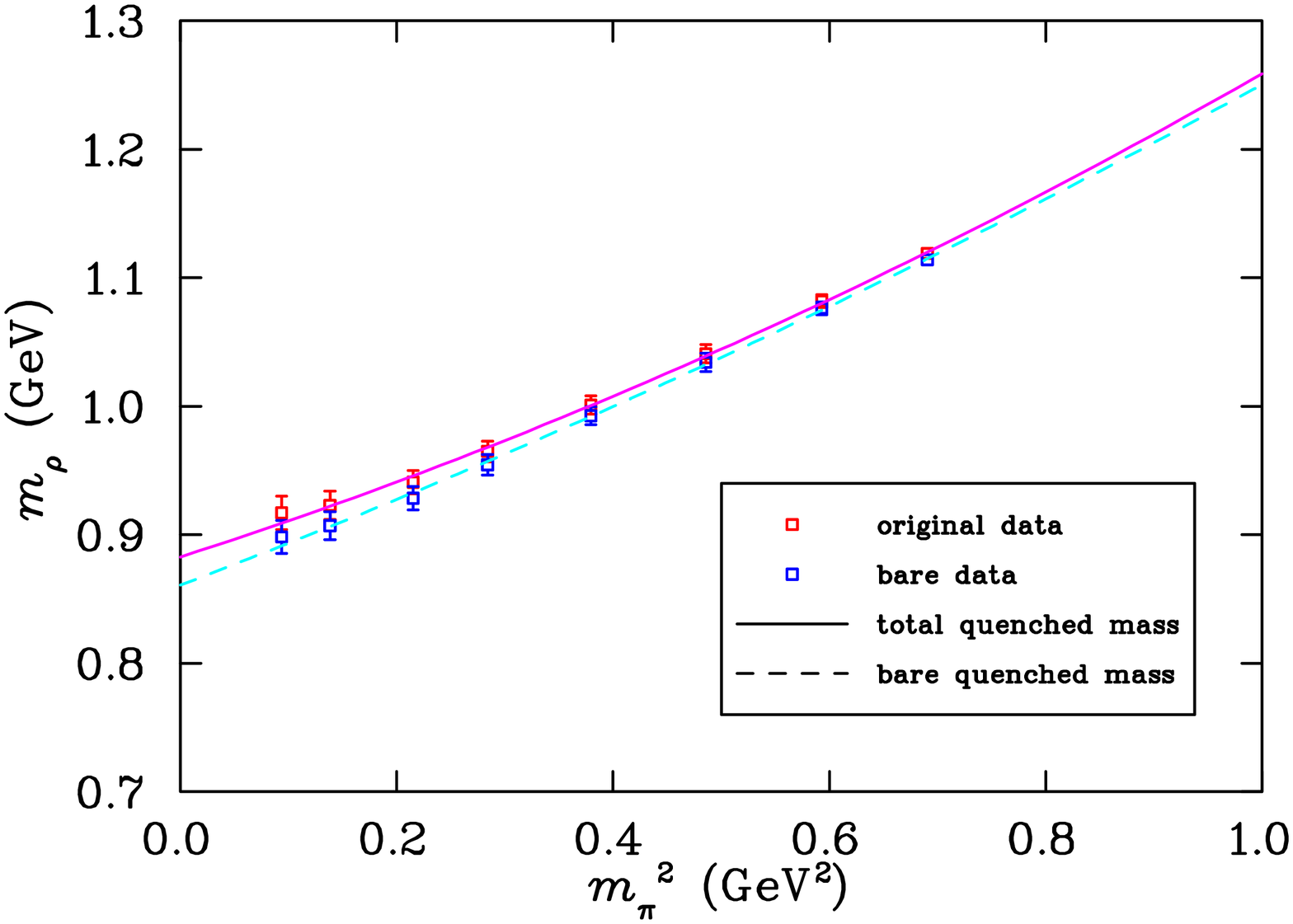}
\caption{\footnotesize{Quenched Fit on Boinepalli \cite{Boinepalli:2004fz} for $\Lambda$ = $0.8$ GeV}}
\label{fig:qL0.8}

\centering
\includegraphics[height=280pt,angle=90]{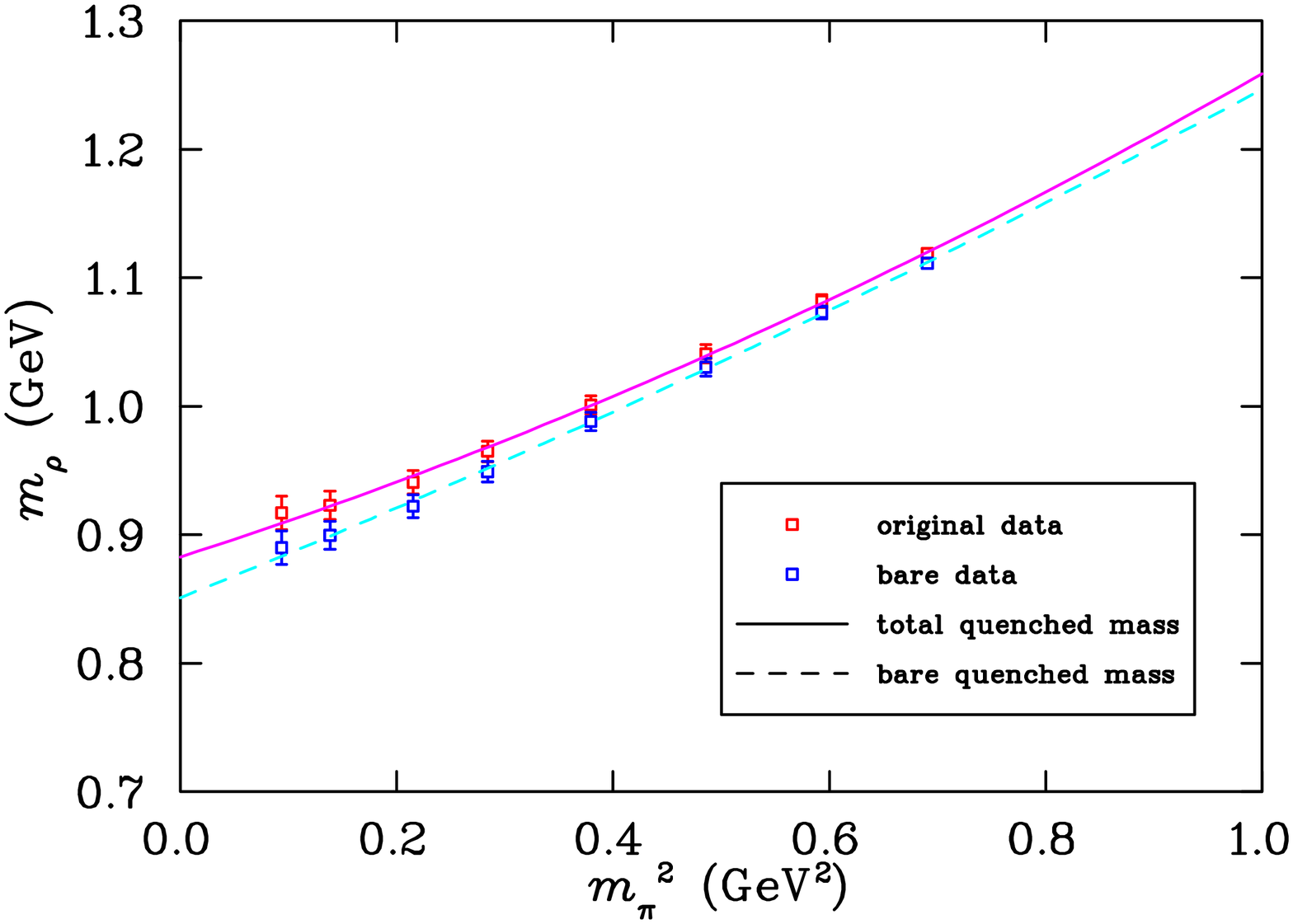}
\caption{\footnotesize{Quenched Fit on Boinepalli for $\Lambda$ = $0.9$ GeV}}
\label{fig:qL0.9}

\centering
\includegraphics[height=280pt,angle=90]{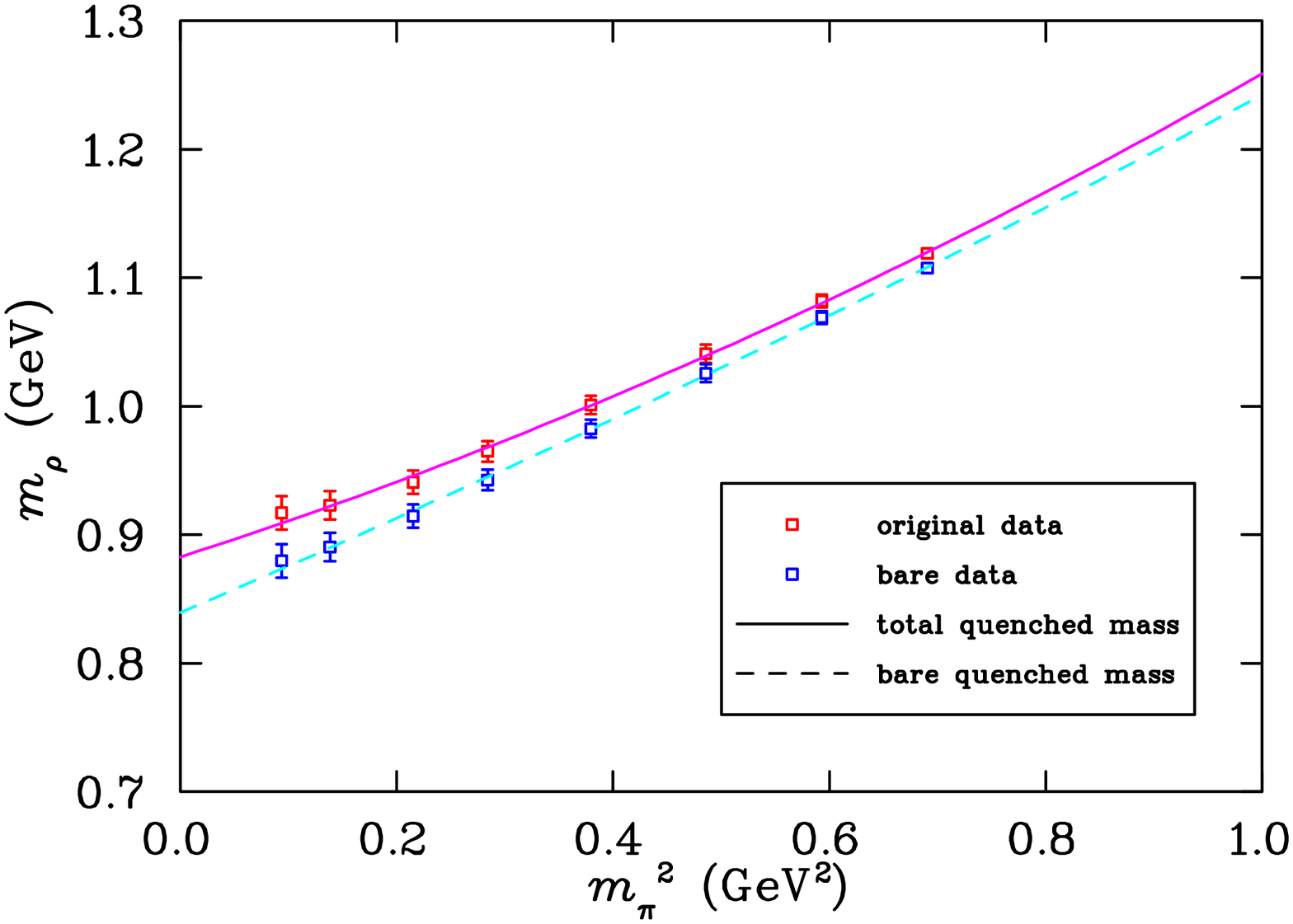}
\caption{\footnotesize{Quenched Fit on Boinepalli for $\Lambda$ = $1.0$ GeV}}
\label{fig:qL1.0}
\end{figure}
\begin{figure}
\centering
\includegraphics[height=280pt,angle=90]{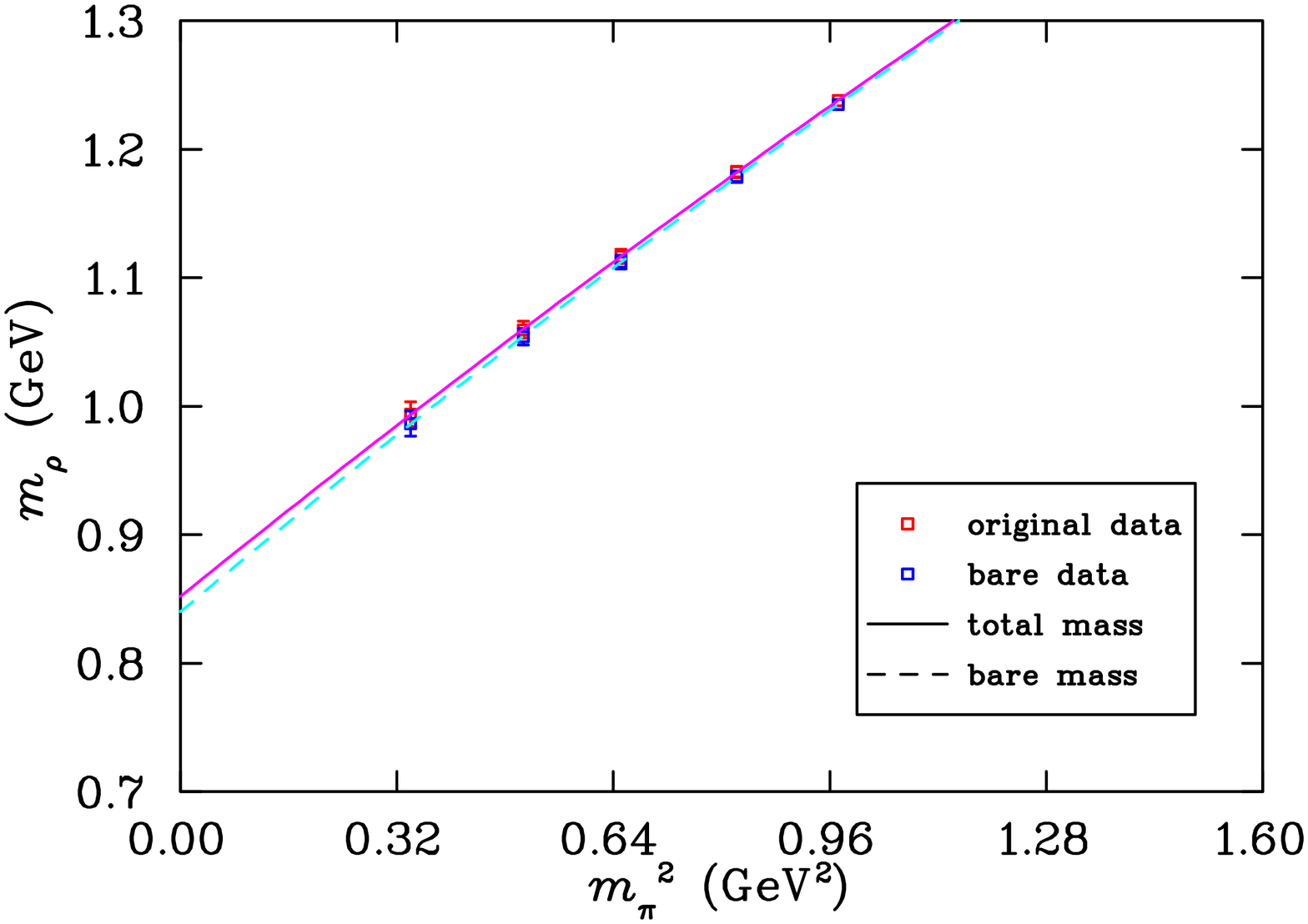}
\caption{\footnotesize{Quenched Fit on Zanotti \cite{Zanotti:2001yb} for $\Lambda$ = $0.8$ GeV}}
\label{fig:q2L0.8}

\centering
\includegraphics[height=280pt,angle=90]{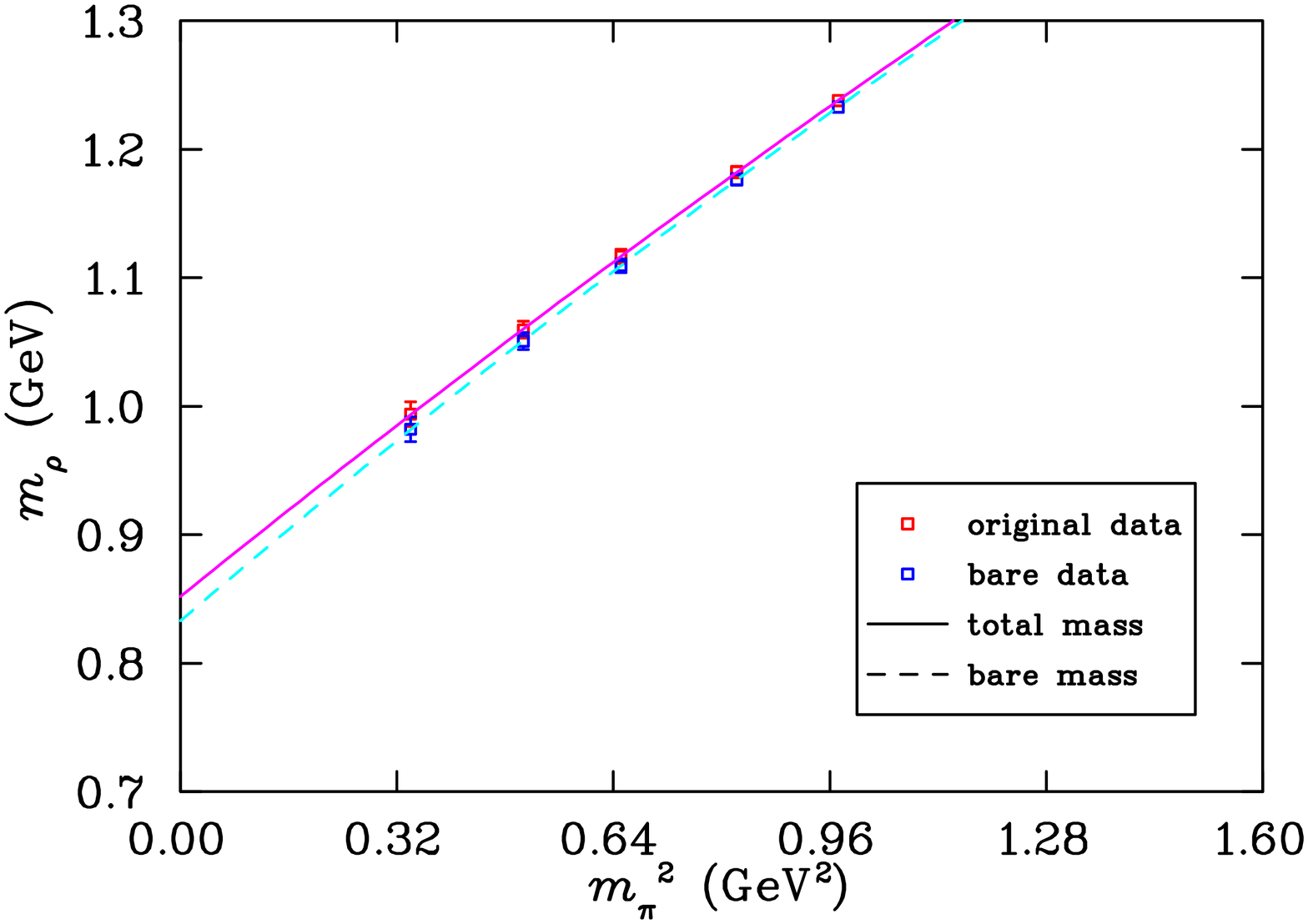}
\caption{\footnotesize{Quenched Fit on Zanotti for $\Lambda$ = $0.9$ GeV}}
\label{fig:q2L0.9}

\centering
\includegraphics[height=280pt,angle=90]{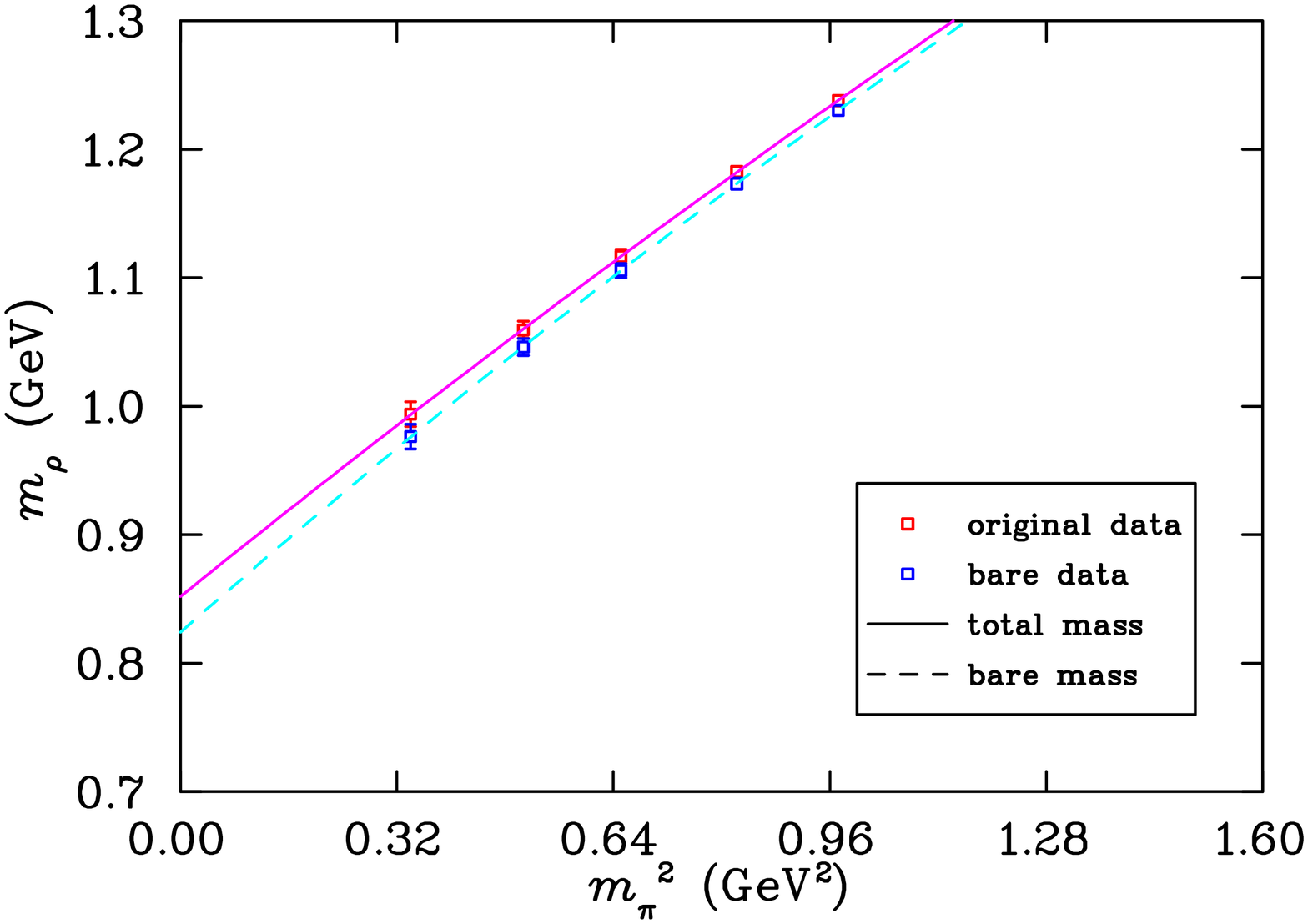}
\caption{\footnotesize{Quenched Fit on Zanotti for $\Lambda$ = $1.0$ GeV}}
\label{fig:q2L1.0}
\end{figure}

 Fig.~\ref{fig:qL0.8} to Fig.~\ref{fig:qL1.0} show the QQCD data by Boinepalli for $m_{\rho}$ against $m_{\pi}^2$ and also the bare $m_{\rho}$ (with finite-volume non-analytic loop contributions subtracted). The polynomial curve calculated was fitted through the latter data. The graph was calculated for multiple values of $\Lambda$. Fig.~\ref{fig:q2L0.8} to Fig.~\ref{fig:q2L1.0} show the QQCD data by Zanotti treated in the same way as the Boinepalli data.

It should be noted that in all graphs the error in the pseudo-scalar mass $m_{\pi}$ was neglected because it corresponds to the maximum signal-to-noise ratio, and is thus minimized.

\begin{figure}
\centering
\includegraphics[height=280pt,angle=90]{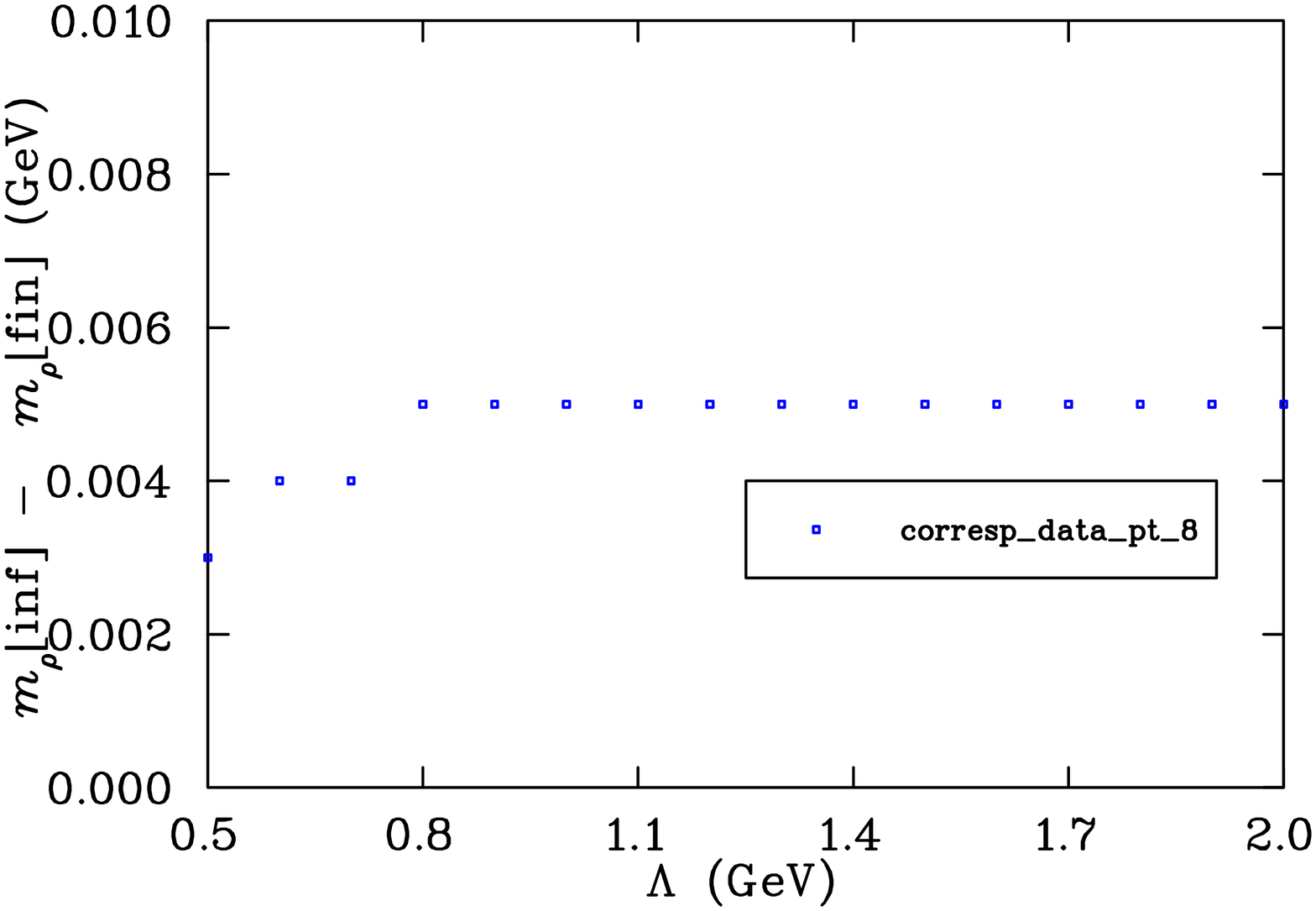}
\caption{\footnotesize{$m_{\rho}$[infinite] - $m_{\rho}$[finite] vs. $\Lambda$ (Data Pt 8) on Boinepalli}}
\label{fig:difflmpi8}

\centering
\includegraphics[height=280pt,angle=90]{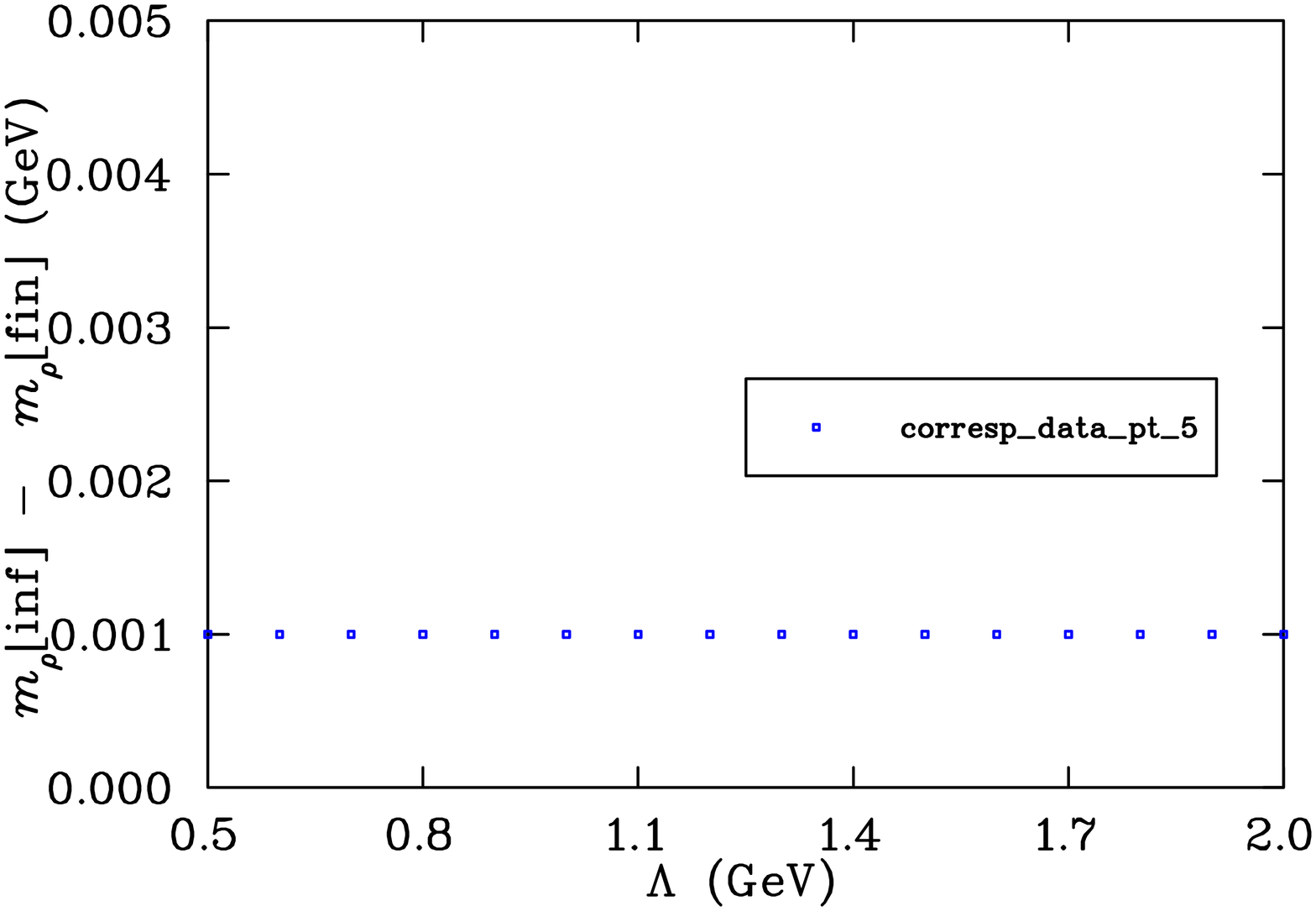}
\caption{\footnotesize{$m_{\rho}$[infinite] - $m_{\rho}$[finite] vs. $\Lambda$ (Data Pt 5) on Zanotti}}
\label{fig:difflmpi5}
\end{figure}

In order to highlight the usefulness of the $\chi$EFT technique, consider the difference between the infinite-volume $m_{\rho}$ and the finite-volume  $m_{\rho}$, for the lightest value of $m_{\pi}^2$ calculated in the two QQCD data sets. Fig.~\ref{fig:difflmpi8} and Fig.~\ref{fig:difflmpi5} show this explicitly for multiple $\Lambda$ values.  The dependence on the parameter $\Lambda$ is slight for small values of $\Lambda$. For values $\Lambda \sim 1$ GeV, the mass difference is independent of $\Lambda$, indicating that real physics, which is not model dependent, is being observed.

\begin{figure}
\centering
\includegraphics[height=280pt,angle=90]{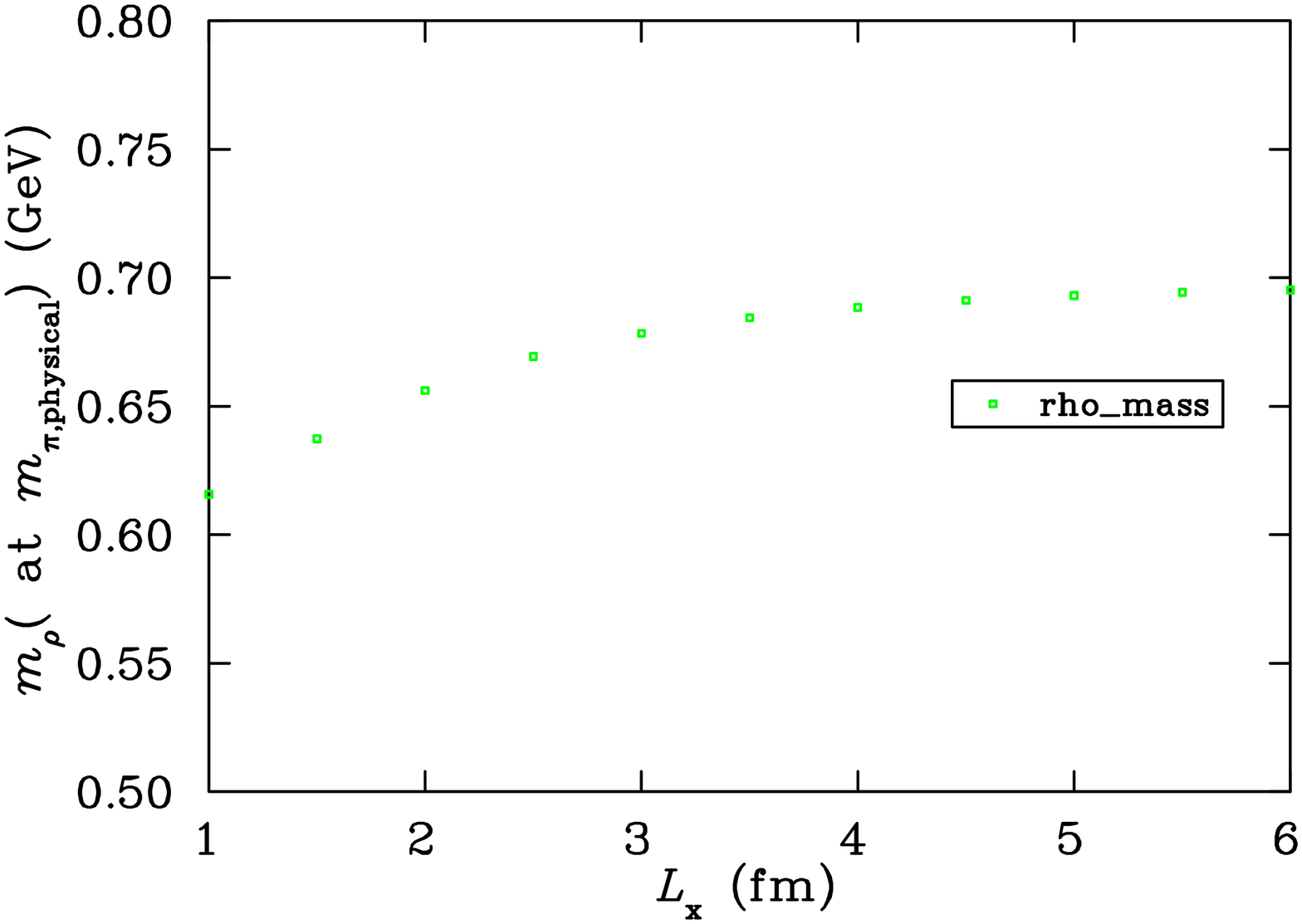}
\caption{\footnotesize{$m_{\rho}$ vs. $L_{x}$  on Boinepalli}}
\label{fig:qL1.0size}

\centering
\includegraphics[height=280pt,angle=90]{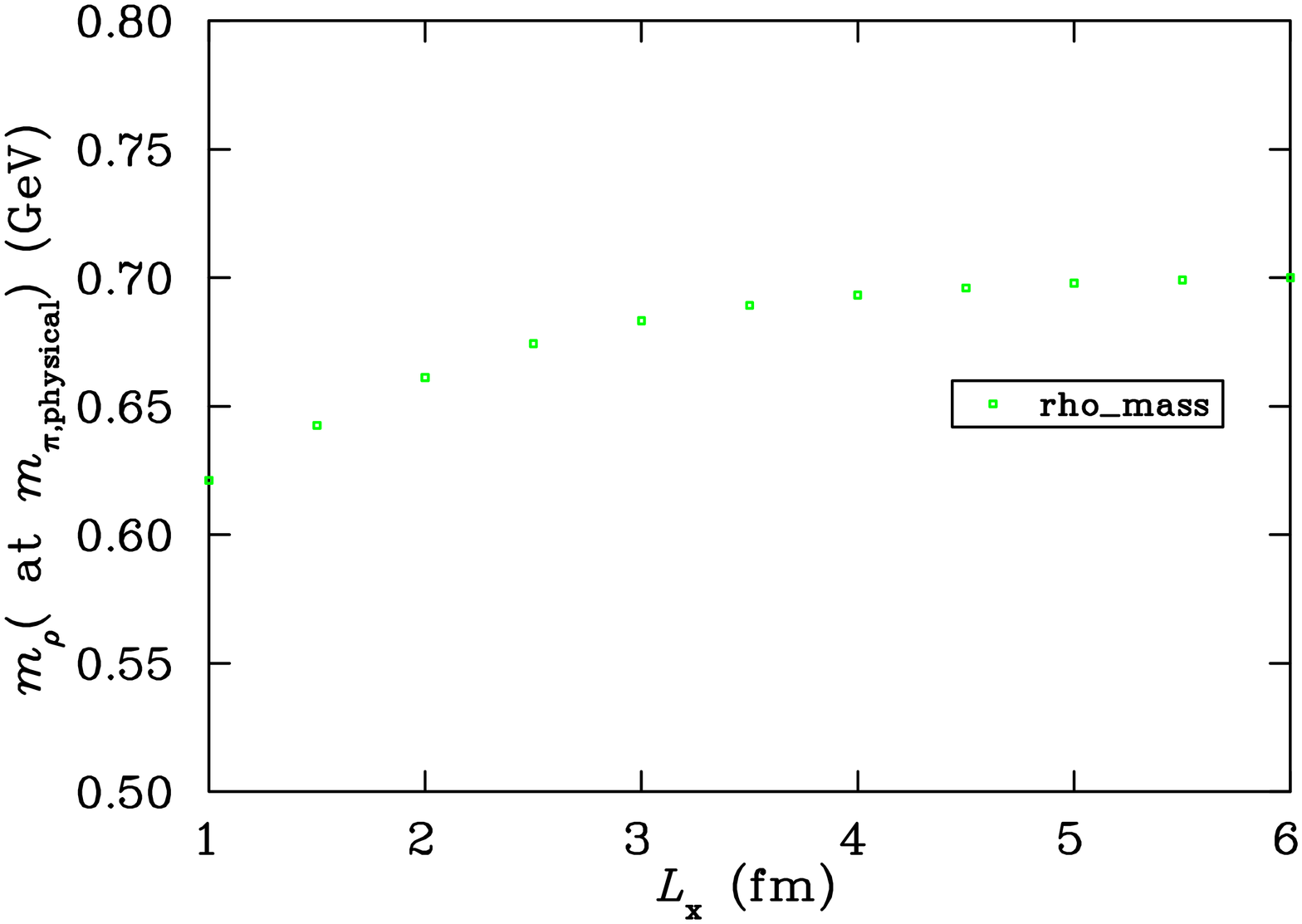}
\caption{\footnotesize{$m_{\rho}$ vs. $L_{x}$ on Zanotti}}
\label{fig:q2L1.0size}
\end{figure}

It might also pay to consider explicitly the dependence of $m_{\rho}$ on the box length $L_{x}$. It is expected that $m_{\rho}$ will be highly dependent on the finite-volume effects for small box sizes, but for box sizes greater than $3$ fermi, $m_{\rho}$ will be independent. The results are shown in Fig.~\ref{fig:qL1.0size} and Fig.~\ref{fig:q2L1.0size}.

\section{Analysis of Full QCD Lattice Data}

The Aoki data for full QCD consist of five values of $m_{\rho}$ and $m_{\pi}^2$. The loop integral contributions for the $\rho$ meson mass in full QCD were subtracted from each data point in the same way as for the QQCD case. The non-analytic terms now correspond to the $\rho \rightarrow \pi\omega$ and $\rho \rightarrow \pi\pi$ processes. The box sizes are different for each data point, so this was taken into account on all graphs.  Data tables with infinite and finite-volume extrapolations of $m_{\rho}$ are provided in Appendix \ref{app:fsigma}. The coefficients from the Aoki data for different values of $\Lambda$ are listed in Table~\ref{table:fcoeffsshort}. 

\begin{figure}
\centering
\includegraphics[height=280pt,angle=90]{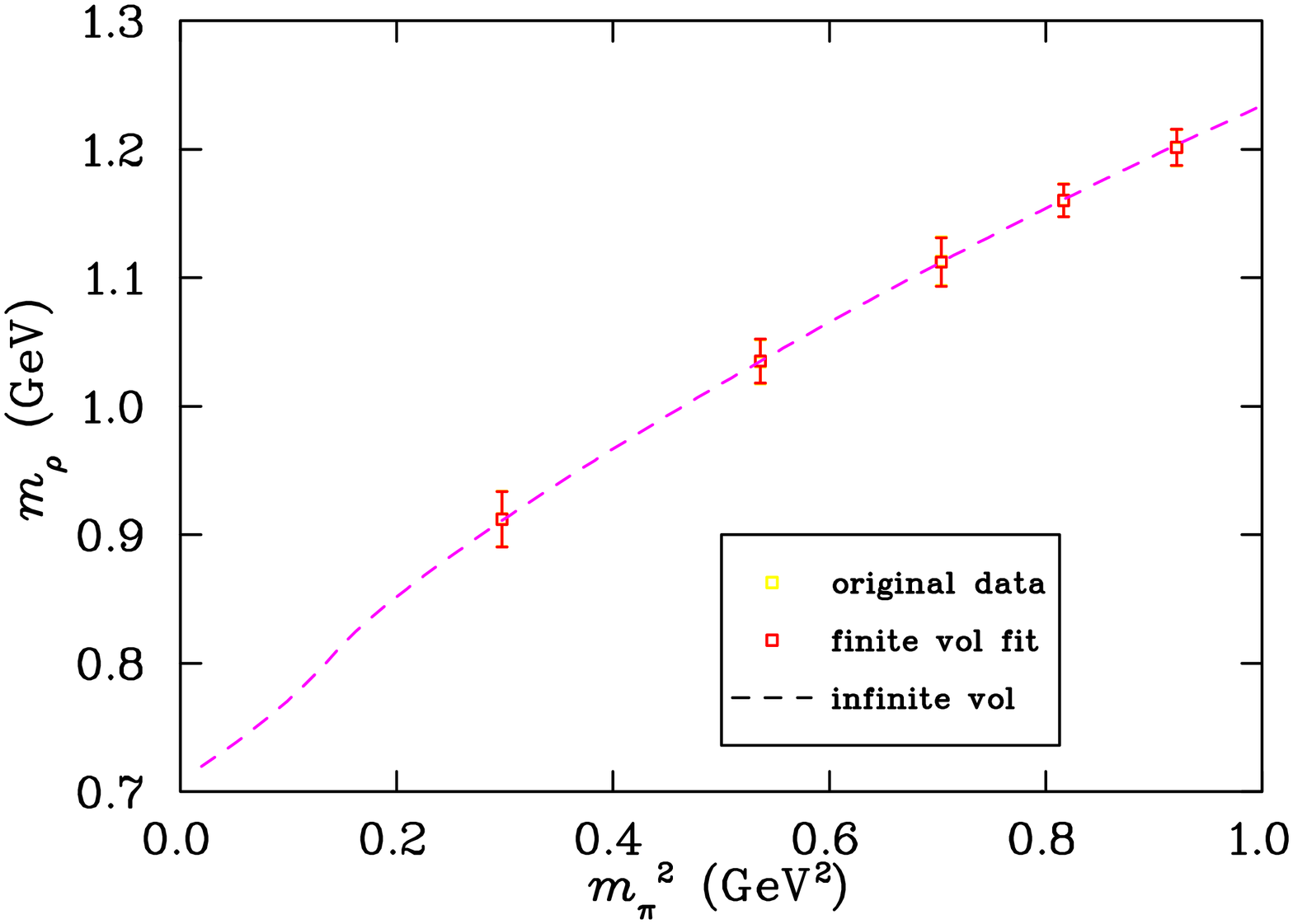}
\caption{\footnotesize{Full QCD Fit on Aoki \cite{Aoki:2002uc} for $\Lambda$ = $0.8$ GeV}}
\label{fig:fL0.8}

\centering
\includegraphics[height=280pt,angle=90]{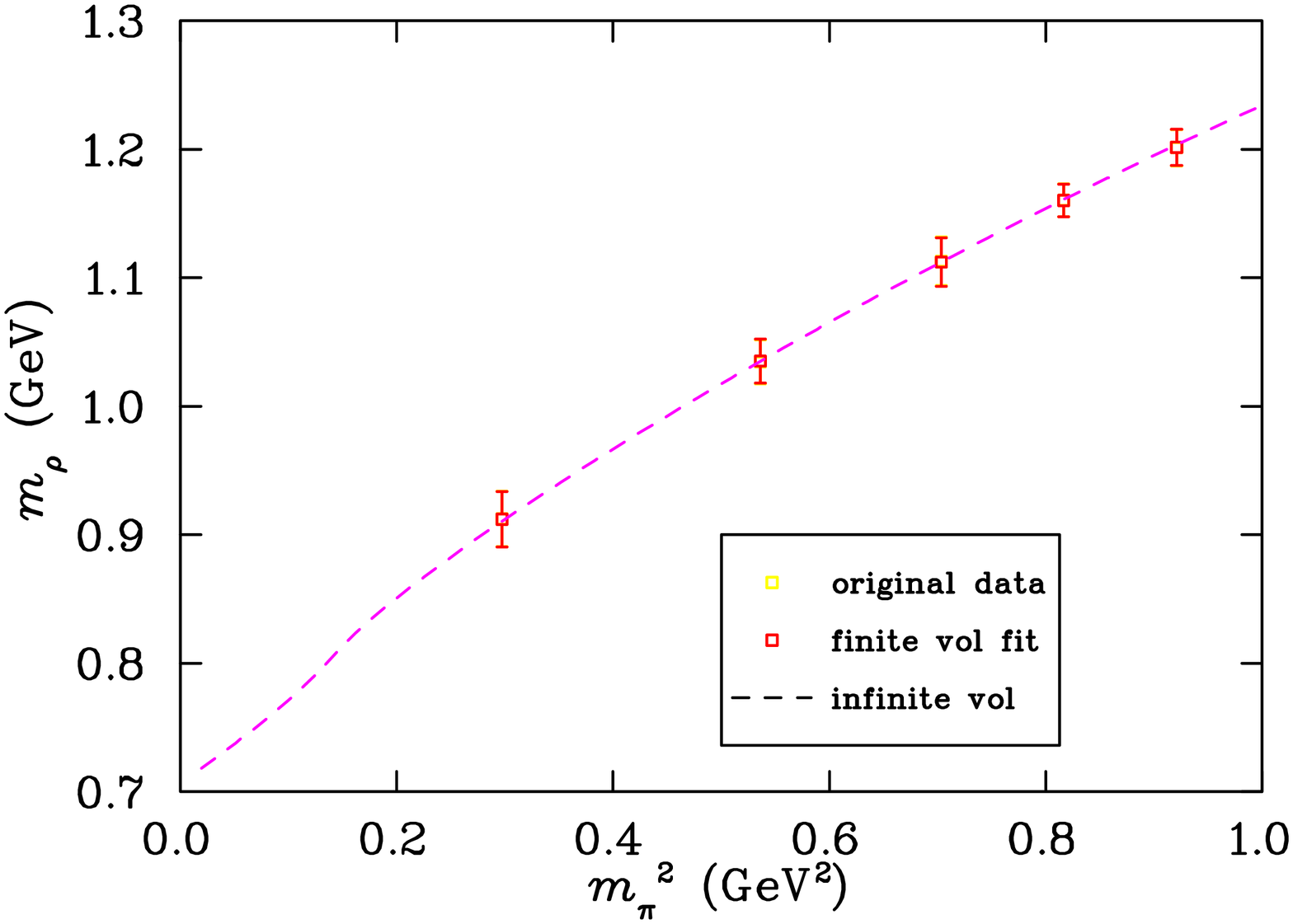}
\caption{\footnotesize{Full QCD Fit on Aoki for $\Lambda$ = $0.9$ GeV}}
\label{fig:fL0.9}

\centering
\includegraphics[height=280pt,angle=90]{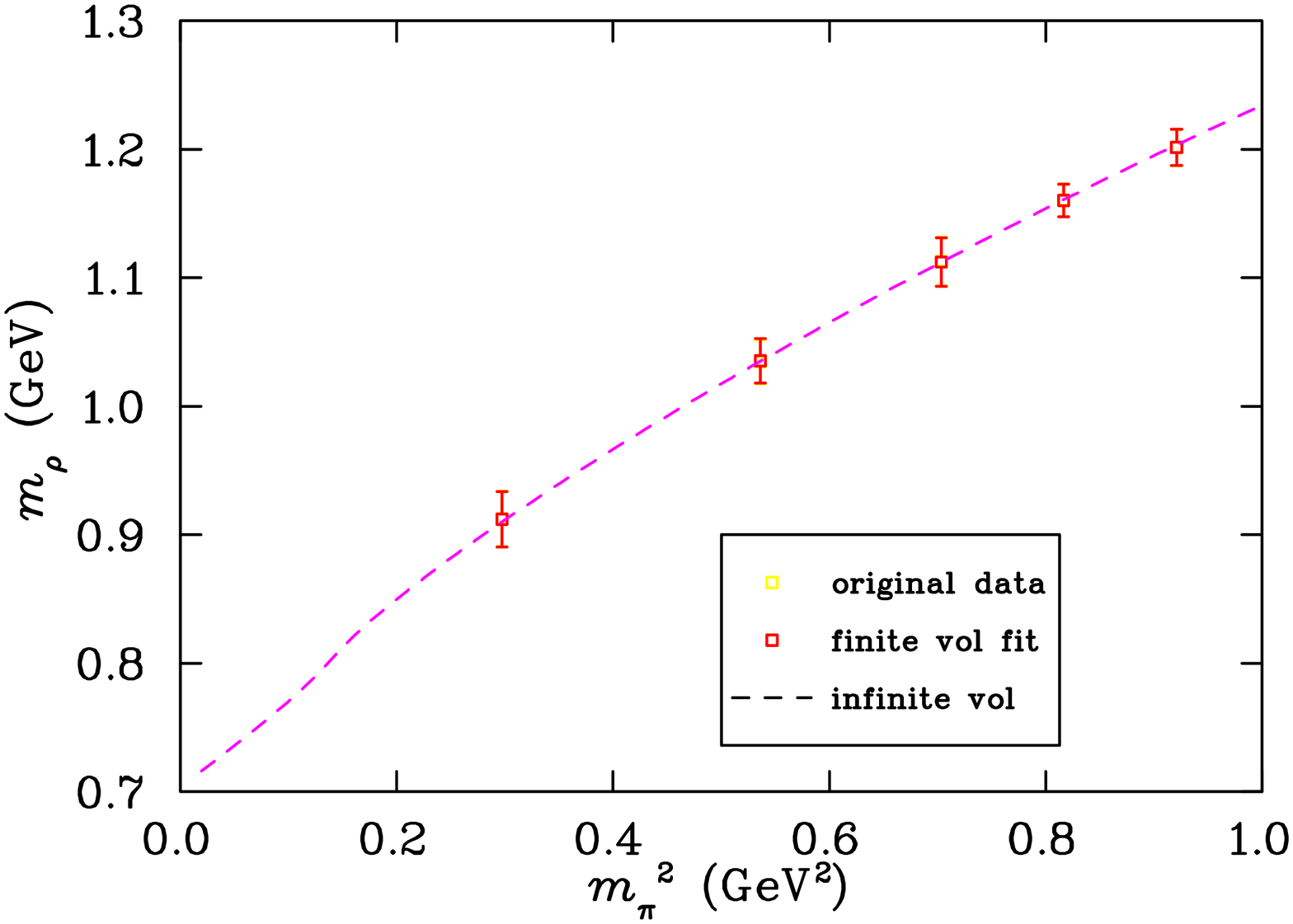}
\caption{\footnotesize{Full QCD Fit on Aoki for $\Lambda$ = $1.0$ GeV}}
\label{fig:fL1.0}
\end{figure}

Coefficients for full QCD seem well correlated with both QQCD data sets for $\Lambda = 0.9$ GeV. Fig.~\ref{fig:fL0.8} to Fig.~\ref{fig:fL1.0} show the full QCD data by Aoki for $m_{\rho}$, and the bare $m_{\rho}$ against $m_{\pi}^2$.

\begin{table}
{\scriptsize
\verbatiminput{Misc/fcoeffsshort.data}
}
 \caption{\footnotesize{Sample coefficients based on Aoki}}
  \label{table:fcoeffsshort}
\end{table}

\begin{figure}
\begin{center}
 \includegraphics[height=280pt,angle=90]{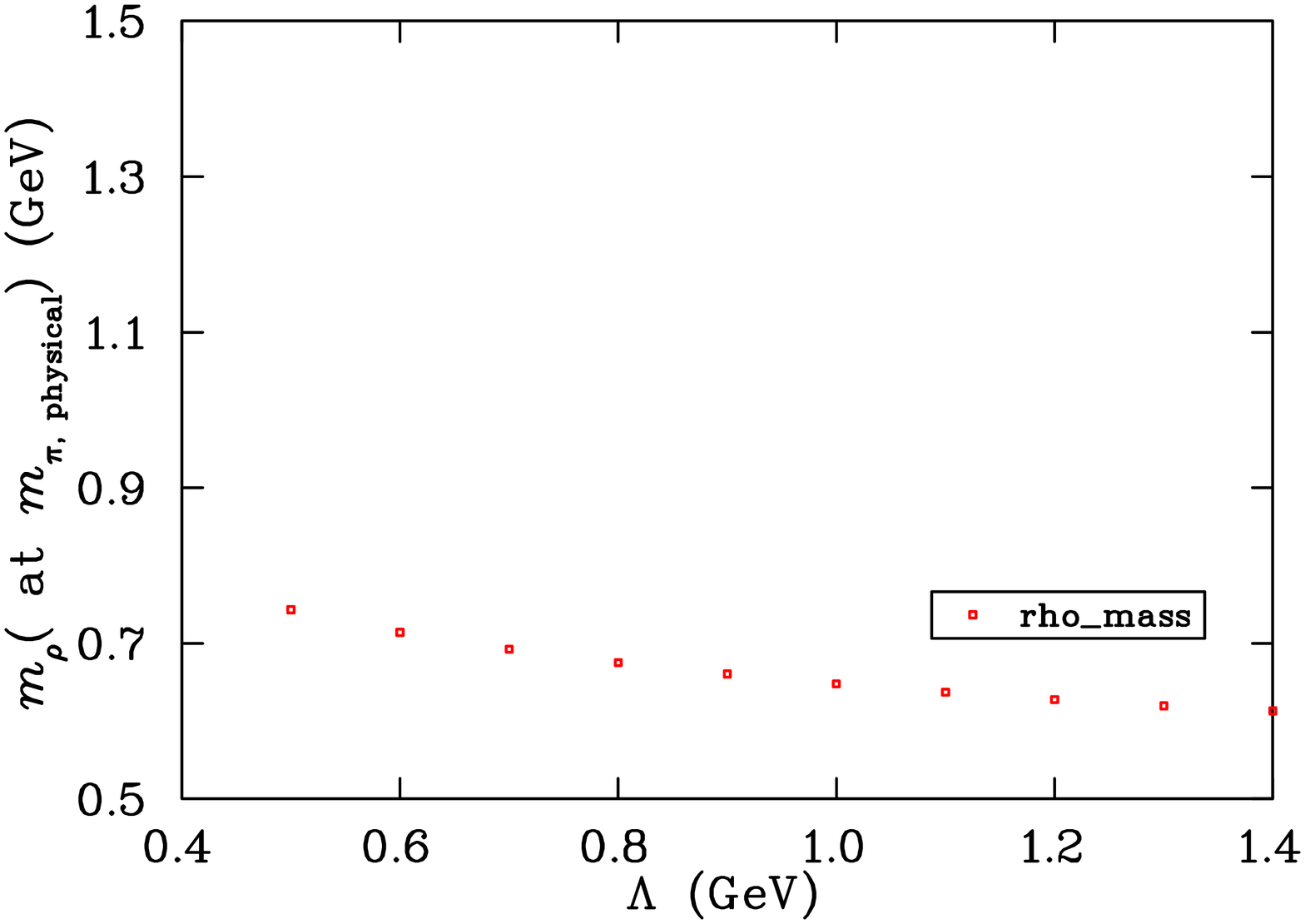}
\caption{\footnotesize{$m_{\rho}$ at $m_{\pi\,,physical}$ vs. $\Lambda$ on Aoki \cite{Aoki:2002uc}}}
\label{fig:mrhoimage}
\end{center}


\centering
\includegraphics[height=280pt,angle=90]{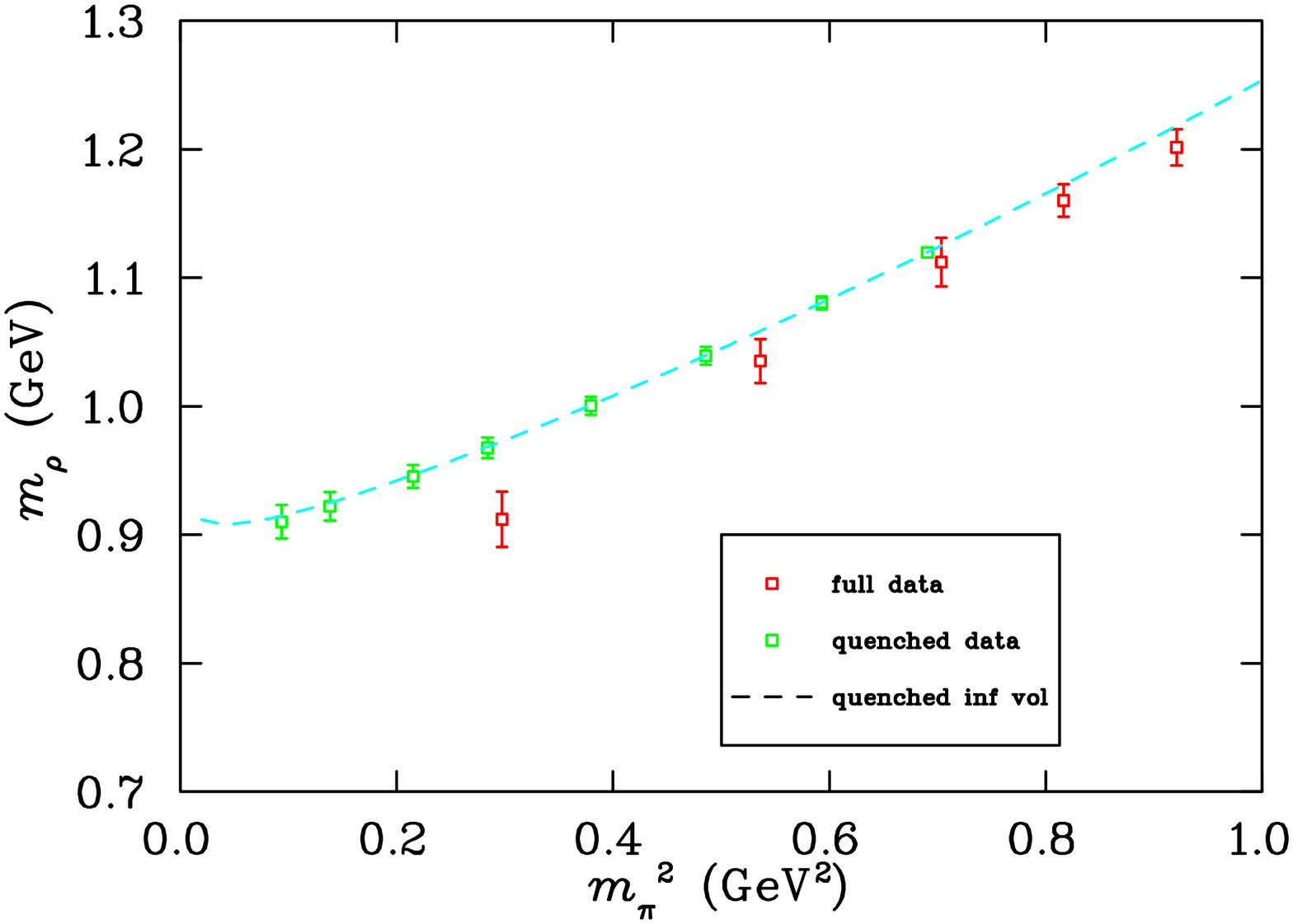}
\caption{\footnotesize{Comparison of Boinepalli and Aoki for $\Lambda$ = $0.8$ GeV}}
\label{fig:bL0.8}
\end{figure}

In full QCD, the value of $m_{\rho}$ at the physical value of $m_{\pi}^2$ will not be strongly $\Lambda$ dependent, except for large values of $\Lambda$ of the order of $10$ GeV or more, as shown in Fig.~\ref{fig:mrhoimage}. This is further evidence from full QCD calculations that using DR in $\chi$PT will over-estimate the $\rho$ mass.

In order to identify a value for $\Lambda$ which corresponds to a stable unquenching procedure, compare the QQCD data from Boinepalli to the full QCD data from Aoki. It is sufficient to compare the Boinepalli data to the Aoki data, since it has been demonstrated already in Fig.~\ref{fig:qL1.0size} and Fig.~\ref{fig:q2L1.0size} that the finite-volume effects are negligible for box sizes for $L_{x} > 3.0$ fermi. The Zanotti data need not be compared to the Aoki data in addition to the Boinepalli data. In addition, all subsequent graphs calculate the integrals for the $\rho \rightarrow \pi\pi$ process using $\Lambda = 0.6$ GeV, in accordance with Allton \cite{Allton:2005fb}.

\begin{figure}
\begin{center}
 \includegraphics[height=280pt,angle=90]{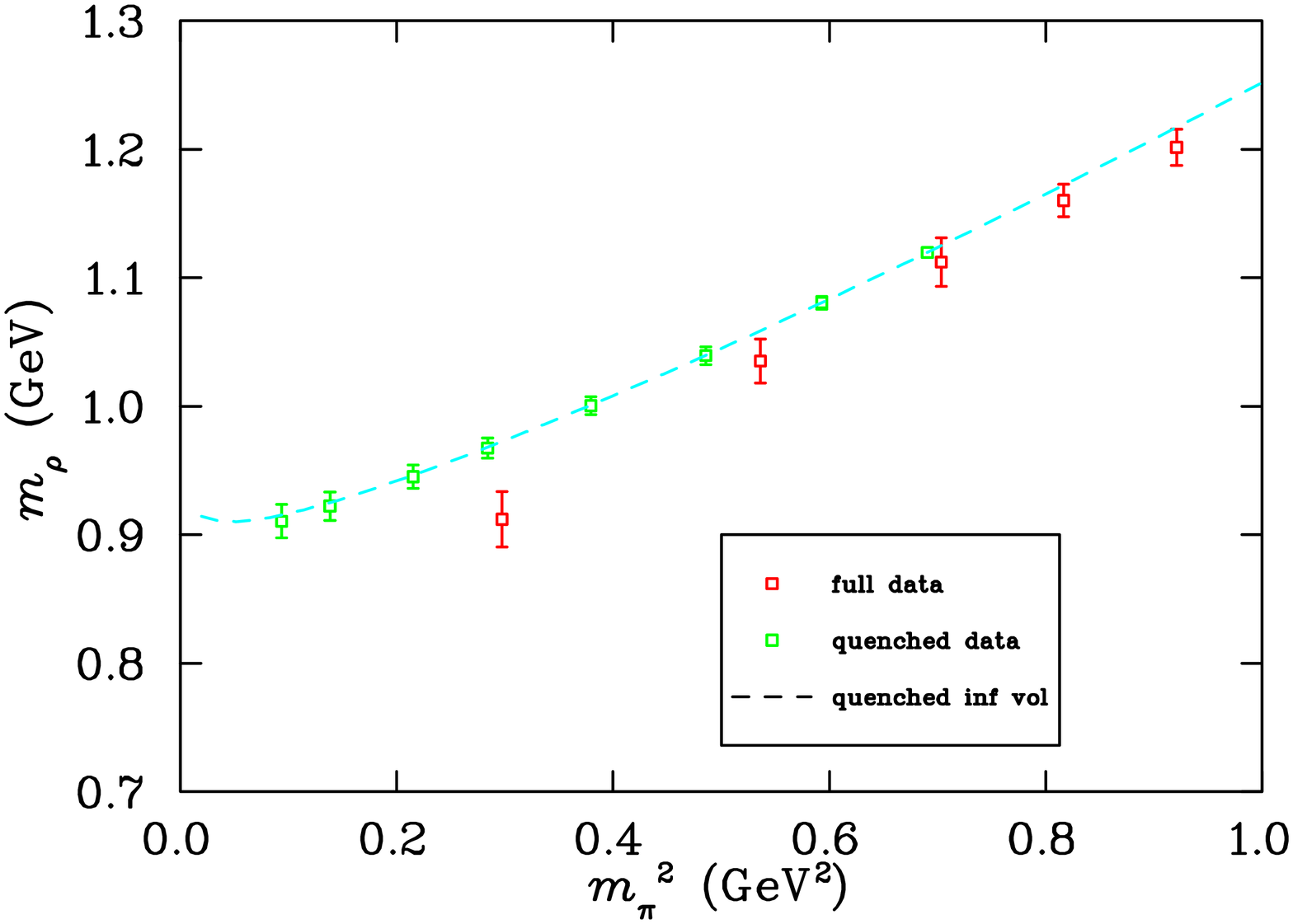}
\caption{\footnotesize{Comparison of Boinepalli and Aoki for $\Lambda$ = $0.9$ GeV}}
\label{fig:bL0.9}
\end{center}

\centering
\includegraphics[height=280pt,angle=90]{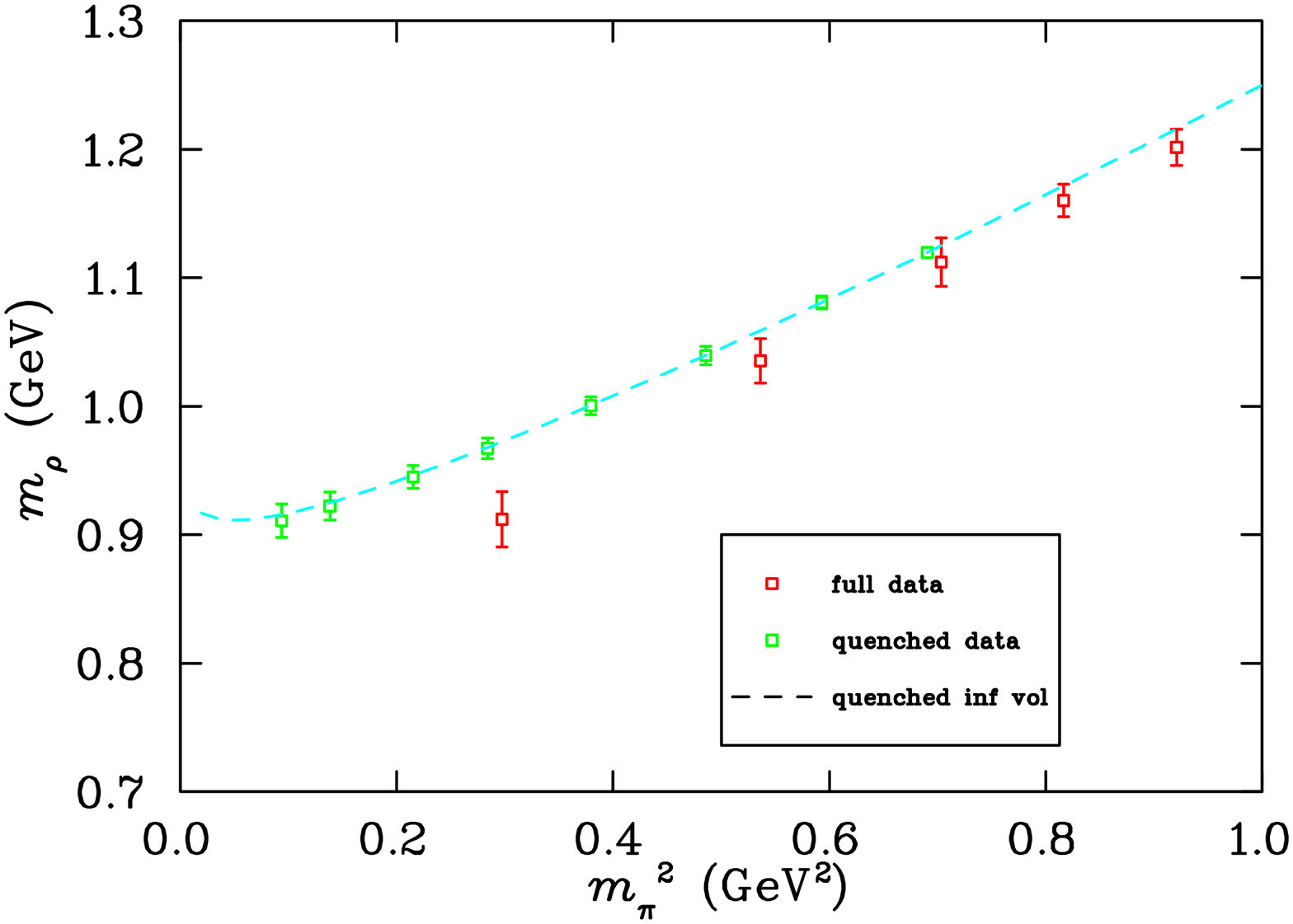}
\caption{\footnotesize{Comparison of Boinepalli and Aoki for $\Lambda$ = $1.0$ GeV}}
\label{fig:bL1.0}



\begin{center}
 \includegraphics[height=280pt,angle=90]{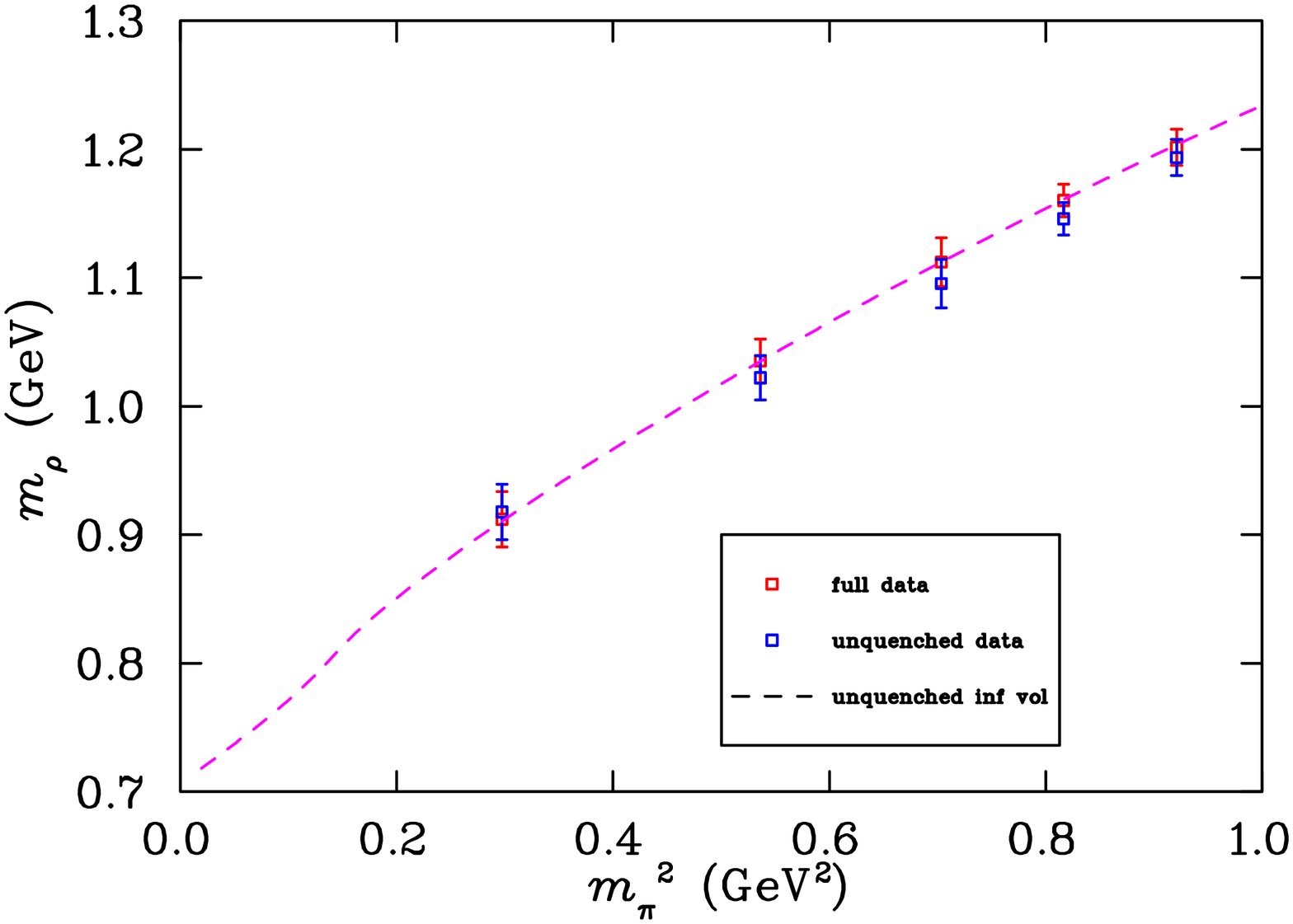}
\caption{\footnotesize{Unquenched Fit against Full QCD Data from Aoki for $\Lambda$ = $0.9$ GeV}}
\label{fig:unquench}
\end{center}
\end{figure}

  Fig.~\ref{fig:bL0.8} to Fig.~\ref{fig:bL1.0}  show the data compared for different values of $\Lambda$. Clearly $\Lambda = 0.9$ GeV is the best match between the QQCD data and the full QCD data. Choosing this value of $\Lambda$ roughly fixes the coefficients in the polynomial expansion for both QQCD and full QCD, as shown in Fig.~\ref{fig:bL0.9}.


In order to demonstrate the usefulness of this value of $\Lambda$, consider the process of \emph{unquenching}. This means the full QCD loop integral contributions are added to quenched data, using the (approximately equal) quenched coefficients at $\Lambda = 0.9$ GeV instead. The unquenched fit is shown in Fig.~\ref{fig:unquench}.


\chapter{Conclusion}
\label{chpt:Conclusion}



\section{Codetta}
\textit{``[A] flash of insight, actually changes the system to make it more coherent.''}
(Bohm, D. 1994 Edition. Thought as a System p.182 \cite{Bohm})

In theoretical physics, alternative insights require a coherent framework in which to place research data. This thesis uses existing theoretical insights presented in the literature review as a basis for realizing new theoretical approaches to research data. The main contributions of this thesis are summarized as follows:

From the literature review, Goldstone's Theorem for chiral symmetry breaking in the context of Quantum Chromodynamics (QCD) was discussed. This led to the formation of a low energy perturbation theory. Finite-Range Regularization was reviewed as a useful scheme and thus Chiral Effective Field Theory was constructed, using a regulator parameter $\Lambda$. Lattice QCD was evaluated as a non-perturbative scheme suitable for large masses, and improved actions were discussed in the context of experimental lattice calculations.

In the research component of the thesis, the Effective Field Theory polynomial expansion of the $\rho$ mass was provided, and the non-analytic contributions to the self-energy were presented: both as diagrams, and as loop integral expressions. Therefore, actual lattice data could be analyzed using this scheme, and coefficients identified.

\begin{figure}
\centering
\includegraphics[height=320pt,angle=90]{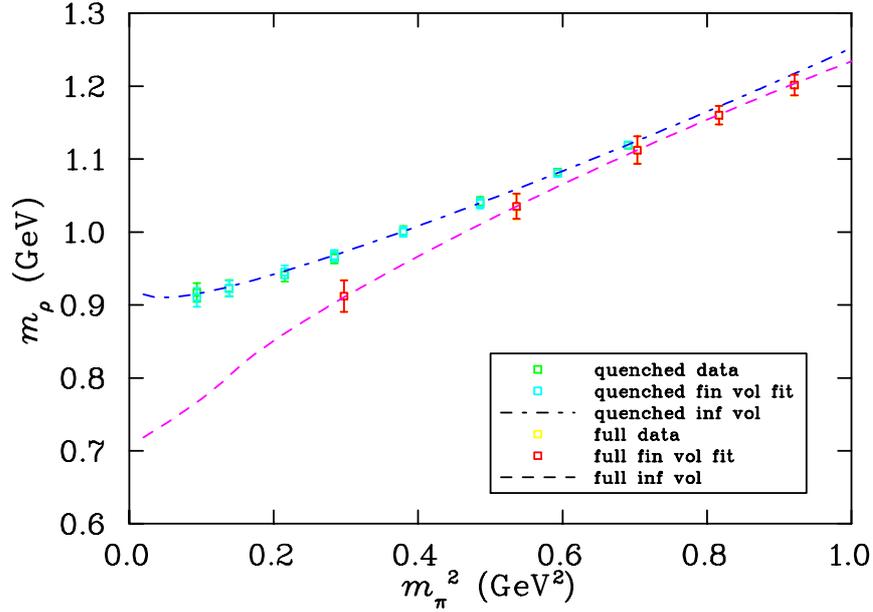}
\caption{\footnotesize{Quenched and Full QCD Data with Infinite Volume Curves}}
\label{fig:finalimage}
\end{figure}

Finite-volume effects on the lattice were quantified for both quenched QCD and full QCD. Curves were fitted to data from Boinepalli \cite{Boinepalli:2004fz}, Zanotti \cite{Zanotti:2001yb} and Aoki \cite{Aoki:2002uc}, based on the coefficients calculated. The Boinepalli data and the Aoki data were compared for different values of the parameter $\Lambda$. A value of $\Lambda$ was chosen so that the data were well correlated. A diagram of the quenched QCD and full QCD data, for $\Lambda = 0.9$ GeV, with finite-volume fitted data points, is shown in Fig.~\ref{fig:finalimage}. This diagram also shows the extrapolated infinite-volume curves. This result shows the finite-volume fitted data exactly coinciding with the actual data. The infinite-volume curves also go through the data, however, the physical value of the $\rho$ meson mass is extrapolated to $0.718$ GeV. The statistical error for the lightest quark mass from Aoki \cite{Aoki:2002uc}, $\delta m_{\rho} = 0.0217$ GeV. Therefore, the extrapolated upper and lower error estimates can be calculated by shifting the smallest data point by this amount before extrapolation. The values obtained are $0.779$ GeV and $0.657$ GeV, respectively. This estimate is valid assuming the data is highly correlated. Since the physical value of the $\rho$ meson is $0.770$ GeV, the unquenching procedure was successful.

The value $\Lambda = 0.6$ GeV was needed for calculations of the $\rho\rightarrow \pi\pi$ process, in order to successfully unquench the results. This was a modelling decision based on Allton \cite{Allton:2005fb}. Thus no Chiral Effective Field Theory exists for unstable particles which involve the $\pi\pi$ decay channel. In the process, $\rho \rightarrow \pi\pi$, the pions can have substantial momenta. In order to construct a Chiral Lagrangian, a reliable low energy expansion is needed, and this cannot be found.

\section{Evaluation of Original Research}

In order to improve the analysis process, more data sets from different research groups could have been examined. This would improve the precision of the final resultant $\Lambda$ value calculated.  More data points in each data set, especially data points near the physical pion mass, would also improve the precision of the final result. Limitations on available time, and the maximum allowable length of the discourse, prevented further data acquisition. Further theoretical knowledge would also have added depth to the interpretation of the final result.

\section{Further Developments}
Future research developments include the generation of lattice data myself, utilizing the super-computer resources at the University of Adelaide. Various other observable quantities of vector mesons, such as width and charge radius, can be examined using the same technique. This would serve to verify the integrity of the research presented in this thesis.

\newpage
\section{Concluding Statement}
Chiral Effective Field Theory was used to calculate the mass of the $\rho$ meson, as a polynomial expansion in $m_{\pi}^2$. A value for the finite-range regulator parameter $\Lambda$ was calculated, so that the coefficients of the expansion in both quenched and full Quantum Chromodynamics, were matched. Thus a successful unquenching procedure was sought to enable quenched lattice calculations to be used to predict the full Quantum Chromodynamics results for the $\rho$ meson mass. This was achieved, but a Chiral Effective Field Theory could not be constructed for such a particle involving unstable processes, and thus the result was model dependent. 

\thispagestyle{plain}

%
%

\appendix

\chapter{Formalism and Convention}
\label{app:formalism}

\section{Full QCD $\rho$ Meson Self Energies}
\label{app:loopint}

Each of the two processes $\rho \rightarrow \pi \omega$ and $\rho \rightarrow \pi \pi$ are important contributions to the $\rho$ meson self energy. It is important to discuss how the final integral expressions used in Chapter \ref{chpt:Results} are derived.  The method will follow that of Wright \cite{Stewart}, and we will explicitly work out the $\rho \rightarrow \pi\pi$ contribution.  The derivation of the $\rho \rightarrow \pi\omega$ process is similar and is not explicitly derived here, although the final result is given.

The contributing interaction terms from the total chiral Lagrangian are:

\begin{eqnarray}
\mathcal{L}_{\rho\pi\omega} &=& g_{\rho\pi\omega}\epsilon_{\mu\nu\alpha\beta}\left ( \partial^{\mu} \omega^{\nu} \right) \left( \partial^{\alpha} \vec{\rho}^{\beta} \right) \cdot \vec{\pi} \, , \\
\nn \\
\mathcal{L}_{\rho\pi\pi} &=& \frac{1}{2}f_{\rho\pi\pi} \vec{\rho}^{\mu} \cdot \left(\vec{\pi} \times \left(\partial_{\mu}\vec{\pi} \right ) - \left(\partial_{\mu}\vec{\pi} \right ) \times \vec{\pi} \right)  \\
\nn \\
&=& f_{\rho\pi\pi}\epsilon_{abc}\rho^{\mu}_{a}\pi_{b} \left(\partial_{\mu}\pi_{c} \right) \, .
\label{eqn:chirallagterms}
\end{eqnarray}

Therefore, the self energy contribution of the $\rho$ meson can be written, using the formalism conventions used by Pichowsky \textit{et al} \cite{Pichowsky:1999mu}:

\begin{eqnarray}
\Pi_{jj'} = \langle\rho_{j'};p_{\rho}',\lambda \mid T \Big\{\frac{i^2}{2} \int \! \ud^4x \ud^4y \left(f_{\rho\pi\pi}\epsilon_{abc}\rho^{\mu}_{a}\pi_{b}(x) \left(\partial_{\mu}\pi_{c}(x) \right) \right ) \nn \\
 \left (f_{\rho\pi\pi}\epsilon_{a'b'c'}\rho^{\nu}_{a'}\pi_{b'}(y) \left(\partial_{\nu}\pi_{c'}(y) \right) \right ) \Big\} \mid \rho_{j};p_{\rho},\lambda\rangle \\
\nn \\
= -\frac{1}{2}f_{\rho\pi\pi}^2\epsilon_{abc}\epsilon_{a'b'c'} \int \! \ud^4x \ud^4y \langle\rho_{j'};p_{\rho}',\lambda \mid T \left \{\rho^{\mu}_{a}\pi_{b}(x) \left(\partial_{\mu}\pi_{c}(x) \right) \right . \nn \\
\left . \rho^{\nu}_{a'}\pi_{b'}(y) \left(\partial_{\nu}\pi_{c'}(y) \right) \right \} \mid \rho_{j};p_{\rho},\lambda\rangle \, .
\label{eqn:selfen}
\end{eqnarray}

Now Wick's Theorem can be employed to write down the possible non-vanishing contractions of this expression. There are four such contractions in total, but two are equal, and the remaining two are also equal to each other.  Therefore, we can write the self energy as twice the sum of only two Wick contractions:

\begin{eqnarray}
\Pi_{jj'} = - f_{\rho\pi\pi}^2\epsilon_{abc}\epsilon_{a'b'c'} \int \! \ud^4x \ud^4y \times \nn \\
\nomathglue{\contraction{  \Big\{ \langle\rho_j';} {p} {{}_{\rho}',\lambda \,|\, } {\rho}
\contraction[2ex]{  \Big\{ \langle\rho_j';p_{\rho}',\lambda  \,|\, \rho^{\mu}_{a}}{\pi}{{}_{b}(x) (\partial_{\mu}\pi_{c}(x) ) \rho^{\nu}_{a'}}{\pi}
\contraction[3ex]{ \Big\{ \langle\rho_j';p_{\rho}',\lambda  \,|\, \rho^{\mu}_{a}\pi_{b}(x) (\partial_{\mu}  }{\pi}{ {}_{c}(x) ) \rho^{\nu}_{a'}\pi_{b'}(y) (\partial_{\nu}}{\pi}
\contraction{  \Big\{ \langle\rho_j';p_{\rho}',\lambda  \,|\, \rho^{\mu}_{a}\pi_{b}(x) (\partial_{\mu}\pi_{c}(x) )} {\rho} {{}^{\nu}_{a'}\pi_{b'}(y) (\partial_{\nu}\pi_{c'}(y) )\,|\,} {\rho}
 \Big\{ \langle\rho_j';p_{\rho}',\lambda \,|\,\rho^{\mu}_{a}\pi_{b}(x) (\partial_{\mu}\pi_{c}(x) ) \rho^{\nu}_{a'}\pi_{b'}(y) (\partial_{\nu}\pi_{c'}(y) )\,|\, \rho_{j};p_{\rho},\lambda\rangle} \nn \\
\nomathglue{\contraction{+   \langle\rho_{j'};} {p} {{}_{\rho}',\lambda \,|\,} {\rho} 
\contraction[2ex]{+   \langle\rho_{j'};p_{\rho}',\lambda \,|\, \rho^{\mu}_{a}} {\pi} {{}_{b}(x) (\partial_{\mu}\pi_{c}(x) ) \rho^{\nu}_{a'}\pi_{b'}(y) (\partial_{\nu} } {\pi}
\contraction[3ex]{+   \langle\rho_{j'};p_{\rho}',\lambda \,|\, \rho^{\mu}_{a}\pi_{b}(x) (\partial_{\mu} } {\pi} { {}_{c}(x) ) \rho^{\nu}_{a'}} {\pi}
\contraction{+   \langle\rho_{j'};p_{\rho}',\lambda \,|\, \rho^{\mu}_{a}\pi_{b}(x) (\partial_{\mu}\pi_{c}(x) )} {\rho} { {}^{\nu}_{a'}\pi_{b'}(y) (\partial_{\nu}\pi_{c'}(y) )\,|\,} {\rho}
+   \langle\rho_{j'};p_{\rho}',\lambda \,|\, \rho^{\mu}_{a}\pi_{b}(x) (\partial_{\mu}\pi_{c}(x) ) \rho^{\nu}_{a'}\pi_{b'}(y) (\partial_{\nu}\pi_{c'}(y) )\,|\, \rho_{j};p_{\rho},\lambda\rangle  \Big\}}
\nn \\
\nn \\
= - f_{\rho\pi\pi}^2\epsilon_{abc}\epsilon_{a'b'c'} \int \! \ud^4x \ud^4y \epsilon^{\mu *}_{j'}(\lambda)\delta_{aj'}\epsilon^{\nu}_{j}(\lambda)\delta_{a'j}e^{ip_{\rho}\cdot x}e^{-ip_{\rho}'\cdot y} \nn \\
\times \left \{ \delta_{bb'}\delta_{cc'}D_{F}(x - y) \partial^{x}_{\mu}\partial^{y}_{\nu}D_{F}(x - y)  \right . \nn \\
\left . + \delta_{bc'}\delta_{cb'}(\partial^{y}_{\nu}D_{F}(x - y))(\partial^{x}_{\mu}D_{F}(x - y))  \right \}\, .
\label{eqn:selfenwick}
\end{eqnarray}

The function $D_{F}(x - y)$ is the Feynman Propagator as usually defined in a Quantum Field Theory: a Green's Function connecting two interaction coordinates. Thus the expression can be simplified further:

\begin{eqnarray}
\Pi_{jj'} = - f_{\rho\pi\pi}^2\epsilon_{abc}\epsilon_{a'b'c'} \int \! \ud^4x \ud^4y \epsilon^{\mu *}_{j'}(\lambda)\delta_{aj'}\epsilon^{\nu}_{j}(\lambda)\delta_{a'j}e^{ip_{\rho}\cdot x}e^{-ip_{\rho}'\cdot y} \nn \\
\times (-1) \left \{ \delta_{bb'}\delta_{cc'} \int \! \frac{\ud^4k_{1} \ud^4k_{2}}{{(2\pi)}^8} \frac{k_{2\mu}k_{2\nu} e^{-i(k_{1}+k_{2})\cdot(x-y)}}{(k_{1}^2 - m_{\pi}^2 + i\epsilon)(k_{2}^2 - m_{\pi}^2 + i\epsilon)} \right . \nn \\
\left . +  \delta_{bc'}\delta_{cb'} \int \! \frac{\ud^4k_{1} \ud^4k_{2}}{{(2\pi)}^8} \frac{k_{1\nu}k_{2\mu} e^{-i(k_{1}+k_{2})\cdot(x-y)}}{(k_{1}^2 - m_{\pi}^2 + i\epsilon)(k_{2}^2 - m_{\pi}^2 + i\epsilon)}  \right \} \, .
\label{eqn:selfenfprop}
\end{eqnarray}

The sum of the two pion momenta is equal to the momentum of the $\rho$ because of the elasticity of the process. Also, the symmetry of the polarization terms can be used in order to simplify the expression further:

\begin{eqnarray}
k_{1} + k_{2} &=& p_{\rho}  \nn \\
\epsilon_{j}^{*} \cdot k_{2} &=& - \epsilon_{j}^{*} \cdot k_{1} \nn \\
\epsilon_{j}     \cdot k_{2} &=& - \epsilon_{j}     \cdot k_{1} \, .
\end{eqnarray}

The integrals of the variables $y$, $x$ and $k_{2}$ now can both be calculated:
\begin{eqnarray}
\int \! \ud^4y e^{-i(p_{\rho} - k_{1} - k_{2})\cdot y} &=& {(2\pi)}^4\delta^{4}(k_{1} + k_{2} - p_{\rho}) \nn \\
&=& {(2\pi)}^4\delta^{4}(p_{\rho}' - p_{\rho}) \,, \\
\nn \\
\int \! \frac{\ud^4k_{2} \ud^4x}{{(2\pi)}^4}e^{-i(p_{\rho} - k_{1} - k_{2})\cdot x}f(k_{2}) &=& \int \! \ud^4k_{2}\delta^{4}(p_{\rho} - k_{1} - k_{2})f(k_{2}) \nn \\
&=& f(p_{\rho}' - k_{1})  \, .
\label{eqn:selfenint}
\end{eqnarray}

Here, $f$ represents the interior of the total integral over the variable $k_{2}$.

The total self energy integral now reduces to a form with only one single integration parameter $k_{1}$, which we will relabel $k$ for simplicity:

\begin{eqnarray}
\Pi_{jj'} = - f_{\rho\pi\pi}^2\epsilon_{abc}\epsilon_{a'b'c'} {(2\pi)}^4\delta^{4}(p_{\rho}' - p_{\rho})(-1)(\delta_{bb'}\delta_{cc'} - \delta_{bc'}\delta_{cb'}) \nn \\
\times \int \! \frac{\ud^4k}{{(2\pi)}^4} \frac{(\epsilon_{j'}^{*}\cdot k)(\epsilon_{j}\cdot k)}{(k^2 - m_{\pi}^2 + i\epsilon)((p_{\rho}' - k)^2 - m_{\pi}^2 + i\epsilon)} \,.
\label{eqn:selfensing}
\end{eqnarray}

The next step is to calculate the Kronecker Delta Function expression:

\begin{eqnarray}
\epsilon_{j'bc}\epsilon_{jb'c'}(\delta_{bb'}\delta_{cc'} - \delta_{bc'}\delta_{cb'}) &=& \epsilon_{j'bc}\epsilon_{jbc} - \epsilon_{j'bc}\epsilon_{jcb} \nn \\
&=& 2 (\delta_{j'j}\delta_{bb} - \delta_{j'b}\delta_{bj}) \nn \\
&=& 4\delta_{j'j} \nn\,.
\label{eqn:selfenkron}
\end{eqnarray}

Thus the following expression represents the self energy of the process $\rho \rightarrow \pi\pi$:
\begin{eqnarray}
\Pi_{jj'} =  {(2\pi)}^4\delta^{4}(p_{\rho}' - p_{\rho})\delta_{j'j} \times \nn \\
4 f_{\rho\pi\pi}^2 \int \! \frac{\ud^4k}{{(2\pi)}^4} \frac{(\epsilon_{j'}^{*}\cdot k)(\epsilon_{j}\cdot k)}{(k^2 - m_{\pi}^2 + i\epsilon)((p_{\rho}' - k)^2 - m_{\pi}^2 + i\epsilon)} \,, \\
\nn \\
\Sigma^{\rho}_{\pi\pi} = i{(2\pi)}^4\delta(p_{\rho}' - p_{\rho}) \Pi_{jj} \,.
\label{eqn:selfenrel}
\end{eqnarray}

\textbf{Non-Relativistic Approximation}

It is often convenient to assume that the $\rho$ meson is at rest ($p_{\rho} = (m_{\rho},\vec{0})$) during the process. It is also a necessary approximation in order to obtain the closed form expression for the self energy contribution we are seeking. 

The fact that relativistic effects will be ignored may be a possible cause for concern. However, it could be argued that a \textit{low} energy Effective Field Theory is being explored. If relativistic effects become significant in some way, it might cast doubt on whether a low energy expansion can be done. Needless to say, small relativistic corrections would be smaller than the higher order terms in the polynomial series which are already estimated in Chapter \ref{chpt:Results}.

The first step is to sum over all polarization vectors $\epsilon_{j}(\lambda)$:

\begin{eqnarray}
\sum_{\lambda}\left( \epsilon_{j}^{*}(\lambda)\cdot k\right)\left( \epsilon_{j}(\lambda)\cdot k\right) &=& k^{\mu}k^{\nu} \left( -g_{\mu\nu} + \frac{p_{\rho\mu }p_{\rho\nu}}{m_{\rho}^2} \right) \nn \\
&=& \vec{k}^2 \,.
\label{eqn:sumpol}
\end{eqnarray}

The denominator of Eq. (\ref{eqn:selfenrel}) can be rewritten in a way that will aid us in our approximation:

\begin{eqnarray}
(p_{\rho} - k)^2 &=& m_{\rho}^2 - 2 p_{\rho} \cdot k + k^2 \nn \\
&=& k_{0}^2 - 2m_{\rho} k_{0} + m_{\rho}^2 - \vec{k}^2 \,.
\label{eqn:denom}
\end{eqnarray}

Now the self energy has been made into a much simpler expression:

\begin{eqnarray}
-i\Sigma^{\rho}_{\pi\pi} = 4 f_{\rho\pi\pi}^2 \int \! \frac{\ud^4k}{{(2\pi)}^4}\vec{k}^2 \frac{1}{k_{0}^2 - m_{\pi}^2 - \vec{k}^2 + i\epsilon} \nn \\
\frac{1}{k-{0}^2 - 2m_{\rho}k_{0} + m_{\rho}^2 - m_{\pi}^2 - \vec{k}^2 + i\epsilon} \,.
\label{eqn:selfensimpler}
\end{eqnarray}

To reach the final form used in  Chapter \ref{chpt:Results}, the integral over the variable $k_{0}$ must be calculated. If an anti-clockwise contour is chosen in the complex plane, Cauchy's Residue Theorem as in Arfken (p.400)\cite{Arfken} yields the result:

\begin{equation}
\int_{-\infty}^{\infty} \frac{\ud k_{0}}{2\pi} f(k_{0}) = i \sum_{k_{0i}}\mathrm{Res}\,f(k_{0i}) \,,
\label{eqn:cauchy}
\end{equation}
where $k_{0i}$ stands for the $i$ possible n-pole isolated singularities of the function $f$.  As can be seen in Eq. (\ref{eqn:selfensimpler}), there are two such poles, one in each denominator.

The first pole occurs in the denominator:

\begin{eqnarray}
k_{0}^2 - m_{\pi}^2 - \vec{k}^2 + i\epsilon &=& \left(k_{0} - \sqrt{m_{\pi}^2 + \vec{k}^2} + i\epsilon\right)\left(k_{0} + \sqrt{m_{\pi}^2 + \vec{k}^2} + i\epsilon\right) = 0 \nn \\
\nn \\
&=& \left(k_{0} - \omega_{\pi} + i\epsilon\right)\left(k_{0} + \omega_{\pi} + i\epsilon\right)\,,
\label{factorized1}
\end{eqnarray}

with contributing pole:
\begin{equation}
k_{0}^2 = - \omega(|\vec{k}|) + i\epsilon \,.
\label{pole1}
\end{equation}

The second pole occurs in the denominator:

\begin{eqnarray}
k_{0}^2 &-& 2m_{\rho}k_{0}  + m_{\rho}^2 - m_{\pi}^2 - \vec{k}^2 + i\epsilon \nn \\ 
\nn \\
&=& \left(k_{0} - m_{\rho} - \sqrt{m_{\pi}^2 + \vec{k}^2} + i\epsilon\right)\left(k_{0} m_{\rho} + \sqrt{m_{\pi}^2 + \vec{k}^2} + i\epsilon\right) = 0 \nn \\
\nn \\
&=& \left(k_{0} -m_{\rho}- \omega_{\pi} + i\epsilon\right)\left(k_{0} -m_{\rho} + \omega_{\pi} + i\epsilon\right)\,,
\label{factorized2}
\end{eqnarray}

with contributing pole:
\begin{equation}
k_{0}^2 = m_{\rho} - \omega(|\vec{k}|) + i\epsilon \,.
\label{pole2}
\end{equation}

Now the self energy in Eq. (\ref{eqn:selfensimpler}) can be written in the form:

\begin{eqnarray}
-i\Sigma^{\rho}_{\pi\pi} = \frac{i f_{\rho\pi\pi}^2}{3} \int \! \frac{\ud^3k}{{(2\pi)}^3}  \frac{\vec{k}^2}{\omega_{\pi}(k) - \left(\omega_{\pi}^2(k) - m_{\rho}^{2}/4\right)} \nn \\
\nn \\
= \frac{i f_{\rho\pi\pi}^2}{6 \pi^2} \int \! \ud k  \frac{k^4}{\omega_{\pi}(k) - \left(\omega_{\pi}^2(k) - m_{\rho}^{2}/4 \right)} \,. \nn \\
\label{eqn:selfenalmost}
\end{eqnarray}

The self energy for the process $\rho \rightarrow \pi\omega$ can be calculated using the same technique.

The final expressions for the respective self energies now have the approximations that the mass $m_{\rho}$ is equal to the physical value $\mu_{\rho}$:

\begin{eqnarray}
\Sigma^{\rho}_{\pi\pi} = - \frac{ f_{\rho\pi\pi}^2}{6 \mu_{\rho} \pi^2} \int_{0}^{\infty} \! \ud k \frac{k^4 u_{\pi \pi}^{2}(k)}{\omega_{\pi}(k)} \nn \\
\left \{ \frac{1}{\omega_{\pi}(k) + \mu_{\rho}/2 - i \epsilon} - \frac{1}{\omega_{\pi}(k) -\mu_{\rho}/2 - i \epsilon} \right\} \nn \\
\nn \\
= - \frac{f_{\rho\pi\pi}^2}{6 \pi^2} \int_{0}^{\infty} \! \ud k \frac{k^4 u_{\pi\pi}^2(k)}{\omega_{\pi}(k) \left ( \omega_{\pi}^2(k) - \mu_{\rho}^{2}/4 \right)} \,,
\label{eqn:selfenRHOPIPI}
\end{eqnarray}

\begin{eqnarray}
\Sigma^{\rho}_{\pi\omega} = - \frac{(\mu_{\rho}g_{\rho\pi\omega})^2}{6 \pi^2} \int_{0}^{\infty} \! \ud k \frac{k^4 u_{\pi\omega}^2(k)}{\omega_{\pi}(k) - \omega_{\omega}(k) + \mu_{\rho}} \nn \\
\left \{ \frac{1}{(\omega_{\pi}(k) - i \epsilon)(\omega_{\pi}(k) + \omega_{\omega}(k) + \mu_{\rho} - i \epsilon)} \right . \nn \\
\left .  -  \frac{1}{(\omega_{\pi}(k) - i \epsilon)(\omega_{\pi}(k) + \omega_{\omega}(k) - \mu_{\rho} - i \epsilon)} \right\} \nn \\
\nn \\
= - \frac{g_{\rho\pi\omega} \mu_{\rho}^2}{12 \pi^2} \int_{0}^{\infty} \! \ud k \frac{k^4 u_{\pi\omega}^2(k)}{\omega_{\pi}(k)\, ( \omega_\pi(k) +  (\mu_{\omega} - \mu_{\rho}) )} \,. 
\label{eqn:selfenRHOPIOMEGA}
\end{eqnarray}

\pagebreak

\section{Dirac and Pauli Spin Matrices}
\label{app:spin}

The Pauli matrices are usually chosen as such:

\begin{eqnarray}
\tau^1 &=& \pmatrix{0&1\cr 1&0} \nn\\
\nn \\
\tau^2 &=& \pmatrix{0&-i\cr i&0} \nn\\
\nn \\
\tau^3 &=& \pmatrix{1&0\cr 0&-1} \nn
\label{eqn:paulispinmatrices}
\end{eqnarray}

The Dirac representation of the Dirac matrices is one among several used, such as the Weyl/Chiral representation and the Majorana representation. The Dirac representation is as follows:

\begin{eqnarray}
\gamma^0 &=& \pmatrix{\mathbb{I}& 0\cr 0& -\mathbb{I}} \nn\\
\nn \\
\gamma^i &=& \pmatrix{0&\sigma^i\cr -\sigma^i&0} \nn\\
\nn \\
\gamma_5 &=& \pmatrix{0&\mathbb{I}\cr \mathbb{I}&0} \quad = \quad i \gamma^0 \gamma^1 \gamma^2 \gamma^3 \,. \nn
\label{eqn:diracrepresentation}
\end{eqnarray}

All representations of these matrices satisfy the requirement of Clifford Algebra due to the conditions imposed in the derivation of the Dirac Equation \cite{P&S}:

\begin{eqnarray}
\left \{\gamma^\mu, \gamma^\nu \right \} &=& 2 g^{\mu\nu}\,,
\nn \\
\left \{\gamma_5, \gamma^\mu \right \} &=& 0 \,.
\end{eqnarray}

For integration over fermion spinor fields $\psi$ and $\bar{\psi}$, the following rules are adopted \cite{Kizilersu}:

\begin{eqnarray}
\int \! \ud \psi &=& \int \! \ud \bar{\psi} \quad = \quad 0 \nn \\
\nn \\
\int \! \ud \psi_{i} \psi_{j} &=& \int \! \ud \bar{\psi_{i}} 
\bar{\psi_{j}} \quad = \quad \delta_{ij}\,.
\end{eqnarray}

The equal-time canonical anti-commutation relations are:

\begin{eqnarray}
\{\psi(x), \bar{\psi}(y) \}_{x_{0} = y_{0}} &=& \hbar \delta^{3}(\vec{x} - \vec{y})\,, \nn \\
\nn \\
\{\psi(x), \psi(y) \}_{x_{0} = y_{0}} &=& 0 \,.
\end{eqnarray}

The fields take the form \cite{P&S}:

\begin{equation}
\psi(x) = \int \!\frac{\ud^3p}{(2\pi)^3}\frac{1}{\sqrt{2\omega_{\vec{p}}}} \sum_{s} \bigg( a_{\vec{p}}^{s} u^{s}(p) e^{-ip\cdot x} + b_{\vec{p}}^{s \dagger} v^{s} (p) e^{ip \cdot x} \bigg)\,,
\end{equation}
and the canonical anti-commutation relations must satisfy:

\begin{equation}
\{ \psi(x), \bar{\psi}(y)\} = (i\slashed{\partial}_{x} + m)i\Delta(x-y\,;m)\,.
\end{equation}

\chapter{Original Data Tables}
\label{app:data}

\section{QQCD Data based on Boinepalli \cite{Boinepalli:2004fz}}

\subsection{Chiral Expansion Coefficients}
\label{app:qcoeffs}

{\scriptsize
\begin{verbatim}
 a0          a2          a4          lambda
0.879       0.280       0.099       0.500
0.875       0.289       0.092       0.600
0.869       0.303       0.082       0.700
0.861       0.319       0.071       0.800
0.851       0.338       0.058       0.900
0.839       0.359       0.044       1.000
0.826       0.383       0.029       1.100
0.810       0.408       0.013       1.200
0.792       0.435       -0.003       1.300
0.772       0.464       -0.021       1.400
0.748       0.496       -0.040       1.500
0.722       0.532       -0.060       1.600
0.692       0.571       -0.082       1.700
0.658       0.615       -0.107       1.800
0.619       0.665       -0.136       1.900
0.575       0.725       -0.171       2.000
\end{verbatim}
}

\subsection{$\rho$ Mass Predictions against $m_{\pi}$ for Multiple Values of  $\Lambda$}
\label{app:qsigma}

{\scriptsize
\begin{verbatim}
 mpisq    fin_mrho     inf_mrho    lambda    diff
 DATA
0.691       1.120       1.120       0.5       0.00015
0.593       1.080       1.081       0.5       0.00019
0.486       1.039       1.039       0.5       0.00027
0.379       1.001       1.001       0.5       0.00041
0.284       0.968       0.969       0.5       0.00065
0.215       0.946       0.947       0.5       0.00100
0.138       0.922       0.924       0.5       0.00187
0.094       0.909       0.912       0.5       0.00304
 
 EXTRAPOLATIONS
0.090       0.908       0.911       0.5       0.00319
0.062       0.901       0.905       0.5       0.00472
0.040       0.895       0.902       0.5       0.00713
0.020       0.889       0.901       0.5       0.01195
 
 DATA
0.691       1.120       1.120       0.6       0.00016
0.593       1.080       1.081       0.6       0.00021
0.486       1.039       1.040       0.6       0.00031
0.379       1.001       1.001       0.6       0.00048
0.284       0.968       0.969       0.6       0.00078
0.215       0.946       0.947       0.6       0.00122
0.138       0.922       0.925       0.6       0.00232
0.094       0.910       0.913       0.6       0.00382
 
 EXTRAPOLATIONS
0.090       0.909       0.913       0.6       0.00400
0.062       0.901       0.907       0.6       0.00595
0.040       0.895       0.904       0.6       0.00895
0.020       0.890       0.905       0.6       0.01479
 
 DATA
0.691       1.120       1.120       0.7       0.00016
0.593       1.080       1.081       0.7       0.00021
0.486       1.039       1.040       0.7       0.00031
0.379       1.001       1.001       0.7       0.00049
0.284       0.968       0.969       0.7       0.00083
0.215       0.946       0.947       0.7       0.00132
0.138       0.922       0.925       0.7       0.00259
0.094       0.910       0.914       0.7       0.00433
 
 EXTRAPOLATIONS
0.090       0.909       0.913       0.7       0.00454
0.062       0.902       0.909       0.7       0.00680
0.040       0.896       0.907       0.7       0.01026
0.020       0.892       0.909       0.7       0.01687
 
 DATA
0.691       1.120       1.120       0.8       0.00014
0.593       1.081       1.081       0.8       0.00019
0.486       1.039       1.040       0.8       0.00028
0.379       1.001       1.001       0.8       0.00047
0.284       0.968       0.968       0.8       0.00082
0.215       0.945       0.947       0.8       0.00134
0.138       0.922       0.925       0.8       0.00272
0.094       0.910       0.915       0.8       0.00463
 
 EXTRAPOLATIONS
0.090       0.909       0.914       0.8       0.00486
0.062       0.902       0.910       0.8       0.00735
0.040       0.897       0.909       0.8       0.01114
0.020       0.893       0.912       0.8       0.01833
 
 DATA
0.691       1.120       1.120       0.9       0.00011
0.593       1.081       1.081       0.9       0.00016
0.486       1.039       1.040       0.9       0.00025
0.379       1.001       1.001       0.9       0.00043
0.284       0.968       0.968       0.9       0.00078
0.215       0.945       0.947       0.9       0.00132
0.138       0.922       0.925       0.9       0.00276
0.094       0.911       0.915       0.9       0.00479
 
 EXTRAPOLATIONS
0.090       0.910       0.915       0.9       0.00504
0.062       0.903       0.911       0.9       0.00769
0.040       0.899       0.910       0.9       0.01173
0.020       0.895       0.914       0.9       0.01934
 
 DATA
0.691       1.120       1.120       1.0       0.00009
0.593       1.081       1.081       1.0       0.00013
0.486       1.040       1.040       1.0       0.00022
0.379       1.000       1.001       1.0       0.00039
0.284       0.967       0.968       1.0       0.00073
0.215       0.945       0.946       1.0       0.00127
0.138       0.922       0.925       1.0       0.00276
0.094       0.911       0.916       1.0       0.00486
 
 EXTRAPOLATIONS
0.090       0.910       0.915       1.0       0.00512
0.062       0.904       0.912       1.0       0.00789
0.040       0.900       0.912       1.0       0.01212
0.020       0.897       0.917       1.0       0.02005
 
 DATA
0.691       1.120       1.120       1.1       0.00007
0.593       1.081       1.081       1.1       0.00011
0.486       1.040       1.040       1.1       0.00019
0.379       1.000       1.001       1.1       0.00035
0.284       0.967       0.968       1.1       0.00068
0.215       0.945       0.946       1.1       0.00122
0.138       0.922       0.925       1.1       0.00273
0.094       0.911       0.916       1.1       0.00489
 
 EXTRAPOLATIONS
0.090       0.910       0.916       1.1       0.00516
0.062       0.905       0.913       1.1       0.00801
0.040       0.901       0.913       1.1       0.01238
0.020       0.898       0.919       1.1       0.02054
 
 DATA
0.691       1.120       1.120       1.2       0.00006
0.593       1.081       1.081       1.2       0.00009
0.486       1.040       1.040       1.2       0.00016
0.379       1.000       1.001       1.2       0.00031
0.284       0.967       0.968       1.2       0.00064
0.215       0.945       0.946       1.2       0.00118
0.138       0.922       0.925       1.2       0.00269
0.094       0.912       0.917       1.2       0.00490
 
 EXTRAPOLATIONS
0.090       0.911       0.916       1.2       0.00517
0.062       0.905       0.914       1.2       0.00809
0.040       0.902       0.915       1.2       0.01255
0.020       0.900       0.921       1.2       0.02090
 
 DATA
0.691       1.120       1.120       1.3       0.00005
0.593       1.081       1.081       1.3       0.00008
0.486       1.040       1.040       1.3       0.00014
0.379       1.000       1.001       1.3       0.00029
0.284       0.967       0.967       1.3       0.00061
0.215       0.945       0.946       1.3       0.00114
0.138       0.923       0.925       1.3       0.00266
0.094       0.912       0.917       1.3       0.00489
 
 EXTRAPOLATIONS
0.090       0.911       0.916       1.3       0.00516
0.062       0.906       0.914       1.3       0.00813
0.040       0.903       0.916       1.3       0.01268
0.020       0.901       0.922       1.3       0.02117
 
 DATA
0.691       1.119       1.120       1.4       0.00004
0.593       1.081       1.081       1.4       0.00007
0.486       1.040       1.040       1.4       0.00013
0.379       1.000       1.001       1.4       0.00027
0.284       0.967       0.967       1.4       0.00058
0.215       0.944       0.945       1.4       0.00110
0.138       0.923       0.925       1.4       0.00263
0.094       0.912       0.917       1.4       0.00488
 
 EXTRAPOLATIONS
0.090       0.912       0.917       1.4       0.00516
0.062       0.907       0.915       1.4       0.00816
0.040       0.904       0.917       1.4       0.01277
0.020       0.903       0.924       1.4       0.02137
 
 DATA
0.691       1.119       1.119       1.5       0.00004
0.593       1.081       1.081       1.5       0.00006
0.486       1.040       1.040       1.5       0.00012
0.379       1.000       1.001       1.5       0.00025
0.284       0.967       0.967       1.5       0.00055
0.215       0.944       0.945       1.5       0.00107
0.138       0.923       0.925       1.5       0.00260
0.094       0.913       0.918       1.5       0.00486
 
 EXTRAPOLATIONS
0.090       0.912       0.917       1.5       0.00514
0.062       0.908       0.916       1.5       0.00818
0.040       0.905       0.918       1.5       0.01283
0.020       0.904       0.926       1.5       0.02152
 
 DATA
0.691       1.119       1.119       1.6       0.00003
0.593       1.081       1.081       1.6       0.00005
0.486       1.040       1.040       1.6       0.00011
0.379       1.000       1.000       1.6       0.00024
0.284       0.966       0.967       1.6       0.00053
0.215       0.944       0.945       1.6       0.00105
0.138       0.923       0.925       1.6       0.00257
0.094       0.913       0.918       1.6       0.00485
 
 EXTRAPOLATIONS
0.090       0.912       0.917       1.6       0.00513
0.062       0.908       0.916       1.6       0.00819
0.040       0.906       0.919       1.6       0.01288
0.020       0.905       0.927       1.6       0.02163
 
 DATA
0.691       1.119       1.119       1.7       0.00003
0.593       1.082       1.082       1.7       0.00005
0.486       1.040       1.040       1.7       0.00010
0.379       1.000       1.000       1.7       0.00023
0.284       0.966       0.967       1.7       0.00052
0.215       0.944       0.945       1.7       0.00103
0.138       0.923       0.925       1.7       0.00255
0.094       0.913       0.918       1.7       0.00483
 
 EXTRAPOLATIONS
0.090       0.913       0.918       1.7       0.00512
0.062       0.909       0.917       1.7       0.00819
0.040       0.907       0.920       1.7       0.01291
0.020       0.907       0.928       1.7       0.02172
 
 DATA
0.691       1.119       1.119       1.8       0.00003
0.593       1.082       1.082       1.8       0.00005
0.486       1.040       1.040       1.8       0.00010
0.379       1.000       1.000       1.8       0.00022
0.284       0.966       0.967       1.8       0.00051
0.215       0.944       0.945       1.8       0.00101
0.138       0.923       0.925       1.8       0.00253
0.094       0.914       0.919       1.8       0.00482
 
 EXTRAPOLATIONS
0.090       0.913       0.918       1.8       0.00511
0.062       0.910       0.918       1.8       0.00819
0.040       0.908       0.921       1.8       0.01294
0.020       0.908       0.930       1.8       0.02179
 
 DATA
0.691       1.119       1.119       1.9       0.00002
0.593       1.082       1.082       1.9       0.00004
0.486       1.040       1.041       1.9       0.00009
0.379       1.000       1.000       1.9       0.00021
0.284       0.966       0.966       1.9       0.00049
0.215       0.944       0.945       1.9       0.00100
0.138       0.923       0.925       1.9       0.00252
0.094       0.914       0.919       1.9       0.00481
 
 EXTRAPOLATIONS
0.090       0.914       0.919       1.9       0.00510
0.062       0.910       0.919       1.9       0.00819
0.040       0.909       0.922       1.9       0.01296
0.020       0.910       0.931       1.9       0.02184
 
 DATA
0.691       1.119       1.119       2.0       0.00002
0.593       1.082       1.082       2.0       0.00004
0.486       1.041       1.041       2.0       0.00009
0.379       1.000       1.000       2.0       0.00021
0.284       0.966       0.966       2.0       0.00049
0.215       0.943       0.944       2.0       0.00099
0.138       0.923       0.925       2.0       0.00250
0.094       0.915       0.919       2.0       0.00480
 
 EXTRAPOLATIONS
0.090       0.914       0.919       2.0       0.00509
0.062       0.911       0.919       2.0       0.00819
0.040       0.910       0.923       2.0       0.01297
0.020       0.911       0.933       2.0       0.02187
 
\end{verbatim}
}

\subsection{$m_{\rho}$[infinite vol.] - $m_{\rho}$[finite vol.] against $\Lambda$ (Data Pt 8)}
\label{app:difflmpi8}

{\scriptsize
\begin{verbatim}
lambda	    diff	lambda	    diff
0.500       0.004	1.300       0.007
0.600       0.005	1.400       0.007
0.700       0.006	1.500       0.007
0.800       0.006	1.600       0.007
0.900       0.007	1.700       0.007
1.000       0.007	1.800       0.007
1.100       0.007	1.900       0.007
1.200       0.007	2.000       0.007
\end{verbatim}
}

\section{QQCD Data based on Zanotti \cite{Zanotti:2001yb}}

\subsection{Chiral Expansion Coefficients}
\label{app:qcoeffs2}

{\scriptsize
\begin{verbatim}
 a0          a2          a4          lambda
0.851       0.428       -0.030       0.500
0.849       0.431       -0.032       0.600
0.845       0.435       -0.034       0.700
0.840       0.442       -0.037       0.800
0.833       0.451       -0.040       0.900
0.824       0.461       -0.045       1.000
0.813       0.474       -0.050       1.100
0.800       0.489       -0.057       1.200
0.785       0.506       -0.063       1.300
0.767       0.525       -0.071       1.400
0.747       0.546       -0.079       1.500
0.724       0.569       -0.089       1.600
0.697       0.595       -0.099       1.700
0.667       0.625       -0.111       1.800
0.633       0.659       -0.124       1.900
0.594       0.697       -0.140       2.000
\end{verbatim}
}

\subsection{$\rho$ Mass Predictions against $m_{\pi}$ for Multiple Values of  $\Lambda$}
\label{app:qsigma2}

{\scriptsize
\begin{verbatim}
 mpisq    fin_mrho     inf_mrho    lambda    diff
 DATA
0.972       1.238       1.238       0.5       0.00021
0.822       1.182       1.182       0.5       0.00028
0.651       1.117       1.117       0.5       0.00041
0.507       1.060       1.061       0.5       0.00060
0.340       0.993       0.994       0.5       0.00106
 
 EXTRAPOLATIONS
0.090       0.890       0.890       0.5       -0.00074
0.062       0.879       0.878       0.5       -0.00073
0.040       0.869       0.869       0.5       -0.00061
0.020       0.861       0.860       0.5       -0.00041
 
 DATA
0.972       1.238       1.238       0.6       0.00031
0.822       1.182       1.183       0.6       0.00041
0.651       1.117       1.117       0.6       0.00060
0.507       1.060       1.061       0.6       0.00088
0.340       0.993       0.995       0.6       0.00158
 
 EXTRAPOLATIONS
0.090       0.891       0.889       0.6       -0.00215
0.062       0.879       0.877       0.6       -0.00220
0.040       0.870       0.868       0.6       -0.00204
0.020       0.862       0.860       0.6       -0.00174
 
 DATA
0.972       1.238       1.238       0.7       0.00039
0.822       1.182       1.183       0.7       0.00051
0.651       1.117       1.117       0.7       0.00075
0.507       1.060       1.061       0.7       0.00112
0.340       0.993       0.995       0.7       0.00203
 
 EXTRAPOLATIONS
0.090       0.891       0.887       0.7       -0.00461
0.062       0.880       0.876       0.7       -0.00476
0.040       0.871       0.867       0.7       -0.00458
0.020       0.863       0.859       0.7       -0.00417
 
 DATA
0.972       1.238       1.239       0.8       0.00043
0.822       1.182       1.183       0.8       0.00057
0.651       1.117       1.118       0.8       0.00085
0.507       1.060       1.061       0.8       0.00128
0.340       0.993       0.996       0.8       0.00236
 
 EXTRAPOLATIONS
0.090       0.893       0.884       0.8       -0.00817
0.062       0.882       0.873       0.8       -0.00848
0.040       0.873       0.864       0.8       -0.00827
0.020       0.865       0.857       0.8       -0.00776
 
 DATA
0.972       1.238       1.239       0.9       0.00043
0.822       1.182       1.183       0.9       0.00058
0.651       1.117       1.118       0.9       0.00089
0.507       1.060       1.061       0.9       0.00136
0.340       0.993       0.996       0.9       0.00257
 
 EXTRAPOLATIONS
0.090       0.894       0.881       0.9       -0.01278
0.062       0.883       0.870       0.9       -0.01327
0.040       0.875       0.862       0.9       -0.01302
0.020       0.867       0.855       0.9       -0.01237
 
 DATA
0.972       1.238       1.238       1.0       0.00041
0.822       1.182       1.183       1.0       0.00057
0.651       1.117       1.118       1.0       0.00088
0.507       1.060       1.061       1.0       0.00138
0.340       0.994       0.996       1.0       0.00268
 
 EXTRAPOLATIONS
0.090       0.895       0.877       1.0       -0.01831
0.062       0.885       0.866       1.0       -0.01898
0.040       0.877       0.858       1.0       -0.01866
0.020       0.870       0.852       1.0       -0.01784
 
 DATA
0.972       1.238       1.238       1.1       0.00038
0.822       1.182       1.183       1.1       0.00053
0.651       1.117       1.118       1.1       0.00084
0.507       1.060       1.061       1.1       0.00136
0.340       0.994       0.996       1.1       0.00271
 
 EXTRAPOLATIONS
0.090       0.897       0.872       1.1       -0.02462
0.062       0.887       0.862       1.1       -0.02545
0.040       0.879       0.854       1.1       -0.02500
0.020       0.872       0.848       1.1       -0.02397
 
 DATA
0.972       1.238       1.238       1.2       0.00034
0.822       1.182       1.183       1.2       0.00049
0.651       1.117       1.117       1.2       0.00080
0.507       1.060       1.061       1.2       0.00131
0.340       0.994       0.996       1.2       0.00269
 
 EXTRAPOLATIONS
0.090       0.898       0.867       1.2       -0.03158
0.062       0.889       0.857       1.2       -0.03254
0.040       0.882       0.850       1.2       -0.03190
0.020       0.875       0.845       1.2       -0.03058
 
 DATA
0.972       1.238       1.238       1.3       0.00030
0.822       1.182       1.183       1.3       0.00044
0.651       1.117       1.117       1.3       0.00074
0.507       1.060       1.061       1.3       0.00125
0.340       0.994       0.996       1.3       0.00265
 
 EXTRAPOLATIONS
0.090       0.900       0.861       1.3       -0.03908
0.062       0.891       0.851       1.3       -0.04011
0.040       0.884       0.845       1.3       -0.03921
0.020       0.878       0.841       1.3       -0.03753
 
 DATA
0.972       1.238       1.238       1.4       0.00027
0.822       1.182       1.183       1.4       0.00040
0.651       1.117       1.117       1.4       0.00069
0.507       1.060       1.061       1.4       0.00119
0.340       0.994       0.996       1.4       0.00259
 
 EXTRAPOLATIONS
0.090       0.902       0.855       1.4       -0.04702
0.062       0.894       0.845       1.4       -0.04808
0.040       0.887       0.840       1.4       -0.04684
0.020       0.881       0.837       1.4       -0.04472
 
 DATA
0.972       1.238       1.238       1.5       0.00023
0.822       1.182       1.183       1.5       0.00036
0.651       1.117       1.117       1.5       0.00064
0.507       1.060       1.061       1.5       0.00113
0.340       0.994       0.996       1.5       0.00253
 
 EXTRAPOLATIONS
0.090       0.904       0.849       1.5       -0.05533
0.062       0.896       0.839       1.5       -0.05635
0.040       0.890       0.835       1.5       -0.05468
0.020       0.885       0.833       1.5       -0.05205
 
 DATA
0.972       1.238       1.238       1.6       0.00021
0.822       1.183       1.183       1.6       0.00033
0.651       1.117       1.117       1.6       0.00059
0.507       1.060       1.061       1.6       0.00108
0.340       0.994       0.996       1.6       0.00247
 
 EXTRAPOLATIONS
0.090       0.906       0.842       1.6       -0.06396
0.062       0.898       0.833       1.6       -0.06488
0.040       0.893       0.830       1.6       -0.06269
0.020       0.888       0.828       1.6       -0.05947
 
 DATA
0.972       1.238       1.238       1.7       0.00019
0.822       1.183       1.183       1.7       0.00030
0.651       1.117       1.117       1.7       0.00056
0.507       1.059       1.061       1.7       0.00103
0.340       0.994       0.997       1.7       0.00242
 
 EXTRAPOLATIONS
0.090       0.908       0.835       1.7       -0.07285
0.062       0.901       0.827       1.7       -0.07359
0.040       0.895       0.825       1.7       -0.07081
0.020       0.891       0.824       1.7       -0.06693
 
 DATA
0.972       1.238       1.238       1.8       0.00017
0.822       1.183       1.183       1.8       0.00028
0.651       1.117       1.117       1.8       0.00053
0.507       1.059       1.060       1.8       0.00099
0.340       0.994       0.997       1.8       0.00237
 
 EXTRAPOLATIONS
0.090       0.910       0.828       1.8       -0.08197
0.062       0.903       0.821       1.8       -0.08247
0.040       0.899       0.820       1.8       -0.07900
0.020       0.895       0.820       1.8       -0.07437
 
 DATA
0.972       1.238       1.238       1.9       0.00015
0.822       1.183       1.183       1.9       0.00026
0.651       1.117       1.117       1.9       0.00050
0.507       1.059       1.060       1.9       0.00095
0.340       0.994       0.997       1.9       0.00232
 
 EXTRAPOLATIONS
0.090       0.913       0.821       1.9       -0.09128
0.062       0.906       0.815       1.9       -0.09145
0.040       0.902       0.815       1.9       -0.08721
0.020       0.899       0.817       1.9       -0.08178
 
 DATA
0.972       1.238       1.238       2.0       0.00014
0.822       1.183       1.183       2.0       0.00024
0.651       1.117       1.117       2.0       0.00048
0.507       1.059       1.060       2.0       0.00092
0.340       0.994       0.997       2.0       0.00229
 
 EXTRAPOLATIONS
0.090       0.915       0.814       2.0       -0.10073
0.062       0.910       0.809       2.0       -0.10050
0.040       0.906       0.810       2.0       -0.09541
0.020       0.903       0.814       2.0       -0.08910
 
\end{verbatim}
}

\subsection{$m_{\rho}$[infinite vol.] - $m_{\rho}$[finite vol.] against $\Lambda$ (Data Pt 5)}
\label{app:difflmpi5}

{\scriptsize
\begin{verbatim}
lambda	    diff	lambda	    diff
0.500       0.001	1.300       0.001	
0.600       0.001	1.400       0.001
0.700       0.001	1.500       0.001	
0.800       0.001	1.600       0.001
0.900       0.001	1.700       0.001	
1.000       0.001	1.800       0.001
1.100       0.001	1.900       0.001	
1.200       0.001	2.000       0.001
\end{verbatim}
}

\section{Full QCD Data based on Aoki \cite{Aoki:2002uc}}

\subsection{Chiral Expansion Coefficients}
\label{app:fcoeffs}

{\scriptsize
\begin{verbatim}
 a0          a2          a4          lambda
0.754       0.570       -0.078       0.500
0.755       0.591       -0.109       0.600
0.766       0.578       -0.106       0.700
0.786       0.546       -0.089       0.800
0.813       0.502       -0.065       0.900
0.847       0.452       -0.040       1.000
0.887       0.397       -0.013       1.100
0.933       0.340       0.012       1.200
0.985       0.284       0.036       1.300
1.043       0.228       0.057       1.400
\end{verbatim}
}

\subsection{$\rho$ Mass Predictions against $m_{\pi}$ for Multiple Values of  $\Lambda$}
\label{app:fsigma}

{\scriptsize
\begin{verbatim}
 mpisq    fin_mrho    inf_mrho    lambda    diff
 DATA
0.921       1.201       1.201       0.5       0.00000014
0.816       1.161       1.161       0.5       0.00000010
0.703       1.113       1.113       0.5       0.00000008
0.536       1.035       1.035       0.5       0.00000008
0.297       0.912       0.912       0.5       0.00000009
 
 EXTRAPOLATIONS
0.090       0.758       0.775       0.5       0.01669347
0.062       0.750       0.758       0.5       0.00786814
0.040       0.759       0.748       0.5       -0.01075067
0.020       0.720       0.743       0.5       0.02349287
 
 DATA
0.921       1.201       1.201       0.6       0.00000013
0.816       1.160       1.160       0.6       0.00000013
0.703       1.112       1.112       0.6       0.00000014
0.536       1.035       1.035       0.6       0.00000015
0.297       0.912       0.912       0.6       0.00000016
 
 EXTRAPOLATIONS
0.090       0.743       0.760       0.6       0.01701886
0.062       0.731       0.739       0.6       0.00807405
0.040       0.736       0.725       0.6       -0.01109930
0.020       0.690       0.714       0.6       0.02447954
 
 DATA
0.921       1.202       1.202       0.7       0.00000020
0.816       1.160       1.160       0.7       0.00000020
0.703       1.112       1.112       0.7       0.00000021
0.536       1.036       1.036       0.7       0.00000024
0.297       0.912       0.912       0.7       0.00000026
 
 EXTRAPOLATIONS
0.090       0.732       0.749       0.7       0.01726917
0.062       0.716       0.725       0.7       0.00823436
0.040       0.718       0.707       0.7       -0.01137376
0.020       0.667       0.693       0.7       0.02527137
 
 DATA
0.921       1.202       1.202       0.8       0.00000031
0.816       1.160       1.160       0.8       0.00000032
0.703       1.112       1.112       0.8       0.00000032
0.536       1.036       1.036       0.8       0.00000035
0.297       0.912       0.912       0.8       0.00000039
 
 EXTRAPOLATIONS
0.090       0.723       0.741       0.8       0.01747573
0.062       0.705       0.713       0.8       0.00836789
0.040       0.704       0.693       0.8       -0.01160458
0.020       0.649       0.675       0.8       0.02594764
 
 DATA
0.921       1.202       1.202       0.9       0.00000043
0.816       1.160       1.160       0.9       0.00000046
0.703       1.112       1.112       0.9       0.00000048
0.536       1.036       1.036       0.9       0.00000050
0.297       0.912       0.912       0.9       0.00000057
 
 EXTRAPOLATIONS
0.090       0.716       0.733       0.9       0.01765303
0.062       0.695       0.704       0.9       0.00848341
0.040       0.692       0.681       0.9       -0.01180597
0.020       0.634       0.661       0.9       0.02654569
 
 DATA
0.921       1.202       1.202       1.0       0.00000060
0.816       1.160       1.160       1.0       0.00000063
0.703       1.112       1.112       1.0       0.00000064
0.536       1.036       1.036       1.0       0.00000068
0.297       0.911       0.911       1.0       0.00000079
 
 EXTRAPOLATIONS
0.090       0.709       0.727       1.0       0.01780782
0.062       0.687       0.695       1.0       0.00858495
0.040       0.682       0.670       1.0       -0.01198448
0.020       0.621       0.648       1.0       0.02708228
 
 DATA
0.921       1.202       1.202       1.1       0.00000082
0.816       1.159       1.159       1.1       0.00000083
0.703       1.112       1.112       1.1       0.00000088
0.536       1.037       1.037       1.1       0.00000091
0.297       0.911       0.911       1.1       0.00000106
 
 EXTRAPOLATIONS
0.090       0.704       0.722       1.1       0.01794326
0.062       0.680       0.688       1.1       0.00867436
0.040       0.673       0.661       1.1       -0.01214299
0.020       0.610       0.637       1.1       0.02756405
 
 DATA
0.921       1.202       1.202       1.2       0.00000104
0.816       1.159       1.159       1.2       0.00000108
0.703       1.112       1.111       1.2       0.00000112
0.536       1.037       1.037       1.2       0.00000124
0.297       0.911       0.911       1.2       0.00000138
 
 EXTRAPOLATIONS
0.090       0.699       0.717       1.2       0.01806095
0.062       0.673       0.682       1.2       0.00875249
0.040       0.666       0.654       1.2       -0.01228274
0.020       0.600       0.628       1.2       0.02799313
 
 DATA
0.921       1.202       1.202       1.3       0.00000135
0.816       1.159       1.159       1.3       0.00000138
0.703       1.111       1.111       1.3       0.00000145
0.536       1.037       1.037       1.3       0.00000157
0.297       0.911       0.911       1.3       0.00000177
 
 EXTRAPOLATIONS
0.090       0.695       0.713       1.3       0.01816202
0.062       0.668       0.677       1.3       0.00881994
0.040       0.660       0.647       1.3       -0.01240452
0.020       0.591       0.620       1.3       0.02837040
 
 DATA
0.921       1.203       1.203       1.4       0.00000165
0.816       1.159       1.159       1.4       0.00000172
0.703       1.111       1.111       1.4       0.00000178
0.536       1.037       1.037       1.4       0.00000195
0.297       0.911       0.911       1.4       0.00000222
 
 EXTRAPOLATIONS
0.090       0.692       0.710       1.4       0.01824761
0.062       0.664       0.673       1.4       0.00887732
0.040       0.654       0.642       1.4       -0.01250923
0.020       0.584       0.613       1.4       0.02869732
 
\end{verbatim}
}

\pagestyle{plain}
\bibliography{references}

\end{document}